\documentclass[longauth]{aa}
\usepackage{graphicx}
\newcommand{\tls}{{\fontfamily{pcr}\selectfont tls}\,}
\newcommand{\allesfitter}{{\fontfamily{pcr}\selectfont allesfitter}\,}
\newcommand{\sherlock}{{\fontfamily{pcr}\selectfont SHERLOCK}\,}
\newcommand{\matrixtk}{{\fontfamily{pcr}\selectfont MATRIX}\,}
\newcommand{\wotan}{{\fontfamily{pcr}\selectfont wotan}\,}
\newcommand{\triceratops}{{\fontfamily{pcr}\selectfont TRICERATOPS}\,}
\usepackage{txfonts}
\usepackage[colorlinks=true,citecolor=blue]{hyperref}
\begin{document}

   \title{Two super-Earths at the edge of the habitable zone of the nearby M dwarf TOI-2095}

   \author{F.~Murgas\inst{ \ref{iac},\ref{ull}}
          \and
          A.~Castro-Gonz\'{a}lez\inst{ \ref{cabinta} }
          \and
          E.~Pall\'{e}\inst{ \ref{iac},\ref{ull}}
          \and
          F.\,J.~Pozuelos\inst{\ref{iaa}}
          \and 
          S.~Millholland\inst{\ref{kavlimit}}
          \and
          O.~Foo\inst{\ref{kavlimit}}
          \and
          J.~Korth\inst{\ref{lundobs}}
          \and
          E.~Marfil\inst{ \ref{iac},\ref{ull},\ref{hs}}
          \and
          P.\,J.~Amado\inst{\ref{iaa}}
          \and
          J.\,A.~Caballero\inst{ \ref{cabinta} }
          \and
          J. L. Christiansen\inst{\ref{caltech}}
          \and
          D.\,R.~Ciardi\inst{\ref{caltech}}
          \and 
          K.\,A.~Collins\inst{\ref{cfaharvard}}
          \and 
          M.~Di\,Sora\inst{\ref{ccao}}
          \and
          A.~Fukui\inst{\ref{utokyo},\ref{iac}}
          \and 
          T.~Gan\inst{\ref{tcabeijing}}
          \and 
          E.\,J.~Gonzales\inst{\ref{ucsc}}
          \and
          Th.~Henning\inst{\ref{mpiaa}}
          \and
          E.~Herrero\inst{\ref{ieec}}
          \and 
          G.~Isopi\inst{\ref{ccao}}
          \and
          J. M. Jenkins\inst{\ref{nasaames}} 
          \and
          J.~Lillo-Box\inst{ \ref{cabinta} }
          \and
          N.~Lodieu\inst{ \ref{iac},\ref{ull}}
          \and
          R.~Luque\inst{\ref{uchicago}}
          \and 
          F.~Mallia\inst{\ref{ccao}}
          \and
          J.~C.~Morales\inst{\ref{ice},\ref{ieec}}
          \and
          G.~Morello\inst{\ref{inst:chalmers},\ref{iac}}
          \and 
          N.~Narita\inst{\ref{utokyo},\ref{osawa},\ref{iac}}
          \and
          J.~Orell-Miquel\inst{ \ref{iac},\ref{ull}}
          \and
          H.~Parviainen\inst{ \ref{iac},\ref{ull}}
          \and
          M.~P\'erez-Torres\inst{\ref{iaa}}
          \and
          A.~Quirrenbach\inst{\ref{lsw}}
          \and
          A.~Reiners\inst{\ref{iag}}
          \and
          I.~Ribas\inst{\ref{ice},\ref{ieec}}
          \and 
          B.\,S.~Safonov\inst{\ref{sai}}
          \and
          S. Seager\inst{\ref{eapmit},\ref{kavlimit},\ref{aamit}} 
          \and
          R.\,P.~Schwarz\inst{\ref{cfaharvard}}
          \and
          A.~Schweitzer\inst{\ref{hs}}
          \and
          M.~Schlecker\inst{\ref{uarizona}}
          \and 
          I.\,A.~Strakhov\inst{\ref{sai}}
          \and
          S.~Vanaverbeke\inst{\ref{Vereniging},\ref{leuvenmat},\ref{astrolab}}
          \and 
          N.~Watanabe\inst{\ref{utokyo2}}
          \and
          J. N. Winn\inst{\ref{princeton}}
          \and
          M.~Zechmeister\inst{\ref{iag}}
          }

  \institute{
Instituto de Astrof\'isica de Canarias (IAC), E-38205 La Laguna, Tenerife, Spain \label{iac} \\
              \email{fmurgas@iac.es} 
\and
Departamento de Astrof\'isica, Universidad de La Laguna (ULL), E-38206 La Laguna, Tenerife, Spain \label{ull}
\and
Centro de Astrobiolog\'{i}a (CSIC-INTA), ESAC campus, 28692 Villanueva de la Ca\~{n}ada (Madrid), Spain \label{cabinta}
\and
NASA Exoplanet Science Institute-Caltech/IPAC, Pasadena, CA 91125, USA\label{caltech}
\and
University of California Santa Cruz, Santa Cruz CA 95065, USA\label{ucsc}
\and
Sternberg Astronomical Institute, M.V. Lomonosov Moscow State University, 13, Universitetskij pr., 119234, Moscow, Russia\label{sai}
\and
Center for Astrophysics \textbar \ Harvard \& Smithsonian, 60 Garden Street, Cambridge, MA 02138, USA\label{cfaharvard}
\and
Department of Astronomy and Tsinghua Centre for Astrophysics, Tsinghua University, Beijing 100084, China\label{tcabeijing}
\and
Campo Catino Astronomical Observatory, Regione Lazio,
Guarcino (FR), 03010 Italy\label{ccao}
\and
Lund Observatory, Division of Astrophysics, Department of Physics, Lund University, Box 43, 22100 Lund, Sweden\label{lundobs}
\and
Komaba Institute for Science, The University of Tokyo, 3-8-1 Komaba, Meguro, Tokyo 153-8902, Japan\label{utokyo}
\and
Astrobiology Center, 2-21-1 Osawa, Mitaka, Tokyo 181-8588, Japan\label{osawa}
\and
Department of Multi-Disciplinary Sciences, Graduate School of Arts and Sciences, The University of Tokyo, 3-8-1 Komaba, Meguro, Tokyo 153-8902, Japan\label{utokyo2}
\and
Instituto de Astrof\'isica de Andaluc\'ia (IAA-CSIC), Glorieta de la Astronom\'ia s/n, 18008 Granada, Spain\label{iaa}
\and
Max-Planck-Institut für Astronomie, Königstuhl 17, D-69117 Heidelberg, Germany \label{mpiaa}
Department of Astronomy and Astrophysics, University of Chicago, Chicago, IL 60637, USA\label{uchicago}
\and
Astrobiology Research Unit, Universit\'e de Li\`ege, 19C All\`ee du 6 Ao\^ut, 4000 Li\`ege, Belgium\label{uliege}
\and
Space Sciences, Technologies and Astrophysics Research (STAR) Institute, Universit\'e de Li\`ege, 19C All\`ee du 6 Ao\^ut, 4000 Li\`ege, Belgium\label{star_belg}
\and
Landessternwarte, Zentrum für Astronomie der Universität Heidelberg, Königstuhl 12, 69117 Heidelberg, Germany\label{lsw}
\and
Institut f\"ur Astrophysik und Geophysik, Georg-August-Universit\"at, Friedrich-Hund-Platz 1, 37077 G\"ottingen, Germany\label{iag}
\and
Institut de Ciències de l’Espai (CSIC), Campus UAB, c/ de Can Magrans s/n, E-08193 Bellaterra, Barcelona, Spain\label{ice} 
\and
Institut d’Estudis Espacials de Catalunya, E-08034 Barcelona, Spain Spain\label{ieec}
\and
Hamburger Sternwarte, Universität Hamburg, Gojenbergsweg 112, 21029 Hamburg, Germany\label{hs}
\and
Department of Physics and Kavli Institute for Astrophysics and Space Research, Massachusetts Institute of Technology, Cambridge, MA 02139, USA\label{kavlimit}
\and
Steward Observatory and Department of Astronomy, The University of Arizona, Tucson, AZ 85721, USA\label{uarizona}
\and
Vereniging Voor Sterrenkunde (VVS), Oostmeers 122 C, 8000 Brugge, Belgium\label{Vereniging}
\and
Centre for Mathematical Plasma-Astrophysics, Department of Mathematics, KU Leuven, Celestijnenlaan 200B, 3001 Heverlee, Belgium\label{leuvenmat}
\and
Public Observatory ASTROLAB IRIS, Provinciaal Domein “De Palingbeek”, Verbrandemolenstraat 5, 8902 Zillebeke, Ieper, Belgium\label{astrolab}
\and
Department of Earth, Atmospheric, and Planetary Sciences, Massachusetts Institute of Technology, Cambridge, MA 02139, USA\label{eapmit}
\and
Department of Aeronautics and Astronautics, Massachusetts Institute of Technology, Cambridge, MA 02139, USA\label{aamit}
\and
NASA Ames Research Center, Moffett Field, CA, 94035, USA\label{nasaames}
\and
Department of Astrophysical Sciences, Princeton University, Princeton, NJ 08544, USA\label{princeton}
\and
Department of Space, Earth and Environment, Chalmers University of Technology, SE-412 96 Gothenburg, Sweden\label{inst:chalmers}
\and 
}

\date{Received / Accepted}


\abstract{The main scientific goal of \textit{TESS} is to find planets smaller than Neptune around stars that are bright enough to allow for further characterization studies. Given our current instrumentation and detection biases, M dwarfs are prime targets in the search for small planets that are in (or near) the habitable zone of their host star. In this work, we use photometric observations and CARMENES radial velocity (RV) measurements to validate a pair of transiting planet candidates found by \textit{TESS}. The data were fitted simultaneously, using a Bayesian Markov chain Monte Carlo (MCMC) procedure and taking into account the stellar variability present in the photometric and spectroscopic time series. We confirm the planetary origin of the two transiting candidates orbiting around TOI-2095 (LSPM J1902+7525). The star is a nearby M dwarf ($d = 41.90 \pm 0.03$\,pc, $T_{\rm eff} = 3759 \pm 87$\,K, $V = 12.6$\,mag), with a stellar mass and radius of $M_\star = 0.44 \pm 0.02 \; M_\odot$ and $R_\star = 0.44 \pm 0.02 \; R_\odot$, respectively. The planetary system is composed of two transiting planets: TOI-2095b, with an orbital period of $P_b = 17.66484 \pm (7\times 10^{-5})$ days, and TOI-2095c, with $P_c = 28.17232 \pm (14\times 10^{-5})$ days. Both planets have similar sizes with $R_b = 1.25 \pm 0.07 \; R_\oplus$ and $R_c = 1.33 \pm 0.08 \; R_\oplus$ for planet b and planet c, respectively. Although we did not detect the induced RV variations of any planet with significance, our CARMENES data allow us to  set stringent upper limits on the masses of these objects. We find $M_b < 4.1 \; M_\oplus$ for the inner and $M_c < 7.4 \; M_\oplus$ for the outer planet (95\% confidence level). These two planets present equilibrium temperatures in the range of 300--350\,K and are close to the inner edge of the habitable zone of their star.}


   \maketitle
%

\section{Introduction}
Due to their relatively small masses and sizes, M dwarfs are prime targets in the search for small ($R_p \sim 1-4 \; R_\oplus$) planets using current detection techniques. Furthermore, it is likely that the only rocky worlds around main sequence stars whose atmospheres we will be able to study over the next decade are transiting planets detected around late-type stars (e.g., \citealp{Batalha2018}). Among the more than 5400 planets detected to date, fewer than 300 have been found to be transiting around M dwarfs\footnote{NASA Exoplanet Archive: \url{https://exoplanetarchive.ipac.caltech.edu/}}. With such a small sample size, it is not surprising to see that there are close to 30 known exoplanets around M dwarfs that are in the habitable zone (HZ) of their star (\citealp{MartinezRodriguez2019}) -- and only about ten of those planets are transiting (e.g., Kepler-186f, \citealp{Quintana2014}; TRAPPIST-1e, -1f, -1g, \citealp{Gillon2016, Gillon2017}; Kepler-1652b, \citealp{Torres2017}; LHS 1140b, \citealp{Dittmann2017}, \citealp{LilloBox2020}; K2-288B b, \citealp{Feinstein2019}; TOI-700d, \citealp{Rodriguez2020}; LP~890-9~c, \citealp{delrez2022}). There is also the late-K-dwarf HZ desert pointed out by \citet{LilloBox2022} that is of interest. We note that although the true habitability of M dwarfs has been highly debated in great part due to their frequent and energetic flares \citep{2016PhR...663....1S}, it has been shown that stellar activity and magnetic flaring is dramatically diminished for stars earlier than M3\,V. We refer to \citealp{2010AJ....140.1402H, Jeffers2018}; and \citealp{Gunther2020} for a more recent analysis using \textit{TESS} data.

Due to the possibility of detecting biomarkers in their atmospheres using transmission spectroscopy, this select group of transiting planets will be the subject of intensive observing campaigns with facilities such as JWST (\citealp{Gardner2006}). Temperate planets located at the predicted edges of the HZ of their host stars are also excellent targets for detailed atmospheric studies to better understand processes such as the runaway greenhouse effect~\citep{Ingersoll1969,Kasting1988} as well as to test the predicted theoretical limits of the habitable zone (e.g., \citealp{Kopparapu2013}, \citealp{Zsom2013}, \citealp{Turbet2019}).

In addition to their prospects for atmospheric characterization, studying small planets can help improve the understanding of the global picture of planet populations found in our galaxy. \cite{Fulton2017} showed that the size distributions of small planets with orbital periods of less than 100 days present a bimodal distribution, presumably representing two general groups: rocky planets with none or small scale atmospheres ($R_p \sim 1 - 2 \; R_\oplus$) and planets with considerable gaseous envelopes and sizes in the range $R_p \sim 2 - 4 \; R_\oplus$. The origin of this separation of planet populations has been attributed to atmospheric mass loss processes (e.g., \citealp{Owen&Wu2013}, \citealp{Lopez&Rice2018}, \citealp{Mordasini2020}). The position of the gap between both distributions (at around $R_p \sim 1.7 \; R_\oplus$) has been found to be dependent on planetary orbital period, the stellar flux received by the planet, and even the stellar type of the host star (e.g., \citealp{Martinez2019}, \citealp{Wu2019}). Recently, \cite{LuquePalle2022} suggested that for planets around M dwarfs, the bimodal distribution of planetary radii is better explained  by density classes, with three populations comprised of rocky, water-rich (i.e., water worlds), and gaseous planets -- rather than by a separation between rocky and gas-rich
planets. To obtain a more comprehensive picture, it is necessary to increase the sample of small transiting planets with mass estimates.

In this work, we report the discovery and validation of two small ($R_p$ $\sim$ $1.2-1.3$ $R_\oplus$) transiting planets around the M dwarf TOI-2095. The planets were found and announced to the community by the Transiting Exoplanet Survey Satellite (\textit{TESS}). The host star is a nearby M dwarf ($d = 41.90 \pm 0.03$ pc, \citealp{Bailer-Jones2021}) that is relatively bright in the near infrared ($V = 12.7$ mag, $J = 9.8$ mag). We use space- and ground-based observations to establish the radius and put upper limits on the masses of both transiting planets. 

This paper is structured as follows. In Section \ref{Sec:TESS_Obs}, we describe \textit{TESS} observations, in Section \ref{Sec:Ground_Obs} we describe the ground-based imaging and spectroscopic observations of the star, and in Section \ref{Sec:StarProperties} we present the stellar properties of the host star. In Section \ref{Sec:AnalysisResults}, we describe our data fitting procedure and the results of our analysis. In Section \ref{Sec:search}, we discuss the search for more transiting planets in \textit{TESS} data, in Section \ref{sec:Dyn}, we discuss the dynamical stability of the system, and in Section \ref{Sec:Discussion}, we present a discussion of the characteristics of the planets around TOI-2095. Finally, in Section \ref{Sec:Conclusions} we present our conclusions.
 
\section{\textit{TESS} photometry}
\label{Sec:TESS_Obs}
\textit{TESS} (\citealp{Ricker2014}) is a NASA space-based observatory dedicated to search the entire sky for new transiting planets around bright stars. The satellite observes an area of the sky of $24^{\circ}\times96^{\circ}$ continuously for $\sim$27 days sending data to Earth every $\sim$13.7 days. The data analysis process of the \textit{TESS} Science Processing Operations Center (SPOC, \citealp{Jenkins2016}) at the NASA Ames Research Center consists of generating light curves using simple aperture photometry (SAP, \citealp{Morris2020}) that are then removed of systematic effects using the Presearch Data Conditioning (PDC) pipeline module (\citealp{Smith2012}, \citealp{Stumpe2012, Stumpe2014}). The resulting photometric time series are then searched to identify transit events with a wavelet-based matched filter (\citealp{Jenkins2002}, \citealp{Li2019}, \citealp{JenkinsJM2020}), and tests are applied to rule out some non-planetary scenarios (\citealp{Twicken2018}). After this process, \textit{TESS} Science Office (TSO) at MIT reviews the vetting reports and which transit-like signatures should be promoted to planet candidate status. On 15 July 2020, the TSO alerted to the community the detection of two distinct transiting signal on TOI-2095. The detected signals have a period of $P = 17.6649$ days with a transit depth of 670 ppm (0.72 mmag) and $P = 28.1723$ days with a depth of 820 ppm (0.89 mmag). The candidates were assigned a \textit{TESS} object of interest (TOI) number of TOI-2095.01 (TOI-2095b, hereafter) and TOI-2095.02 (TOI-2095c from now on) for the $P \sim 17$ day and $P \sim 28$ day signals, respectively. 

In this work, we analyzed a total of 22 \textit{TESS} sectors (PDC-SAP photometry) taken with a 2 minute cadence for the star TOI-2095 (TIC 235678745). The data were taken from 18 July 2019 to 4 July 2020 (Sectors 14 to 26), from 28 June 2021 to 20 August 2021 (Sectors 40 and 41), and from 30 December 2021 to 1 September 2022 (Sectors 47 to 55). We excluded  data from Sector 54  from the analysis,
 since the PDCSAP light curve presented a poor correction of the instrumental systematics when compared to the rest of available observations. The target pixel files showing the \textit{TESS} apertures used to compute the photometric time series are shown in Fig.~\ref{Fig:TESS_TPFs}.

\section{Ground-based observations}
\label{Sec:Ground_Obs}

\subsection{Seeing-limited photometry}

\subsubsection{Long-term photometric monitoring}

TOI-2095 was observed using the 0.8\,m Joan Or\'o Telescope (TJO) at the Observatori Astron\`{o}mic del Montsec (OAdM), Sant Esteve de la Sarga, Catalonia, Spain as a part of a photometric monitoring campaign of \textit{TESS} targets. The objective of the observations was to establish the rotational period of the star from photometric variations. To obtain the data, we used TJO LAIA 4k$\times$4k CCD camera with a field of view of 30$\arcmin$ (pixel scale of 0.4$\arcsec$ pixel$^{-1}$). The photometric observations were obtained using the Johnson $R$ filter, with an exposure time of 120 sec. A standard data reduction was performed with the \texttt{icat} pipeline (\citealp{Colome2006}) and differential aperture photometry was done with \texttt{AstroImageJ} (\citealp{Collins2017}). From TJO, we collected a total of 759 epochs with a time baseline of $\sim$589 days.

We also searched for archival data on public photometric data bases. We obtained Zwicky Transient Facility (ZTF, \citealp{Masci2019}) observations of TOI-2095 taken with the 48-inch Samuel Oschin Telescope at the Palomar Observatory, USA. It uses an array of 16 6k$\times$6k to cover a field of view of 47 square degrees (pixel scale of 1.0$\arcsec$ pixel$^{-1}$) and the data are taken with the Sloan $z$ filter. After removing some outliers using a sigma-clipping procedure (3$\sigma$ threshold), we analyzed a total of 390 epochs spanning a time baseline of $\sim$1333 days.

\subsubsection{OACC transit photometry}
An ingress of TOI-2095b was observed on 23 August 2020 with the 0.35\,m Planewave telescope at the remotely controlled Campo Catino Rodeo Observatory (OACC-Rodeo) located in Rodeo, New Mexico, USA. The telescope is equipped with a FLI KAF50100 camera with $8176 \times 6132$ pixels and a field of view of $17\arcmin$ (pixel scale of 0.48$\arcsec$\,pixel$^{-1}$). The observations were taken with a clear filter and with an exposure time of 180 s. The data reduction and aperture photometry was done using \texttt{AstroImageJ} (\citealp{Collins2017}). The data covered the ingress and almost the full predicted transit. The transit event was too shallow to be detected but the observations discarded eclipsing binaries (EBs) around the target star in a 3$\arcmin$ radius as possible sources of the transit signal detected by \textit{TESS}.

\subsubsection{LCOGT transit photometry}
Two transit events were observed by Las Cumbres Observatory Global Telescope Network (LCOGT, \citealp{Brown2013}) using its 1\,m telescopes. We used the {\tt TESS Transit Finder}, which is a customized version of the {\tt Tapir} software package \citep{Jensen2013}, to schedule our transit observations. LCOGT 1\,m telescopes are equipped with Sinistro cameras with a field of view of $26\arcmin \times 26 \arcmin$ and a pixel scale of 0.389$\arcsec$\,pixel$^{-1}$. A transit of TOI-2095c was observed at the LCOGT McDonald Observatory node in Jeff Davis County, Texas, USA, on the night of 6 September 2020. The observations were made using the Sloan $i'$ filter with an exposure time of 35\,s (Fig. \ref{Fig:Transit_TOI2095c_GroundBased}). A transit of TOI-2095b was observed from Observatorio del Teide, Tenerife, Spain on 21 June 2022. The data were acquired using the Sloan $i'$ band and with an exposure time of 40\,s (Fig. \ref{Fig:Transit_TOI2095b_GroundBased}). The raw images were processed with LCOGT's pipeline {\tt BANZAI} \citep{McCully2018}, which performs standard dark and flat-field corrections. The aperture photometry used to produce the differential light curves was done with \texttt{AstroImageJ}. 

Due to the shallow depth of the transits, the LCOGT light curves could only provide tentative detections of the events. In particular, the transit observations of TOI-2095c present a likely detection, with a transit depth consistent with \textit{TESS} measurements. However, these observations also were useful to rule out the presence of EBs located near the position of TOI-2095 as the origin of the detected signals found by \textit{TESS}. The LCOGT light curves are available in the Exoplanet Follow-up Observation Program (ExoFOP) website\footnote{\url{https://exofop.ipac.caltech.edu/tess/target.php?id=235678745}}.

\subsubsection{MuSCAT3 transit photometry}
A transit event of TOI-2095b was observed with the Multicolor Simultaneous Camera for studying Atmospheres of Transiting exoplanets (MuSCAT3) instrument (\citealp{Narita2020}) mounted at the 2\,m Faulkes Telescope North (FTN) at Haleakal\=a Observatory on Maui, Hawai'i, USA. The instrument has a $9.1\arcmin \times 9.1\arcmin$ field of view (0.27$\arcsec$\,pixel$^{-1}$ scale) and is capable of taking simultaneous images in the Sloan filters $g'$, $r'$, $i'$, and $z_s'$. On the night of 29 May 2021 a transit of TOI-2095b was observed using all the MuSCAT3 filters, the exposure times were set to 15\,s, 35\,s, 30\,s, and 20\,s for $g'$, $r'$, $i'$, and $z_s'$, respectively. The raw science images were calibrated with {\tt BANZAI} and we performed aperture photometry using a custom MuSCAT3 pipeline. Although the transit event could not be detected with a significant degree of certainty, the data were useful to rule out EBs inside \textit{TESS} pixel (Fig.~\ref{Fig:Transit_TOI2095b_GroundBased}).

\subsection{High-resolution imaging}

\subsubsection{Near-infrared adaptive optics imaging}

Observations of TOI-2095 were taken by the NIRC2 instrument on Keck II telescope \citep{wizinowich2000} in the narrow-angle mode with a full field of view of $\sim10\arcsec$ and a pixel scale of approximately $0.0099442\arcsec$ per pixel, on 9 September 2020 in the standard three-point dither pattern. The dither pattern step size was $3\arcsec$ and was repeated twice, with each dither having an offset from the previous dither by $0.5\arcsec$. Observations were taken with the narrow band $K$ filter $(\lambda_o = 2.196; \Delta\lambda = 0.336\,\mu$m) with an exposure time of 0.181\,s.

\begin{figure*}
    \centering
    \includegraphics[width=\hsize]{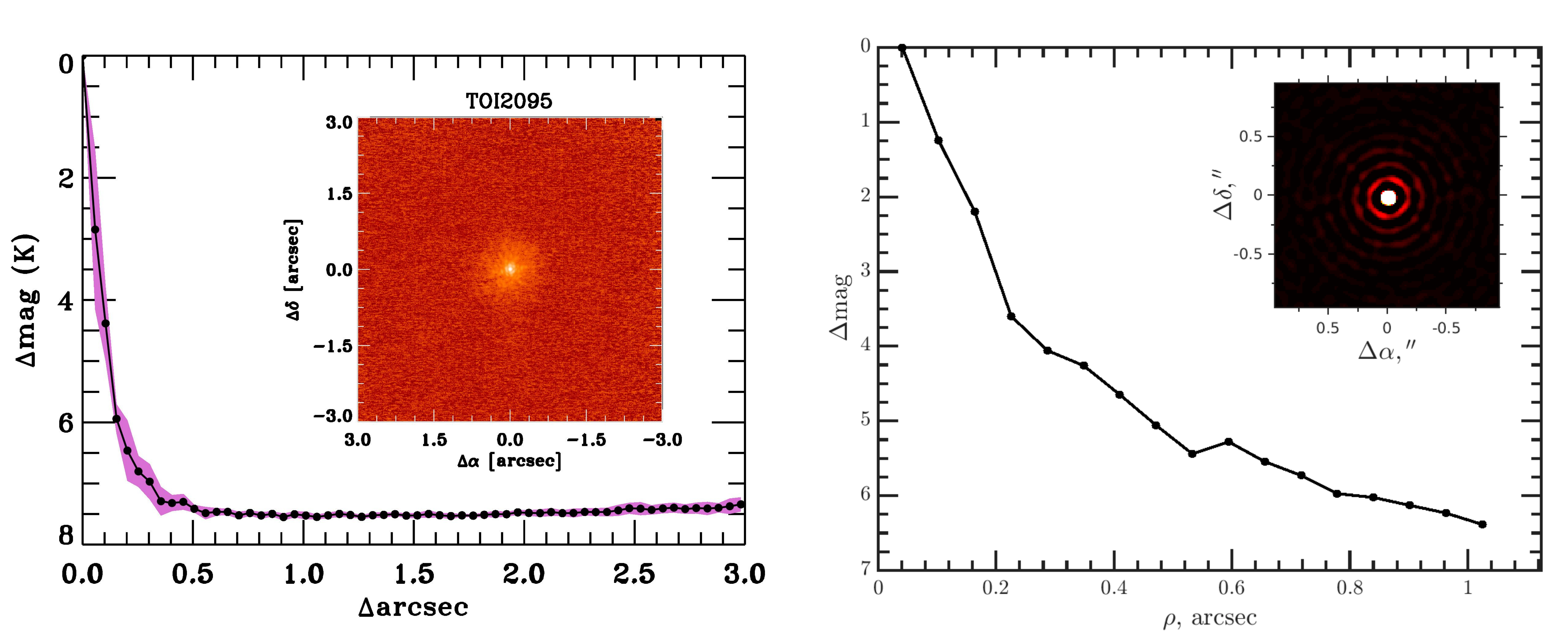}
    \caption{High-resolution imaging of TOI-2095. \textit{Left panel:} Keck NIR AO imaging and sensitivity curve for TOI-2095 taken in the Br-$\gamma$ filter. The image reaches a contrast of $\sim 7.4$ magnitudes fainter than the host star within 0.\arcsec5. {\it Inset:} Image of the central portion of the data. \textit{Right panel:} Speckle sensitivity curve of TOI-2095 taken with the SPeckle Polarimeter (SPP) at the Caucasian Observatory of Sternberg Astronomical Institute (SAI). {\it Inset:} Speckle autocorrelation function.}
    \label{Fig:TOI2095_HRImaging}
\end{figure*}

The sensitivities of the final combined adaptive optics image were determined by azimuthally injecting simulated sources around the primary target every $20^\circ$ at separations of integer multiples of the central source's full width at half maximum (FWHM, \citealp{furlan2017}). The final sensitivity curve is shown in Fig.~\ref{Fig:TOI2095_HRImaging} (left panel). No close-in ($\lesssim 1\arcsec$) stellar companions were detected within the sensitivity limits.

\subsubsection{Optical speckle imaging}
TOI-2095 was observed in $I_c$ band on 26 December 2020 UT with the SPeckle Polarimeter (SPP; \citealp{Safonov2017}) on the 2.5\,m telescope at the Caucasian Observatory of Sternberg Astronomical Institute (SAI) of Lomonosov Moscow State University. As its detector, SPP uses  an electron-multiplying CCD named Andor iXon 897. The detector has a pixel scale of 0.0206$\arcsec$\,pixel$^{-1}$ and a field of view of $5\arcsec \times 5\arcsec$ centered on the star, and the angular resolution was 0.089$\arcsec$. The power spectrum was estimated from 4000 frames with 30\,ms exposure. No stellar companions brighter than $\Delta I_c$ = 3.0\,mag and 6.4\,mag at 0.2$\arcsec$ and 1.0$\arcsec$, respectively, were detected. The obtained sensitivity curve is shown in Fig.~\ref{Fig:TOI2095_HRImaging} (right panel).

\subsubsection{Gaia assessment}
\label{sec:gaia_assesment}

We used \textit{Gaia} DR3 data \citep{GaiaDR3} in order to complement our high-resolution imaging by searching for possible contaminating background, foreground, or bound stars inside or surrounding the SPOC photometric aperture. To do so, we used the \texttt{tpfplotter} package \citep{Aller2020}, which plots the nearby \textit{Gaia} DR3 stars over the TESS target pixel files. In Fig.~\ref{Fig:TESS_TPFs}, we show the \texttt{tpfplotter} outcome for each sector. There are two faint stars that fall inside several SPOC photometric apertures: \textit{Gaia} DR3 2268372103913342080 (star $\#$2, $G = 20.4$) and \textit{Gaia} DR3 2268372477573039744 (star $\#$3, $G = 20.1$). Having a magnitude difference of $\Delta G \sim$ 8\,mag with TOI-2095 in both the \textit{Gaia} and \textit{TESS} bandpasses, each star dilutes the transit depth by a factor of $\sim6\times10^{-4}$, which ensures that none of those stars can be the origin of the transit-like signal. Although negligible in practice, these dilutions are taken into account by the SPOC pipeline, which estimates an overall contamination ratio of $\sim1.08\times10^{-3}$. We also checked the parallaxes and proper motions of the TOI-2095 nearby stars in order to identify possible bound members of the system, without finding matches \citep[see also][]{mugrauer2020,mugrauer2021}.

Additionally, the \textit{Gaia} DR3 astrometry provides information on the possibility of inner companions that may have gone undetected by either \textit{Gaia} or the high-resolution imaging. The \textit{Gaia} renormalised unit weight error (RUWE) is a metric whereby values that are $\lesssim 1.4$ indicate that the \textit{Gaia} astrometric solution is consistent with the star being single, whereas RUWE values $\gtrsim 1.4$ may indicate an astrometric excess noise, possibly caused by the presence of an unseen companion \citep[e.g., ][]{ziegler2020}. TOI-2095 has a \textit{Gaia} DR3 RUWE value of 1.13, meaning that the astrometric fit is consistent with a single star model.

\subsection{CARMENES radial velocity monitoring}

TOI-2095 was observed by the CARMENES spectrograph at the 3.5\,m telescope at the Calar Alto Observatory in Almer\'{i}a, Spain, as part of a follow-up program of \textit{TESS} candidates for validation (22B-3.5-006; PI: E.~Pall\'{e}). CARMENES has two channels: the visible one (VIS), which covers the spectral range $0.52-0.96$\,$\mu$m and the near-infrared one (NIR), which covers $0.96-1.71$\,$\mu$m; with an average spectral resolution for both channels of $R = 94\,600$ and $R = 80\,400$, respectively (\citealp{CARMENES,CARMENES18}).

The star was observed by CARMENES from 1 April 2021 until 7 November 2021, collecting a total of 44 spectra with a time baseline of 219 days. The exposure time used to acquired the radial velocities (RVs) was of 1800\,s. The data reduction was done with the \texttt{CARACAL} pipeline (\citealp{Caballero2016}). \texttt{CARACAL} performs basic data reduction (bias, flat, and cosmic ray corrections) and extracts the spectra using the {\tt FOX} optimal extraction algorithm (\citealp{Zechmeister2014}). The wavelength calibration is done following the algorithms described by \citet{Bauer2015}. The RV measurements were computed with \texttt{SERVAL}\footnote{\url{https://github.com/mzechmeister/serval}}\citep{Zechmeister2018}. \texttt{SERVAL} produces a template spectrum by co-adding and shifting the observed spectra and computes the RV shift relative to this template using a $\chi^2$ minimization with the RV shift as a free parameter. The RV measurements were corrected for barycentric motion, secular acceleration, instrumental drifts, and nightly zero points (for details see, e.g., \citealp{Trifonov2018}, \citealp{Luque2018}). 

\texttt{SERVAL} also computes several activity indices, such as spectral line indices (Na~{\sc i} doublet $\lambda\lambda$589.0\,nm, 589.6\,nm; H$\alpha$ $\lambda$656.2\,nm; and the Ca~{\sc ii} infrared triplet $\lambda\lambda\lambda$849.8\,nm, 854.2\,nm, 866.2\,nm), and other indicators such as the RV chromatic index (CRX) and the differential line width (dLW). These indicators are useful to monitor the stellar activity of the star and its effect on the RV measurements \citep[cf.][]{Zechmeister2018}. Additionally we computed the activity indices from the cross-correlation function (CCF) using the \texttt{raccoon}\footnote{\url{https://github.com/mlafarga/raccoon}} pipeline (\citealp{Lafarga2020}). \texttt{Raccoon} uses weighted binary masks to compute the CCF and obtains RVs and the associated CCF activity indicators: FWHM, contrast (CON), and bisector (BIS).

The median overall signal-to-noise ratio (S/N) for the spectra taken with the visible channel was 297 with a standard deviation of 74 (minimum S/N $=79$, maximum S/N $=394$); for the near-infrared spectra the median S/N was 338 with a standard deviation of 104 (min. S/N $=84$, max. S/N $=469$). The median uncertainty values of the measured RVs were of $3.5$\,m\,s$^{-1}$ ($\sigma_{\rm dev} = 1.5$\,m\,s$^{-1}$) for the VIS channel and $11.7$\,m\,s$^{-1}$ ($\sigma_{\rm dev}=7.3$\,m\,s$^{-1}$) for the NIR channel. The visible and near-infrared \texttt{SERVAL} RV measurements are shown in Fig.~\ref{Fig:TOI2095_RVs_VIS_NIR}. In this work we used the RV data extracted using the visible channel, as this data set presented a lower scatter than the near infrared measurements. Extensive comparisons of the CARMENES VIS and NIR channel RVs have been provided elsewhere, especially by \citet{Reiners2018} and \citet{Bauer2020}. The RV and activity indicators measured with \texttt{SERVAL} can be found in the Appendix (Tables \ref{Tab:CARMENES_RVs} and \ref{Tab:CARMENES_Activ}) and the CCF activity indicators are given in Table \ref{Tab:CARMENES_Activ_CCF}.

\section{Stellar properties}
\label{Sec:StarProperties}

\begin{table*}
\centering
\small
\caption{Stellar parameters of TOI-2095.} \label{Tab:Star}
\begin{tabular}{lcr}
\hline\hline
\noalign{\smallskip}
Parameter                               & Value                 & Reference \\ 
\hline
\noalign{\smallskip}
\multicolumn{3}{c}{Name and identifiers}\\
\noalign{\smallskip}
LSPM                            & J1902+7525                    & {\citet{Lepine2005}}      \\    
Karmn                           & J19025+754                    & {\citet{2016csss.confE.148C}}      \\    
TOI                             & 2095                           & {\it TESS} Alerts      \\  
TIC                             & 235678745                     & {\citet{Stassun2018}}      \\
{\it Gaia} DR3                       & 2268372099615724288      & {\citet{GaiaDR3}}      \\
2MASS & J19023192+7525070           & {\citet{2MASS}}      \\
  
\noalign{\smallskip}
\multicolumn{3}{c}{Coordinates and spectral type}\\
\noalign{\smallskip}
$\alpha$ (J2000)                        & 19:02:31.93           & {\it Gaia} DR3     \\
$\delta$ (J2000)                        & +75:25:06.9           & {\it Gaia} DR3     \\
Spectral type                           & M2.5\,V                & {This work}             \\
\noalign{\smallskip}
\multicolumn{3}{c}{Magnitudes}\\
\noalign{\smallskip}
$B$ [mag]                               & $14.364\pm0.010$       & UCAC4       \\
$g$ [mag]                               & $13.597\pm0.010$       & UCAC4       \\
$G_{BP}$ [mag]                          & $13.111\pm0.003$      & {\it Gaia} DR3 \\
$V$ [mag]                               & $12.676\pm0.010$       & UCAC4       \\
$G$ [mag]                               & $12.086\pm0.003$      & {\it Gaia} DR3 \\
$r$ [mag]                               & $12.099\pm0.010$       & UCAC4       \\
$i$ [mag]                               & $11.205\pm0.010$       & UCAC4       \\
$G_{RP}$ [mag]                          & $11.080\pm0.004$      & {\it Gaia} DR3 \\
$J$ [mag]                               & $9.797\pm0.020$        & 2MASS       \\
$H$ [mag]                               & $9.186\pm0.015$       & 2MASS       \\
$K_s$ [mag]                             & $8.988\pm0.015$       & 2MASS       \\
$W1$ [mag]                              & $8.868\pm0.025$       & AllWISE       \\
$W2$ [mag]                              & $8.766\pm0.024$       & AllWISE       \\
$W3$ [mag]                              & $8.670\pm0.024$       & AllWISE       \\
$W4$ [mag]                              & $8.709\pm0.269$       & AllWISE       \\
\noalign{\smallskip}
\multicolumn{3}{c}{Parallax and kinematics}\\
\noalign{\smallskip}
$\pi$ [mas]                             & $23.86\pm0.01$       & {\it Gaia} DR3             \\
$d$ [pc]                                & $41.90^{+0.02}_{-0.03}$      & {\citet{Bailer-Jones2021}}             \\
$\mu_{\alpha}\cos\delta$ [$\mathrm{mas\,yr^{-1}}$]  & $+203.47 \pm 0.02$ & {\it Gaia} DR3          \\
$\mu_{\delta}$ [$\mathrm{mas\,yr^{-1}}$]            & $-21.40 \pm 0.02$ & {\it Gaia} DR3          \\
$\gamma$ [$\mathrm{km\,s^{-1}}]$           & $-19.94 \pm 0.52$  & {\it Gaia} DR3  \\
$U$ [$\mathrm{km\,s^{-1}}]$             & $-2.43 \pm 0.17$   & {This work}  \\
$V$ [$\mathrm{km\,s^{-1}}]$             & $-1.72 \pm 0.55$   & {This work}  \\
$W$ [$\mathrm{km\,s^{-1}}]$             & $-45.38 \pm 0.27$  & {This work}   \\
\noalign{\smallskip}
\multicolumn{3}{c}{Photospheric parameters}\\
\noalign{\smallskip}
$T_{\mathrm{eff}}$ [K]                      & $3759 \pm 87$         & {This work}   \\
$\log g$                                    & $5.11 \pm 0.12$       & {This work}   \\
{[Fe/H]}                                    & $-0.24 \pm 0.04$      & {This work}   \\
\noalign{\smallskip}
\multicolumn{3}{c}{Physical parameters}\\
\noalign{\smallskip}
$M_\star$ [$M_{\odot}$]                       & $0.44 \pm 0.02$     & {This work}       \\
$R_\star$ [$R_{\odot}$]                       & $0.44 \pm 0.02$     & {This work}       \\
$L$ [$10^{-4}\,L_\odot$]                & $347.6 \pm 1.2$       & {This work}       \\
$P_{\rm rot}$ [days] & $40^{+0.2}_{-0.4}$ & {This work} \\
Age [Gyr] & $>1$ &  {This work} \\
\noalign{\smallskip}
\hline
\end{tabular}
\tablebib{
    {\it Gaia} DR3: \citet{GaiaDR3};
    UCAC4: \citet{UCAC4};
    2MASS: \citet{2MASS}.
    AllWISE: \citet{CutriWISECat2014}
}
\end{table*}

\subsection{Stellar parameters}
Table \ref{Tab:Star} presents the coordinates, photometric and astrometric properties, and stellar parameters of TOI-2095. Throughout the text we use the TESS Object of Interest nomenclature, but the star was discovered almost twenty years ago by \citet{2005AJ....129.1483L} in the Digitized Sky Surveys because of its high proper motion.
Afterwards, the star has been tabulated in a number of catalogs of relatively bright M dwarfs and potential targets for habitable planet surveys (e.g., \citealp{2011AJ....142..138L, 2013MNRAS.435.2161F, 2019ApJ...874L...8K, 2022A&A...665A..30S}).
We derived the effective temperature ($T_{\rm eff}$), surface gravity ($\log g$), and iron abundance ([Fe/H]) of TOI-2095 with the \texttt{SteParSyn}\footnote{\url{https://github.com/hmtabernero/SteParSyn/}} code (\citealp{Tabernero2022}) using the line list and model grid described by \cite{Marfil2021}. The line list of \cite{Marfil2021} makes use of both visible and near-infrared wavelength ranges available in CARMENES data. To obtain a conservative error estimate we used the systematic offset to interferometric effective temperatures of 72\,K given by \cite{Marfil2021} as our systematic error, and added this to the measured stellar effective temperature uncertainty of 15\,K. We estimated the target's spectral type with $\pm$0.5\,dex accuracy from the colour-, absolute magnitude-, and luminosity-spectral type relations of \cite{Cifuentes2020}. The stellar luminosity ($L$) was computed following \cite{Cifuentes2020} and the stellar mass ($M_\star$) and radius ($R_\star$) were determined following \cite{Schweitzer2019}. We found TOI-2095 to be an M dwarf with $T_{\rm eff} = 3759 \pm 87$\,K, and a stellar mass and radius of $M_\star = 0.44 \pm 0.02 \; M_\odot$ and $R_\star = 0.44 \pm 0.02 \; R_\odot$, respectively. 

We used the astrometric properties and RV measurements of TOI-2095 from \textit{Gaia} DR3 (\citealp{GaiaDR3}) to compute the galactocentric velocities $(U,V,W)$ of the star following \cite{CortesContreras2017}. The velocities, presented in Table \ref{Tab:Star}, were computed using a right-handed system and not corrected by the solar motion. The derived galactic velocities indicate that TOI-2095 belongs to the thin disk (\citealp{Montes2001}), suggesting an age $> 1$\,Gyr. According to \texttt{BANYAN $\Sigma$}\footnote{\url{https://www.exoplanetes.umontreal.ca/banyan/banyansigma.php}} (\citealp{Gagne2018}), the $U,V,W$ values indicate that TOI-2095 is not associated to any young moving group, supporting a relatively old age for this star.

Our derived iron abundance for TOI-2095 is $\mathrm{[Fe/H]} = -0.24 \pm 0.04$\,dex. This value is in agreement to what is found for local stars, although TOI-2095 seems to be in the metal poor end compared to the median iron abundances of the Solar neighborhood. \cite{Casagrande2011} found a metallicity distribution function (MDF) slightly subsolar with a median $\mathrm{[Fe/H]} \sim -0.05$\,dex, but the same work pointed out that other studies found an MDF peak in the --0.2 to --0.1\,dex range. For M stars, there have been studies pointing to an MDF with median values close to solar metaillicity (e.g., \citealp{Bonfils2005, Casagrande2008}), although planet-hosting M dwarfs typically appear to be  metal-rich (e.g., \citealp{RojasAyala2010, Terrien2012, Hobson2018, Passegger2018}).

\subsection{Stellar rotation from seeing-limited photometry}
To obtain the rotation period of TOI-2095, we analyzed the photometric time series taken with ZTF ($z$ band) and TJO ($R$ band). We modeled the photometry using a linear function and a periodic term using Gaussian Processes (GP). To model the photometric variations, we used the GP package \texttt{george} (\citealp{Ambikasaran2015}) and chose a quasi-periodic kernel of the form:
\begin{equation}
    k_{ij\; \mathrm{QP}} = A \exp \left[ \frac{-(|t_i-t_j|)^2}{2 l^2} - \Gamma^2 \sin^2\left( \frac{\pi |t_i-t_j|}{P_{\rm rot}} \right) \right]
\label{Eq:Prot_GP}
,\end{equation}
where $|t_i-t_j|$ is the difference between two epochs or observations, the parameters $A$, $l$, and $\Gamma$ are constants, and $P_{\rm rot}$ is the period of the sinusoidal variation (i.e. the rotational period of a star). 

For each photometric time series, we set as free parameters the two terms of the linear function and the constants describing the shape of the kernel from Eq. \ref{Eq:Prot_GP}, while imposing $P_{\rm rot}$ as a common term for both time series. We started the fitting procedure by optimizing a posterior probability function using \texttt{PyDE}\footnote{\url{https://github.com/hpparvi/PyDE}} and we used the optimal set of parameters to start exploring the parameter space using \texttt{emcee} (\citealp{ForemanMackey2013}).

\begin{figure*}[htbp]
   \centering
   \includegraphics[width=\hsize]{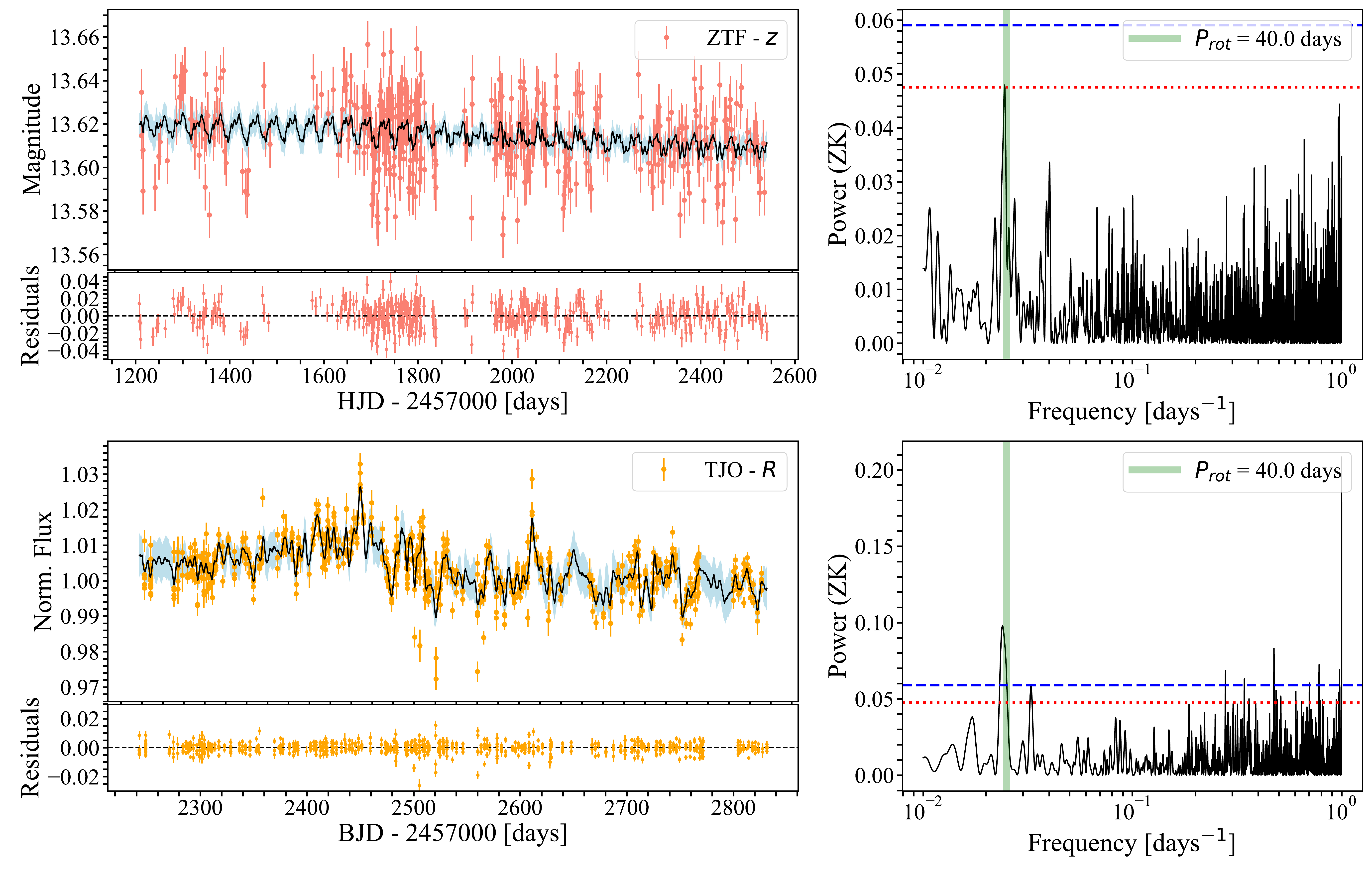}
   \caption{Long-term photometric follow-up of TOI-2095 from ZTF ($z$ band) and TJO ($R$ band). \textit{Left panels:} Photometric time-series and joint best fitted model computed using GP adopting a quasi-periodic kernel (black line). The 1$\sigma$ uncertainty regions of the fit are shown in light blue. The residuals of the fit are shown at the bottom panels. \textit{Right panels:} GLS periodogram (\citealp{Zechmeister2009}) for each photometric data set. The vertical green line shows the fitted rotational period of the star. The horizontal lines represent the FAP levels of 10\% (red dotted line) and 1\% (blue dashed line).}
    \label{Fig:TOI2095_Prot}
\end{figure*}

Figure \ref{Fig:TOI2095_Prot} shows the photometric time series from ZTF and TJO and their respective periodograms; the black line represents the model produced using the median of the posterior distributions of the fitted parameters and the blue shaded area the 1$\sigma$ uncertainty of the model. The periodograms presented in this work were computed using the Generalized Lomb-Scargle (GLS) periodogram implementation by \citet{Zechmeister2009}. We used a \texttt{Python} implementation of GLS\footnote{\url{https://github.com/mzechmeister/GLS}} that includes a subroutine to compute the false alarm probability (FAP) levels. This subroutine computes the FAP levels analytically using Eq.~(24) of \citet{Zechmeister2009}.

The root mean square of the residuals of the fit are $rms_{\mathrm{ztf}} = 0.014$\,mag for ZTF and $rms_{\mathrm{tjo}} = 0.004$\,mag for the TJO data. From the time series and periodograms, ZTF-$z$ data present a larger scatter compared to TJO-$R$ ones and there is no clear detection of a periodic signal in the photometry for the former data set. On the other hand, the periodogram for the TJO observations presents a significant peak around 40 days. An individual fit for both data sets using the previously described kernel resulted in a bimodal posterior distribution for ZTF with peaks around $P_{\rm rot} \sim 40$ and $\sim 80$ days, while the TJO fit delivered a rotation period of $P_{\rm rot} = 40.2^{+2.1}_{-3.8}$ days. Using our joint fit, we found a rotational period of $P_{\rm rot} = 40.0^{+0.2}_{-0.4}$ days for TOI-2095, which will be our final adopted value. Based on the gyrochronology relations from open clusters, an early-M dwarf with a rotation period of 40 days is consistent with typical old field dwarf with ages of a few Gyr (\citealp{Curtis2020}). Based on these relations we conclude that TOI-2095 is likely older than the age of Praesepe (670\,Myr old).

\section{Analysis and results}
\label{Sec:AnalysisResults}

\subsection{Radial velocity and activity indices periodogram analysis}
\label{Sec:rvs}
\begin{figure}
    \centering
    \includegraphics[width=\hsize]{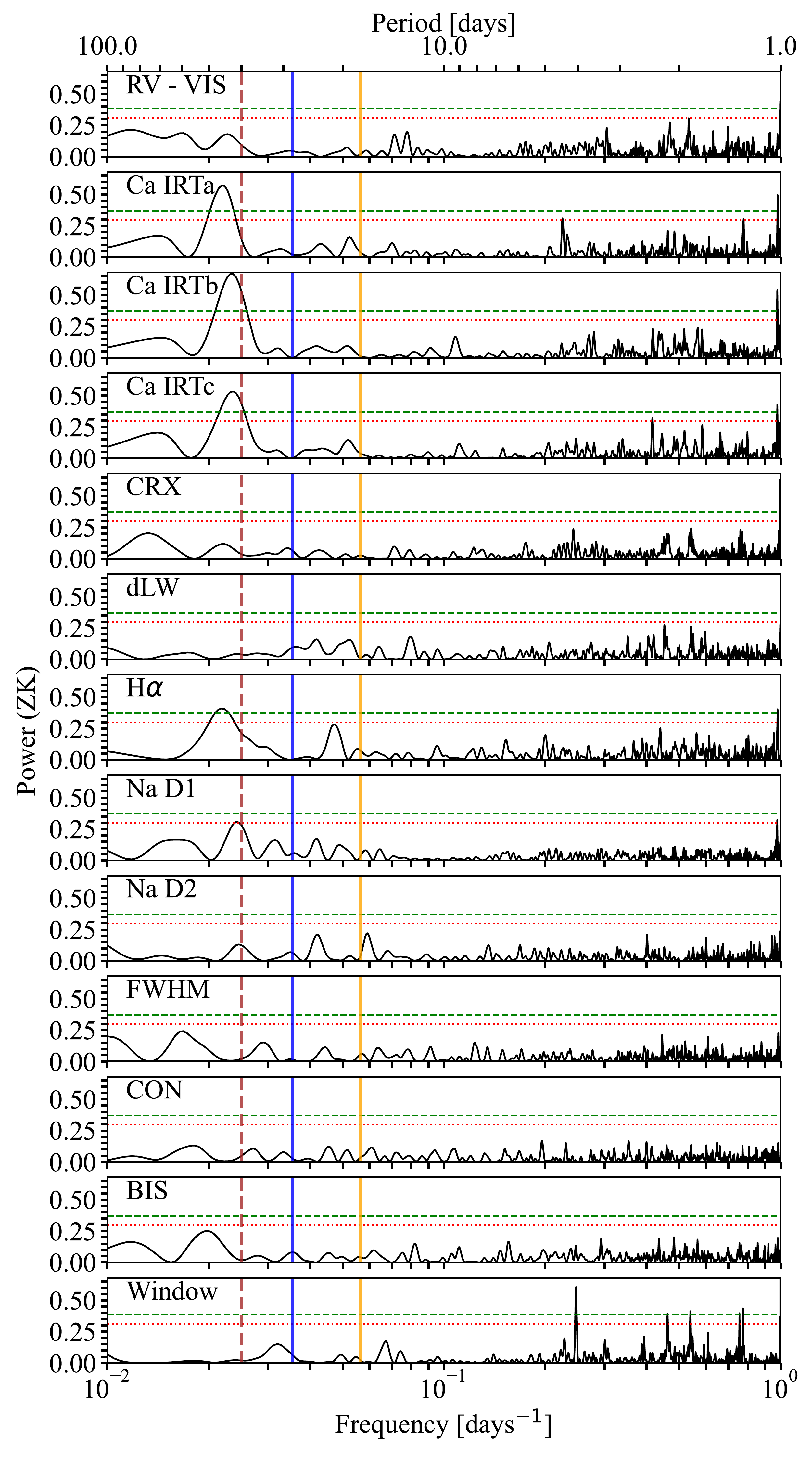}
    \caption{GLS periodograms of the CARMENES RV measurements taken with the optical channel, spectral activity indices, and window function. The horizontal lines represent the false alarm probability (FAP) levels of 10\% (red dotted line) and 1\% (green dash line). The vertical lines mark the period of planets TOI-2095b ($P=17.66$ days, solid orange line), TOI-2095c ($P=28.17$ days, solid blue line), and the measured rotation period of the star from photometric observations ($P=40$ days: dashed brown line).}
    \label{Fig:RVActiv_Periodogram}
\end{figure}

To search for the periodic variations in the RV of the star induced by the two transiting planet candidates, we computed the GLS periodogram. Figure~\ref{Fig:RVActiv_Periodogram} shows the periodograms for the CARMENES RV measurements (visible channel) and the activity indices provided by \texttt{SERVAL}. We marked the orbital periods of the two planet candidates using \textit{TESS} ExoFOP ephemeris, the TOI-2095b period is represented by the orange vertical line, while TOI-2095c is highlighted by the blue vertical line. The measured rotation period based on ground-based photometric monitoring of the star is marked by the brown vertical line. 

From Fig.~\ref{Fig:RVActiv_Periodogram} we note that several activity indices present a peak around $P\sim40$ days, in agreement with the measured rotation period derived from ground-based photometric observations. The RV periodogram shows no evidence of significant peaks around the periods of the transiting candidates detected by \textit{TESS} nor at the rotation period of the star. As a test, we removed the $\sim$40 day rotation period from the RVs by fitting the data using a quasi-periodic kernel to model the stellar variability with a normal prior based on our derived value for the rotation period of the star. However, the power peaks at the expected orbital periods of the planets did not become significant in the periodogram. The non-detection of both transiting planets in the RV measurements is expected since the predicted RV amplitude for both planets is around $\sim$0.9\,m\,s$^{-1}$\footnote{From \textit{TESS} follow-up recon spectroscopy group (SG2), based on predicted masses values computed using \cite{Chen&Kipping2017}}, close to the precision limit of current spectrographs. 
However, reaching near 1\,m\,s$^{-1}$ RV precision with CARMENES is possible. For example, \cite{Kossakowski2023} detected an Earth-mass planet with a period of 15.6 days around the M dwarf Wolf 1069b. The induced RV semi-amplitude of this planet is $1.07\pm0.17$\,m\,s$^{-1}$. This detection was possible due to the large number of observation (262 RV measurements, close to four years of baseline) and the low levels of activity of the star. With our RV data (44 measurements) and the stellar activity variability in the observations, we can only use the RV measurements to put upper limits on the masses of the transiting objects to validate them as planets.

\subsection{\textit{TESS} and CARMENES joint fit}
\label{Sec:joint}

We used the available \textit{TESS} photometry and CARMENES RV measurements to perform a joint fit of the data. We decided to use \textit{TESS} transit observations since the ground-based data presented only tentative transit detections of the transit events. We followed the method outlined in \cite{Murgas2021}, which we briefly describe here. The transits of both planets were modeled using \texttt{PyTransit}\footnote{\url{https://github.com/hpparvi/PyTransit}} (\citealp{Parviainen2015}), we adopted a quadratic limb darkening (LD) law and used \texttt{LDTK}\footnote{\url{https://github.com/hpparvi/ldtk}} (\citealp{Parviainen2015b}) to compare the fitted LD coefficients with the expected values (based on the stellar parameters presented in Table \ref{Tab:Star}). To model the planet-induced RV variations, we used \texttt{RadVel}\footnote{\url{https://github.com/California-Planet-Search/radvel}} (\citealp{Fulton2018}). The RV measurements were affected by the activity of the star and presented a moderate correlation with some of the stellar activity indices measured with \texttt{SERVAL}. 
In our case the activity index that presented the strongest correlation with the RVs was the Ca~{\sc ii} IRT line, with a Pearson's $r = 0.46$ (i.e., moderate correlation; see Fig. \ref{Fig:RV_vs_CaIndex} and Fig. \ref{Fig:RV_vs_ActivIndx} for the correlation for the other activity indices). To account for this correlation, we fit a linear function using the Ca~{\sc ii} IRT index as an independent variable and corrected the RV values using the fitted linear trend before performing the joint fit. With this approach, the stellar variability contribution (or part of) is taken into account and not modeled as a red-noise component only.

As a test, we fit the RV data using non-circular orbits for both planets and imposing normal priors in the orbital period and central time of the transits, using the values found by \textit{TESS} while modeling the systemaic noise with GPs. We found that the orbital eccentricity ($e$) and argument of the periastron ($\omega$) were unconstrained by the RV measurements, hence, we decided to adopt circular orbits for both planet candidates for the joint fit. The free parameters used in our modeling were the planet-to-star radius ratio $R_p/R_\star$, the quadratic LD coefficients, $q_1$ and $q_2$ (using \citealp{Kipping2013}), the central time of the transit, $\mathrm{T}_{\mathrm{c}}$, the planetary orbital period, $P$, the stellar density, $\rho_\star$, the transit impact parameter, $\mathrm{b}$, the RV semi-amplitude, $K_\mathrm{RV}$, the host star systemic velocity, $\gamma$, and the RV jitter ($\sigma_{\mathrm{RV\; jitter}}$). 
We modeled the \textit{TESS} correlated noise through a GP with a Mat\'ern 3/2 kernel. 
This kernel has covariance properties that make it very appropriate to model \textit{TESS} data in which the photometric variability is barely influenced by the stellar rotation and, instead, residual short-term red-noise structures dominate the correlated noise \citep{2020AJ....160..259S,2023A&A...675A..52C}. In the case of TOI-2095, we ran the GLS periodogram over the individual and joint \textit{TESS} sectors and found no significant periodicities. Since those structures can vary from one sector to another, we modeled them with different Mat\'ern 3/2 kernels, each of them taking the form:

\begin{equation}
    k_{ij\; \mathrm{TESS}} = c^2_k \left( 1 + \frac{\sqrt{3} |t_i-t_j|}{\tau_k}\right) \exp\left(-\frac{\sqrt{3} |t_i-t_j|}{\tau_k}\right)
\label{Eq:TESS_GPKernel}
,\end{equation}
where $|t_i-t_j|$ is the time between epochs in the series, and the hyperparameters, $c_k$ and $\tau_k$, were allowed to be free, with $k$ indicating the \textit{TESS} sector. 

For the RV measurements we modeled the stellar variability using an exponential squared kernel (i.e., a Gaussian kernel)

\begin{equation}
    k_{ij\; \mathrm{RV}} = c^2_{rv} \exp \left(  - \frac{ (t_i-t_j)^2}{\tau^2_{rv}} \right)
\label{Eq:RV_GPKernel}
,\end{equation}
where $t_i-t_j$ is the time between epochs in the series, and the hyperparameters, $c_{rv}$ and $\tau_{rv}$, were set free.

We fit a total of 61 free parameters to model the photometric and RV measurements considering systematic noise. The fitting procedure started with the optimization of a joint posterior probability function with \texttt{PyDE}. We used the results of the optimization to start a Markov chain Monte Carlo (MCMC) procedure using \texttt{emcee} (\citealp{ForemanMackey2013}). We ran the MCMC using 250 chains for 15\,000 iterations as a burn-in stage to ensure the convergence of the parameters, and then ran the main MCMC for another 16\,000 iterations. We computed the final values of our fitted parameters using the median and 1$\sigma$ limits of the posterior distributions. 

\begin{figure*}
   \centering
   \includegraphics[width=\hsize]{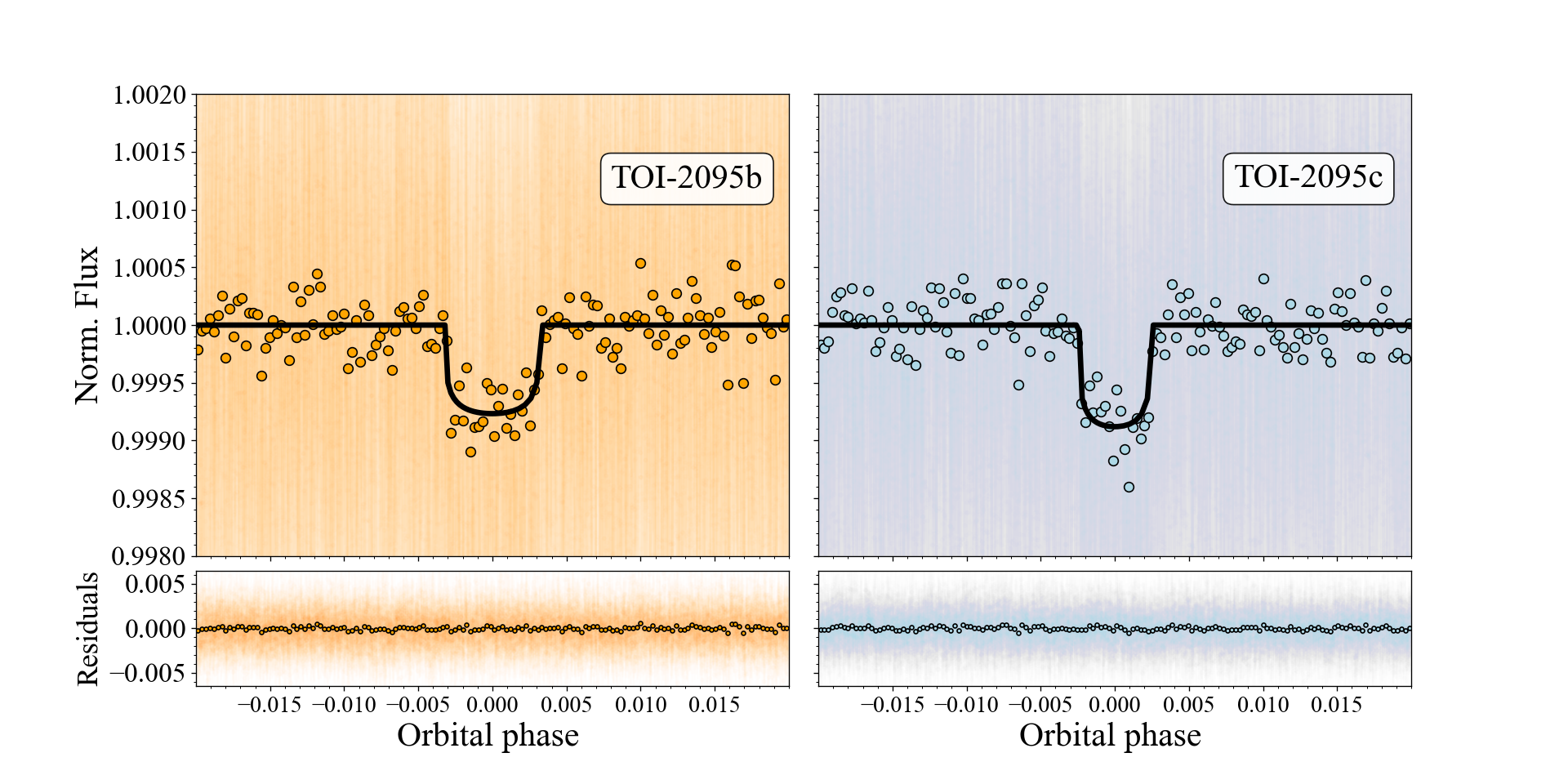}
   \caption{\textit{TESS} phase-folded light curves after subtracting the photometric variability for TOI-2095b (top left) and TOI-2095c (top right). The best-fit model is shown in black, the circles are \textit{TESS} binned data points, and the points are individual \textit{TESS} observations. \textit{Bottom panels:} Residuals of the fit.}
    \label{Fig:TESS_PhFoldedfit}
\end{figure*}

\begin{figure*}
   \sidecaption
   \includegraphics[width=12cm]{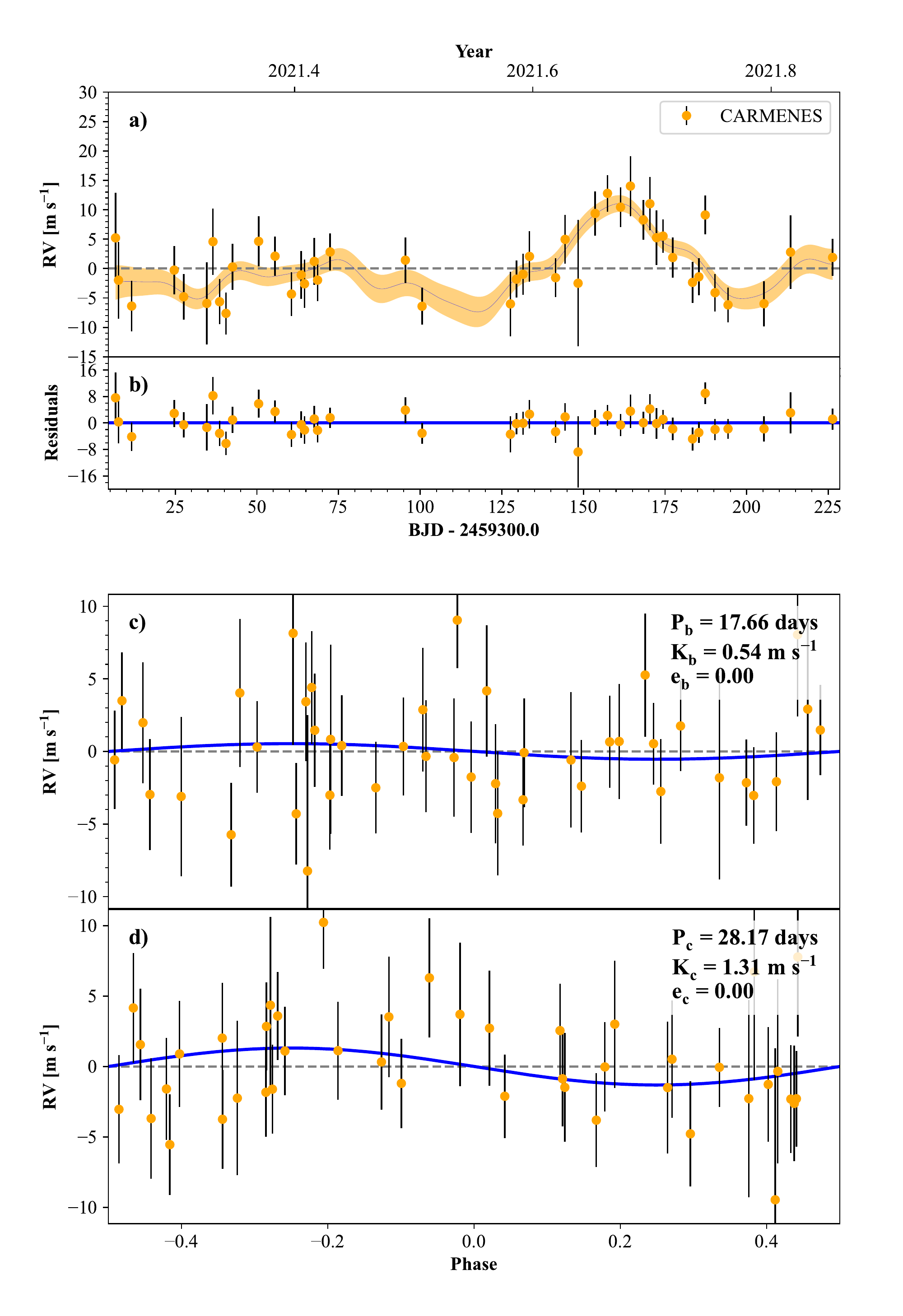}
   \caption{Radial velocity measurements of TOI-2095 taken with CARMENES. \textit{(a):} RV time series and best-fitting model (blue line) including red noise. The plotted RV model was computed using the median values of the posterior distribution for each fitted parameter. The shaded area around the blue line represent the 1$\sigma$ uncertainty levels of the fitted model. \textit{(b):} Residuals of the fit after subtracting the two planet models. \textit{(c)} and \textit{(d):} Phase-folded RV measurements after subtracting the red noise and best-fit model (blue line).}
    \label{Fig:CARMENES_RVfit}
\end{figure*}

The fitted and derived parameters values can be found in Table \ref{Tab:Planet_parameters} (see Fig. \ref{Fig:Fit_ParamDistr_CornerPlot} for the posterior distribution correlation plot of the parameters). The phase folded and systematic-free transit model and data are shown in Fig.~\ref{Fig:TESS_PhFoldedfit}, the individual \textit{TESS} photometry and transit model including red noise is shown in the Appendix (see Fig. \ref{Fig:Fit_AllTESS_TransitFit}). Figure \ref{Fig:CARMENES_RVfit} presents the CARMENES RV measurements and the RV model of the data using the results of the joint fit.

\begin{figure*}
   \includegraphics[width=\hsize]{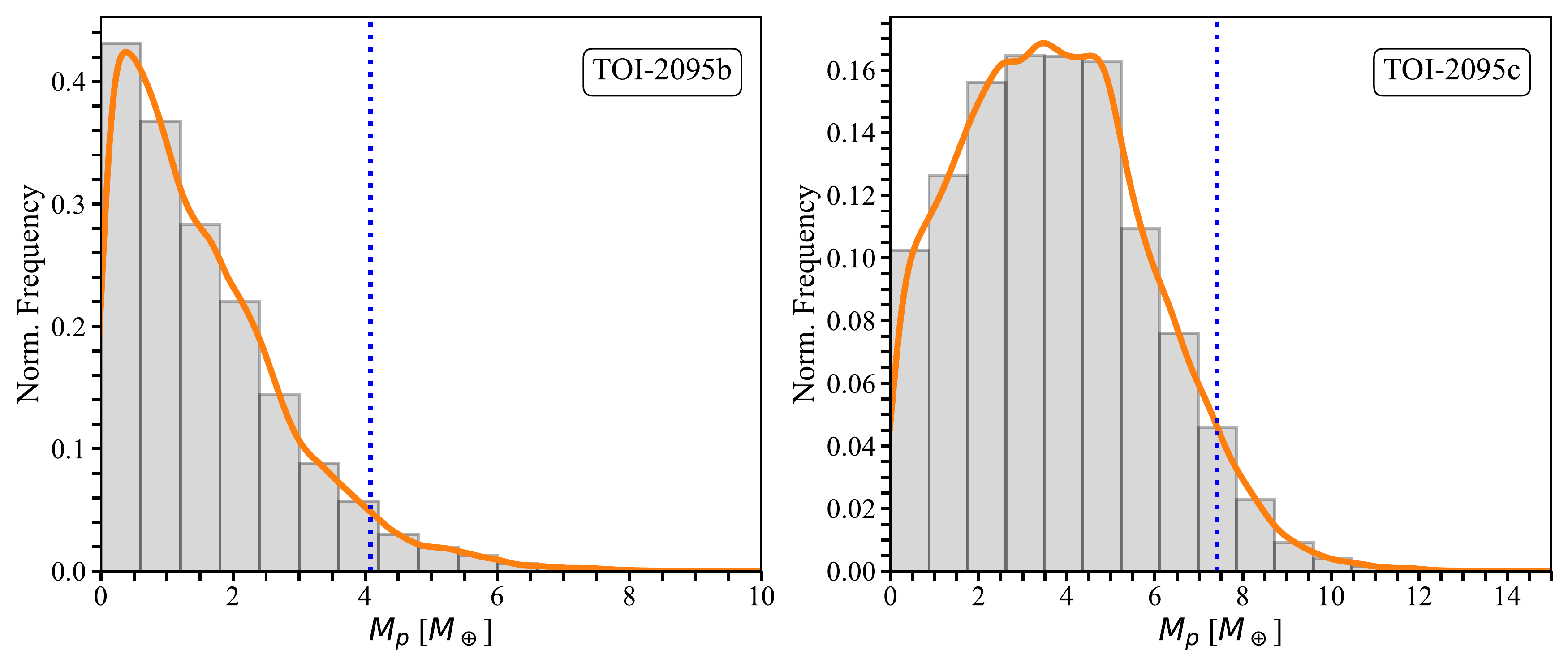}
   \caption{Posterior distributions for the planetary masses of TOI-2095b and TOI-2095c. The vertical dashed blue line represents the 95\% confidence limit for each planet; for TOI-2095b, we find a mass upper limit of $M_b < 4.1 \; M_\oplus$ and for TOI-2095c, it is  $M_c < 7.4 \; M_\oplus$.}
    \label{Fig:TOI2095_MassDistri}
\end{figure*}

From the analysis of all available data on TOI-2095, we find that both planets have similar sizes, with $R_b = 1.25 \pm 0.07 \; R_\oplus$ for TOI-2095b and $R_c = 1.33 \pm 0.08 \; R_\oplus$ for TOI-2095c. Due to their relatively long orbital periods ($P_b = 17.665$ days and $P_c = 28.172$ days) and the small RV amplitude induced by the transiting planets, we do not find significant evidence of the presence of the planets in our RV data. Additionally, the stellar activity of the star dominates the RV variations. 
We computed the mass upper limits using the RV equation from \citet[][ and we refer to \citealt{Perryman2011} for a complete derivation]{Cumming1999} assuming $M_\star \gg M_p$ and $e=0$. We set $M_\star = 0.44 \; M_\odot$ and used the posterior distributions of the orbital period, orbital inclination, and RV amplitudes from the joint fit. The final upper limits were computed using the 95\% percentile limit of the distribution of masses. Using this procedure, we found upper mass limits of $M_b < 4.1 \; M_\oplus$ and $M_c < 7.4\; M_\oplus$ for TOI-2095b and TOI-2095c (see Fig. \ref{Fig:CARMENES_RVfit} and \ref{Fig:TOI2095_MassDistri}), respectively. However, these upper mass limits place both transiting candidates well into the planetary mass regime.

\begin{table*}[t]
\centering
\caption{Fitted and derived parameters of TOI-2095b and TOI-2095c.}
\label{Tab:Planet_parameters}
\begin{tabular}{l c c}
\hline 
\hline
\noalign{\smallskip}
Parameter & TOI-2095b & TOI-2095c \\
\noalign{\smallskip}
\hline
\noalign{\smallskip}
\multicolumn{3}{c}{Fitted transit and orbital parameters} \\
\noalign{\smallskip}
$\rho_*$ [g\,cm$^{-3}$] & $7.23 \pm 0.22$ & $7.23 \pm 0.22$ \\
$R_{p}/R_{\star}$ & $0.0260 \pm 0.0010$ & $0.0277 \pm 0.0011$ \\
$T_{\rm c}$  [BJD] & $2459240.4115 \pm 0.0016$ & $2459239.5625 \pm 0.0024$ \\
$P$ [days] & $17.66484 \pm 0.00007$ & $28.17232 \pm 0.00014$ \\
$b = a/R_{*}\cos(i)$ & $0.285^{+0.085}_{-0.136}$ & $0.104^{+0.108}_{-0.072}$ \\
$e$  & 0 (fixed) & 0 (fixed) \\
$\gamma_0$ [m\,s$^{-1}$] & $-0.55 \pm 3.06$ & $-0.55 \pm 3.06$ \\
$K$ [m\,s$^{-1}$] & $0.54^{+0.63}_{-0.38}$ & $1.31^{+0.84}_{-0.79}$ \\
$\sigma_{\rm RV}$ [m\,s$^{-1}$] & $1.51 \pm 0.95$ & $1.51 \pm 0.95$ \\
\noalign{\smallskip}
\multicolumn{3}{c}{Derived orbital parameters} \\
\noalign{\smallskip}
$a/R_*$ & $49.22 \pm 0.53$ & $67.19 \pm 0.72$ \\
$i$ [deg] & $89.67^{+0.16}_{-0.10}$ & $89.91^{+0.06}_{-0.09}$ \\
\noalign{\smallskip}
\multicolumn{3}{c}{Derived planet parameters} \\
\noalign{\smallskip}
$R_{p}$ [R$_{\oplus}$] & $1.25 \pm 0.07$ & $1.33 \pm 0.08$ \\ 
 $M_{p}$ [M$_{\oplus}$] & $<4.1$ (95\%) & $<7.4$ (95\%) \\ 
$a$ [au] & $0.101 \pm 0.005$ & $0.137 \pm 0.006$ \\
$\langle F_{p} \rangle$ [$10^3$\,W\,m$^{-2}$] & $4.67 \pm 0.42$ & $2.50 \pm 0.23$ \\
$S_{p}$ [$S_\oplus$] & $3.43 \pm 0.31$ & $1.84 \pm 0.17$ \\
$T_{\rm eq}$ ($A_{\rm Bond} = 0.3$) [K] & $347 \pm 9$ & $297 \pm 8$ \\
\noalign{\smallskip}
\multicolumn{3}{c}{Fitted LD coefficients} \\
\noalign{\smallskip}
$q_{1\;TESS}$ & $0.282 \pm 0.003$ & $0.282 \pm 0.003$ \\
$q_{2\;TESS}$ & $0.228 \pm 0.004$ & $0.228 \pm 0.004$ \\
\noalign{\smallskip}
\multicolumn{3}{c}{Derived LD coefficients} \\
\noalign{\smallskip}
$u_{1\;TESS}$ & $0.242 \pm 0.003$ & $0.242 \pm 0.003$ \\
$u_{2\;TESS}$ & $0.289 \pm 0.005$ & $0.289 \pm 0.005$ \\
\noalign{\smallskip}
\hline
\end{tabular}
\tablefoot{$A_{\rm Bond}$ is the Bond albedo.}
\end{table*}

\begin{figure}
    \centering
    \includegraphics[width=\hsize]{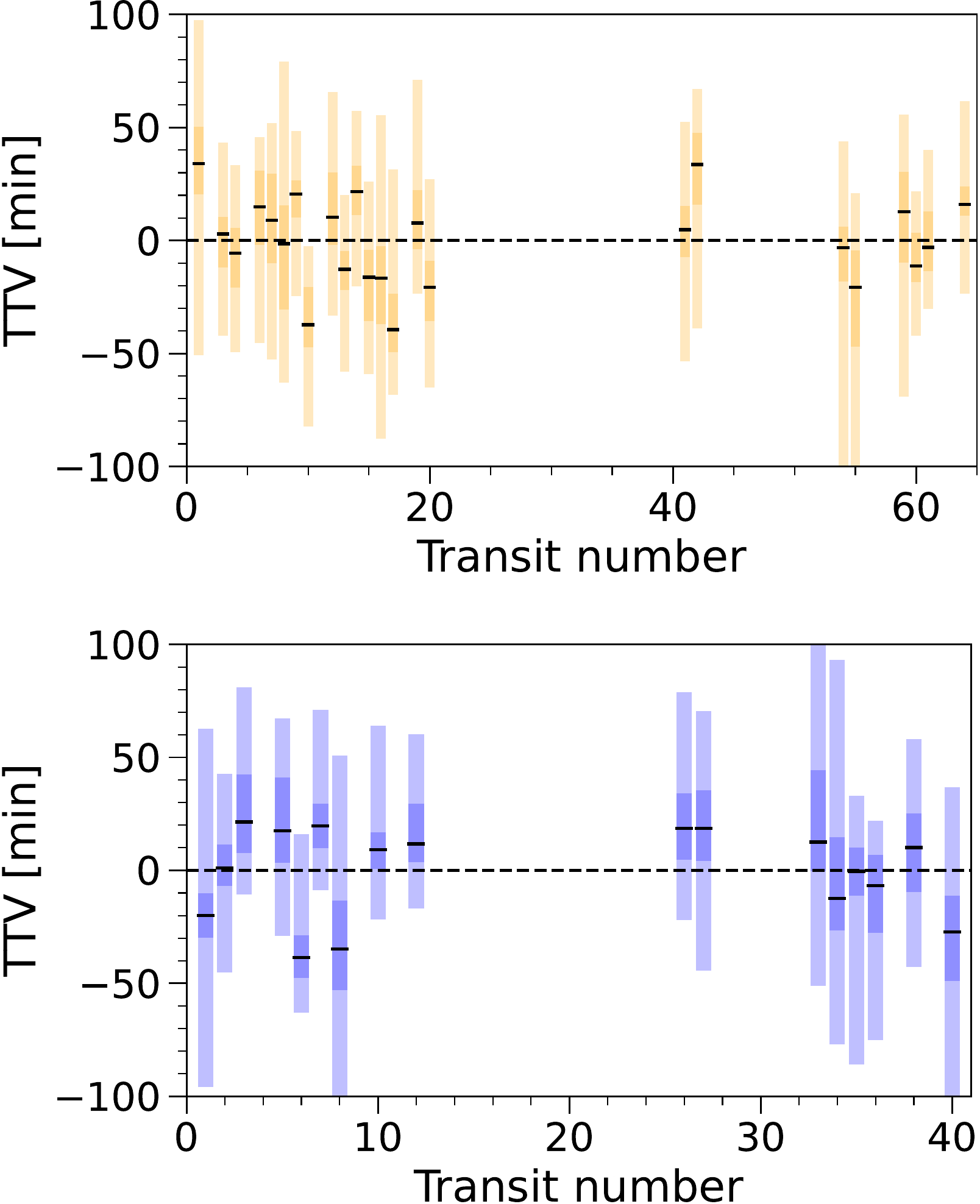}
    \caption{Central mid-transit times measured using \textit{TESS} data for TOI-2095b (top panel) and TOI-2095c (bottom panel). The shaded areas represent the 1$\sigma$ and 3$\sigma$ uncertainty limits for the central time of the transit measurements.}
    \label{Fig:TTV_plot}
\end{figure}

We checked whether we could detect any transit timing variations (TTVs) using \textit{TESS} photometric data. We fit the transits with \texttt{PyTTV} (Korth et al. in prep), allowing each central transit time to be free (values and 1$\sigma$ uncertainties presented in Tables \ref{Tab:TTV_TOI2095b} and \ref{Tab:TTV_TOI2095c}). Unfortunately, the individual transit events of both planets are too shallow to constrain the central times with the precision needed to detect TTVs -- if they do, in fact, exist in this system (see Fig. \ref{Fig:TTV_plot}).

\subsection{Statistical validation}

We also performed a statistical validation analysis for TOI-2095b and TOI-2095c based on TESS photometry, CARMENES spectroscopy, Keck NIR AO imaging, and SPP optical speckle imaging. To do so, we used the \texttt{vespa} package \citep{2012ApJ...761....6M, 2015ascl.soft03011M} to compute false positive probabilities (FPPs), that is, the probabilities of planet candidates being astrophysical false positives: eclipsing binaries (EBs), background (or foreground) EBs blended with the target star (BEBs), or hierarchical triple systems (HEBs). The most commonly adopted threshold to consider a candidate as statistically validated planet is to have a FPP lower than 1$\%$ \citep[FPP < 0.01; e.g.,][]{2014ApJ...784...45R,2015ApJ...809...25M,2020MNRAS.499.5416C,2021MNRAS.508..195D,2022AJ....163..244C}.  However, relying on the FPP alone can lead to misclassifications \citep{2018AJ....156..277L,2018AJ....155..136M}. Hence, before assigning the candidate disposition, we checked that our system meets the following conditions: transit S/N $>$ 10, odd-even mismatch of transit depths < 3$\sigma$, \textit{Gaia} RUWE $\lesssim 1.4$, and the absence of contaminant sources within the photometric aperture (Section \ref{sec:gaia_assesment}). The latter condition has been shown to be of crucial importance, since the re-analysis of early validation works in which it had not been taken into account led to the deprecation of several validated planets such as K2-78b, K2-82b, K2-92b \citep{2017A&A...606A..75C}, and K2-120b \citep{2022MNRAS.509.1075C}.

In the following, we describe the information supplied to \texttt{vespa} for the statistical validation analysis of TOI-2095b and TOI-2095c. We input the target coordinates and parallax from \textit{Gaia} DR3 \citep{GaiaDR3}, \textit{HJK} photometry from 2MASS \citep{2MASS}, and the CARMENES spectroscopic parameters $T_{\rm eff}$, [Fe/H], and log $g$ (Table \ref{Tab:Star}). We also supplied the orbital periods and planet-to-star radius ratios from our fit (Table \ref{Tab:Planet_parameters}). We constrained the maximum allowed depth for a secondary eclipse to be thrice the standard deviation of the out-of-transit region of the TESS light curve. We also constrained the maximum aperture radius from which the signal is expected to come from by circularizing the largest SPOC aperture as $\sqrt{A \pi^{-1}}$ following \citet{2022MNRAS.509.1075C}, being $A$ the maximum aperture area. Finally, we input the \textit{TESS} light curve folded to the candidate period as well as the Keck and SPP contrast curves, which significantly decrease the FPP by discarding the presence of stars above certain brightness at a certain projected distance.

\begin{table}[]
\caption{Posterior probabilities for the EB, BEB, HEB, and planet scenarios for TOI-2095b and TOI-2095c as computed with \texttt{vespa}.}
\renewcommand{\arraystretch}{1.5} 
\setlength{\tabcolsep}{3.6pt}
\begin{tabular}{ccccc}
\hline
\hline
\noalign{\smallskip}
Planet     & $\mathcal{P}$(EB)      & $\mathcal{P}$(BEB)    & $\mathcal{P}$(HEB)    & $\mathcal{P}$(Pl) \\ 
\noalign{\smallskip}
\hline
\noalign{\smallskip}
TOI-2095b & $5.71 \times 10^{-10}$ & $1.56 \times 10^{-4}$ & $2.15 \times 10^{-7}$ & 0.9998             \\
TOI-2095c & $2.82 \times 10^{-9}$  & $1.38\times 10^{-3}$  & $4.09 \times 10^{-9}$ & 0.9986           \\ 
\noalign{\smallskip}
\hline
\end{tabular}
\label{Tab:vespa}
\end{table}

In Table \ref{Tab:vespa} we show the \texttt{vespa} posterior probabilities for TOI-2095b and TOI-2095c. In both cases, the FPP is lower than 1$\%$, so both planets are independently validated. Moreover, planet candidates located in multi-transiting systems are more likely to be genuine planets than those in single planet candidate systems \citep{2011ApJ...732L..24L,2011ApJS..197....8L}. Hence, we followed the statistical framework introduced by \citet{2012ApJ...750..112L} in order to compute the multiplicity-corrected FPP as $\rm FPP_{2}$ = $1-P_{2}$, that is,\ 
\begin{equation}
     P_{2} \approx \frac{X_{2} P_{1}}{X_{2}P_{1} + (1-P_{1})},
\end{equation}
where $P_{1} = 1-\rm FPP$ and $X_{2}$ is the {``multiplicity boost''} for systems with two planet candidates. Similar to \citet{2021ApJS..254...39G}, we compute for the \textit{TESS} postage stamps a $X_{2}$ of 44 for two-planet systems with small planet candidates (R < 6 $R_{\oplus}$). As a result, we obtain multiplicity-corrected FPPs of $\rm FPP_{2,b}$ =  $3.54\times10^{-6}$ and $\rm FPP_{2,c}$ =  $3.14\times10^{-5}$, which supports the planetary nature of TOI-2095b and TOI-2095c.

\section{Planet searches and detection limits from the TESS photometry}
\label{Sec:search}

Besides the two planets found by SPOC orbiting TOI-2095, we wondered if extra transiting planets might exist in the system and remain unnoticed. 
To this end, we searched for hints of new planetary candidates using the \sherlock\footnote{\sherlock (\textbf{S}earching for \textbf{H}ints of \textbf{E}xoplanets f\textbf{R}om \textbf{L}ightcurves \textbf{O}f spa\textbf{C}e-based see\textbf{K}ers) code is fully available on GitHub: \url{https://github.com/franpoz/SHERLOCK}} pipeline \citep[see, e.g.,][]{pozuelos2020,demory2020}.

\sherlock allows for the exploration of TESS data to recover known planets, candidates and to search for new signals that might be attributable to planets.
The pipeline combines different modules to (1) download and prepare the light curves from their repositories, (2) search for planetary candidates using the \tls algorithm \citep{tls}, 
(3) perform an in-depth vetting of interesting candidates to rule out any potential systematic origin of the detected signals, (4) conduct a statistical validation using the \triceratops package \citep{triceratops}, 
(5) model the signals to refine their ephemerides through the \allesfitter code \citep{allesfitter}, and (6) compute observational windows from user-specific ground-based observatories to trigger a follow-up campaign.    

The transit search is optimized by removing any undesired trends in the data, such as instrumental drifts, implementing a multi-detrend approach by means of the \wotan package \citep{wotan}. This strategy consists of detrending the nominal PDCSAP light curve 
several times using a biweight filter by varying the window size. In our case, we performed 20 detrendings with window sizes ranging from 0.20 to 1.30~d. 

Then, \sherlock simultaneously conducts the transit search in each new detrended light curve jointly with the nominal PDCSAP flux. Once the transit search is done, \sherlock combines all the results to choose the most promising signal to be a planetary candidate. This selection is performed considering many aspects, such as the detrend dependency of a given signal, the S/N, the signal-detection-efficiency (SDE), and the number of times that a given signal is happing in borders, among others.   

The entire search process follows a search-find-mask loop until no more signals are found above user-defined S/N and SDE thresholds; that is, once a signal is found, it is stored and masked, and then the search continues.  
Due to the number of sectors available for TOI-2095, we split our search into two independent blocks: first, considering all the sectors observed during the primary mission, that is, sectors 14, 15, 16, 17, 18, 19, 20, 21, 22, 23, 24, and 26; secondly, the sectors
observed during the extended mission, that is, 40, 41, 47, 48, 49, 50, 51, 52, 53, 54, and 55. The motivation to follow this strategy is twofold: on the one hand, the larger the number of sectors we inspect, the higher the computational cost. On the other hand, any actual planetary signal should be recovered independently in both experiments; otherwise, the credibility of the detected signal is lower, hinting that the real source of the signal might be related to some instrumental issues.

We focused our search on orbital periods ranging from 0.5 to 60~d, where a minimum of two transits was required to claim a detection. We recovered the TOI-2095.01 and .02 signals in the first and second runs, respectively. 
In the subsequent runs, we did not find any other signal that hinted at the existence of extra transiting planets. We found other signals, but they were too weak to be considered planetary candidates or were attributable to variability, noise, or systematics. Hence, our search for extra planets yielded negative results, confirming only the two  detections reported by SPOC. 

As asserted in previous studies \cite[see, e.g.,][]{wells2021,schanche2022, delrez2022}, the lack of extra signals might be due to one of the following scenarios: 
(1) no other planets exist in the system; (2) they do exist, but they do not transit; (3) they do exist and transit but have orbital periods longer than the ones explored during our search process; or (4) they do exist and transit, but the photometric precision of the data is not accurate enough to detect them. 

Scenarios (1) and (2) might be further explored by conducting an intense RV follow-up campaign. Unfortunately, with our current RV data described in Sect.~\ref{Sec:rvs}, we cannot constrain if other smaller or longer-orbital period planets are present in the system. Scenario (3) can be tested by adding a more extended time baseline, for example, using the data from Sectors 56, 57, 58, 59, and 60. However, accumulating new sectors would not significantly increase our detection capability due to the already large data set used in this study.   
Finally, to evaluate scenario (4), we studied the detection limits of the TESS photometry performing injection-and-recovery experiments with the \matrixtk code\footnote{{The \matrixtk (\textbf{M}ulti-ph\textbf{A}se \textbf{T}ransits \textbf{R}ecovery from \textbf{I}njected e\textbf{X}oplanets) code is open access on GitHub: \url{https://github.com/PlanetHunters/tkmatrix}}} \citep{matrix}.

\matrixtk allows the user to inject synthetic planets over the PDCSAP light curve and perform a transit search following a similar strategy to \sherlock. 
In this study, due to the relatively long orbital periods of both planets, $\sim$17.6~d for planet b and $\sim$28.2~d for planet c, there is an extensive range of orbital periods where a small innermost planet may orbit. Furthermore, assuming a nearly coplanar configuration for planetary orbits, the transit probability for any hypothetical inner planet would be higher than for planets beyond planet c. For this reason, we focused our injection-and-recovery experiment on planets with orbital periods shorter than planet b; that is, we explored the $R_{\mathrm{planet}}$--$P_{\mathrm{planet}}$ parameter space in the ranges of 0.5--3.0\,R$_{\oplus}$ with steps of 0.28\,R$_{\oplus}$, and 0.5--16.0 days with steps of 0.26 days.  

For each combination of $R_{\mathrm{planet}}$--$P_{\mathrm{planet}}$, \matrixtk explores five phases, that is, different values of $T_{0}$. Then, in total we explored 3000 scenarios. For simplicity, the injected planets have eccentricities and impact parameters equal to zero. Once the synthetic planets are injected, \matrixtk detrends the light curves using a bi-weight filter with a window size of 0.75~d, which was found to be optimal during the \sherlock search and masked the transits corresponding to TOI-2095b, and c. A synthetic planet is recovered when its epoch matches the injected epoch with 1~hour accuracy, and its period is within 5\,\% of the injected period. 

\begin{figure}
\includegraphics[width=\columnwidth]{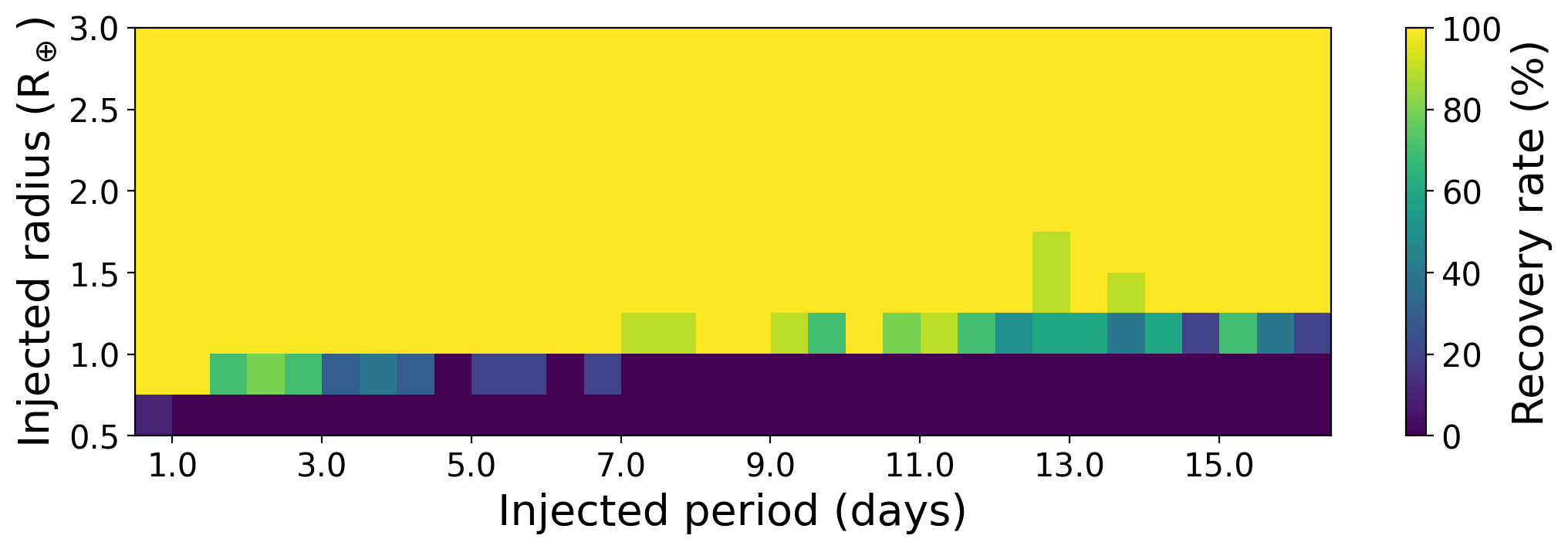}
\caption{Injection-and-recovery experiment performed to test the detectability of inner planets in the system using the TESS sectors described in Sect.~\ref{Sec:TESS_Obs}. We explored a total of 3000 different scenarios. Each pixel evaluated about eight scenarios, that is, eight light curves with injected planets having different $P_{\mathrm{planet}}$, $R_{\mathrm{planet}}$, and T$_{0}$. Higher recovery rates are presented in yellow and green colors, while lower recovery rates are shown in blue and darker hues. Planets smaller than 1.0~R$_{\oplus}$ would remain undetected for the explored periods.} \label{fig:recovery}
\end{figure}

The results obtained during these injection-and-recovery experiments are displayed in Fig.~\ref{fig:recovery}. We found that Earth- and sub-Earth size planets would remain unnoticed almost for the complete set of periods explored. However, planets with sizes larger than Earth seem to be easy to find, with recovery rates > 80\%. Hence, we can rule out the presence of inner transiting planets to TOI-2095b with sizes > 1~R$_{\oplus}$.

\section{Dynamical analysis}
\label{sec:Dyn}
\subsection{Stability-constrained characterization}

The current data on the TOI-2095 system does not allow for strong constraints on the masses and eccentricities of the two super-Earths. We explored the possibility of obtaining tighter constraints on these parameters through dynamical stability considerations. Similar to many previous works \citep[see, e.g.,][]{delrez2021, delrez2022, demory2020, pozuelos2020,jenkins2019}, we made use of the Mean Exponential Growth factor of Nearby Orbits ({\tt MEGNO}; \citealt{2000A&AS..147..205C, 2003PhyD..182..151C}) chaos indicator implemented within the {\tt REBOUND} $N$-body integration software package \citep{2012A&A...537A.128R}. We used the WHFast integration scheme \citep{2015MNRAS.452..376R}, an implementation of the Wisdom-Holman symplectic mapping algorithm \citep{1991AJ....102.1528W}. The time-averaged {\tt MEGNO} value, $\langle Y(t) \rangle$, resulting from an integration indicates the divergence of the planets' trajectories after small perturbations of their initial conditions. For chaotic motion, $\langle Y(t) \rangle$ grows without bound as $t \rightarrow \infty$, whereas for regular motion, $\langle Y(t) \rangle \rightarrow 2$ for $t \rightarrow \infty$.

We constructed two sets of {\tt MEGNO} stability maps to explore the dynamical behavior in the $M_b - M_c$ and $e_b - e_c$ parameter spaces. Based on the current constraints for the planetary masses (Fig.~\ref{Fig:TOI2095_MassDistri}), we explored values ranging from $0 - 7 \; M_{\oplus}$ for TOI-2095b and $0 - 12 \; M_{\oplus}$ for TOI-2095c. As for the eccentricity map, we considered values ranging from $0 - 0.2$ for both planets. We constructed $20 \times 20$ grids in the parameters for both the $M_b - M_c$ and $e_b - e_c$ {\tt MEGNO} maps. For each grid cell, we randomly sampled ten different initial conditions for the mean anomalies and arguments of periastron, ($M, \omega \sim \mathrm{Unif}[0^{\circ}, 360^{\circ}]$). The values represented in each grid cell are the average of the $10$ different initial conditions for $M$ and $\omega$. In the $M_b - M_c$ map, we froze the values of all the other parameters, including eccentricity, to their nominal values. Similarly, when exploring the eccentricity space, we froze the values of all other parameters, including mass, to their nominal values. The integration time and timestep for each realization were set to $10^6$ orbits of TOI-2095c and $5 \%$ of the orbital period of TOI-2095b, respectively.

\begin{figure}[]
    \includegraphics[width=\hsize]{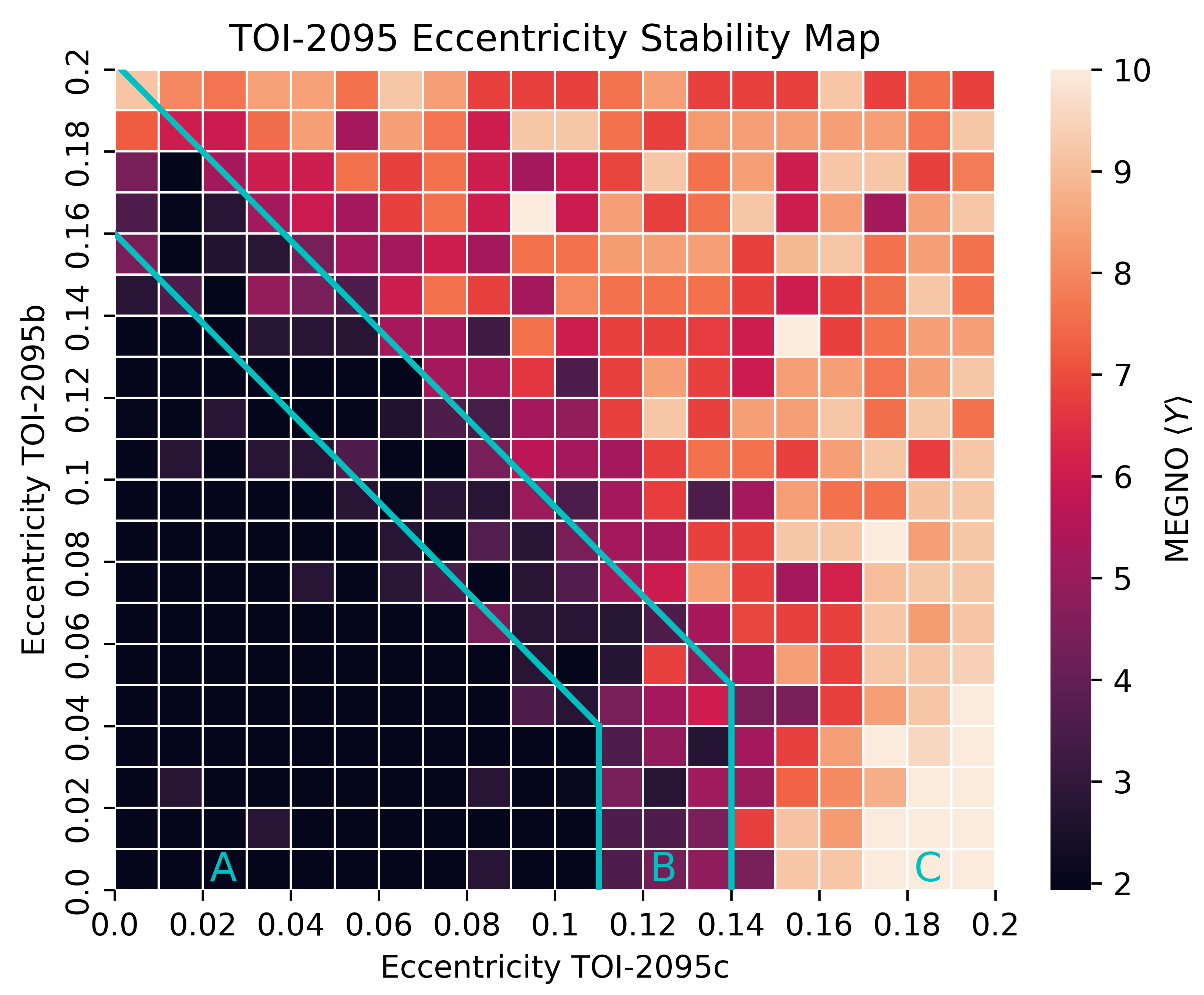}
    \caption{Stability map of the TOI-2095 system in the $e_b - e_c$ parameter space based on the {\tt MEGNO} chaos indicator. The map consists of a $20 \times 20$ grid, where each grid cell is the average of $10$ realizations with randomized mean anomalies and arguments of periastron. The three regions of differing stability (A, B, C) are marked with letters and are described by Eqs.~\ref{eq:regionA} and \ref{eq:regionC}.}
    \label{Fig:TOI2095_EccentricityMap}
\end{figure}

After running the $4000$ realizations for the $M_b - M_c$ parameter space, we found that the system is fully stable through the entire range of masses studied. Therefore, we could not further constrain the masses of TOI-2095b and TOI-2095c using stability considerations. The $e_b - e_c$ parameter space, on the other hand, reveals a much larger variation. With sufficiently large eccentricities for one or both planets, the {\tt MEGNO} value indicates the system's chaotic behavior (Fig.~\ref{Fig:TOI2095_EccentricityMap}). 

Similar to what was found by \cite{demory2020}, we determined three regions of differing stability: A (fully stable), B (transition region), C (fully unstable). When the system is fully stable (A), the eccentricities obey the following inequality:
\begin{equation}
\label{eq:regionA}
    \begin{cases}
        e_b + 1.09e_c < 0.16 & \text{if } e_b > 0.04\\
        e_c < 0.11 & \text{if } e_b < 0.04\
    \end{cases}
,\end{equation}
When the system is fully unstable (C), the eccentricities satisfy:
\begin{equation}
\label{eq:regionC}
    \begin{cases}
        e_b + 1.083e_c > 0.2016 & \text{if } e_b > 0.05\\
        e_c > 0.14 & \text{if } e_b < 0.05\
    \end{cases}
,\end{equation}
The transition region falls between the lines indicated by the aforementioned inequalities.

\subsection{Constraints on potential additional planets}

The TOI-2095 system may have additional planets other than planets b and c. Short-period super-Earths are often found in tightly-spaced, high-multiplicity systems, particularly around M dwarfs \citep[e.g.,][]{2014ApJ...790..146F, 2015ApJ...807...45D}. Here, we explore the parameter space in which an additional third planet in the system could exist in stable orbits. We used the Stability Orbital Configuration Klassifier ({\tt SPOCK}; \citealt{2020PNAS..11718194T}), which is a machine-learning model trained on $\sim 100\,000$ orbital configurations of three-planet systems, to classify the long-term orbital stability of a given planetary system. 

\begin{figure}[]
\includegraphics[width = \hsize]{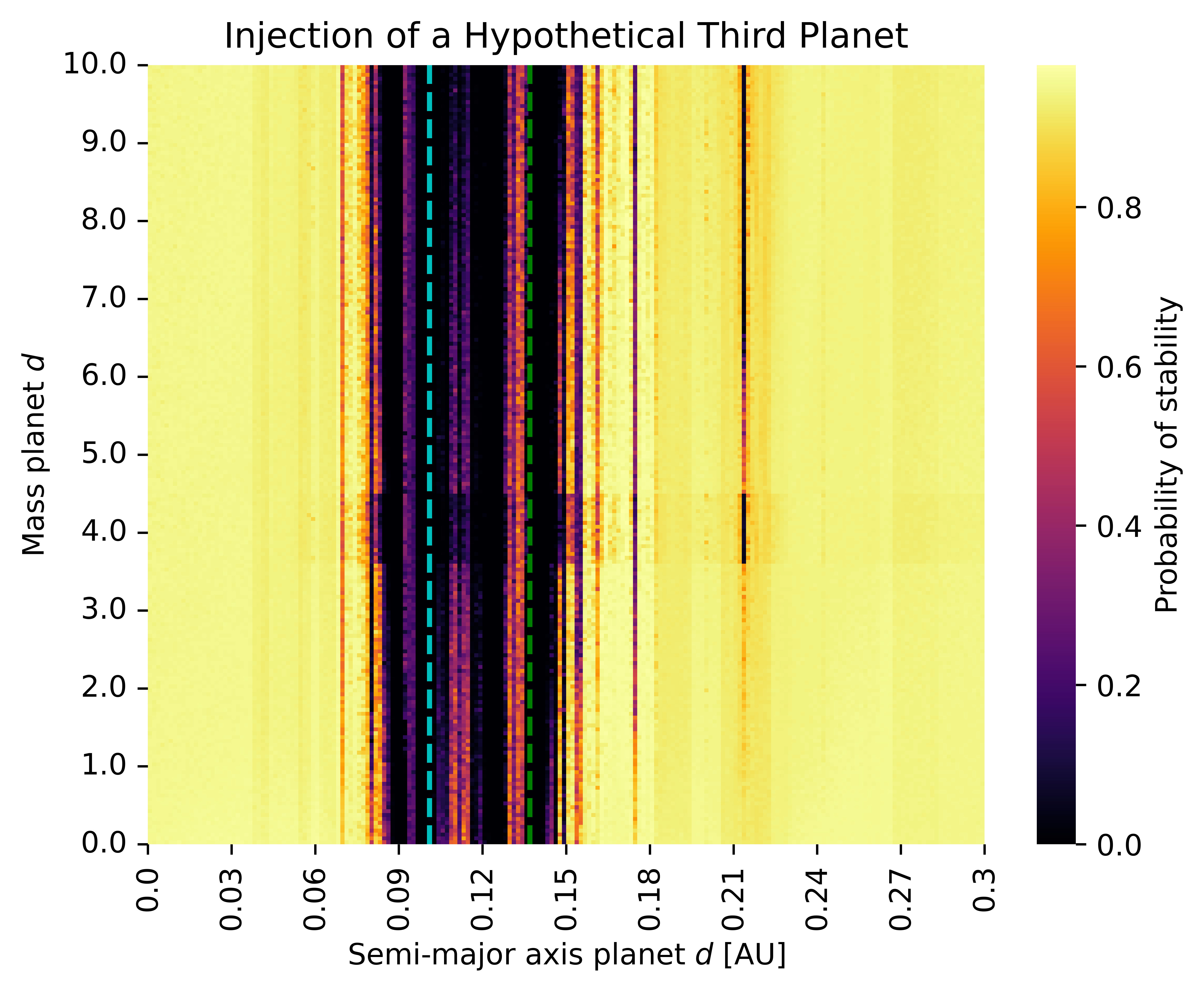}
\includegraphics[width = \hsize]{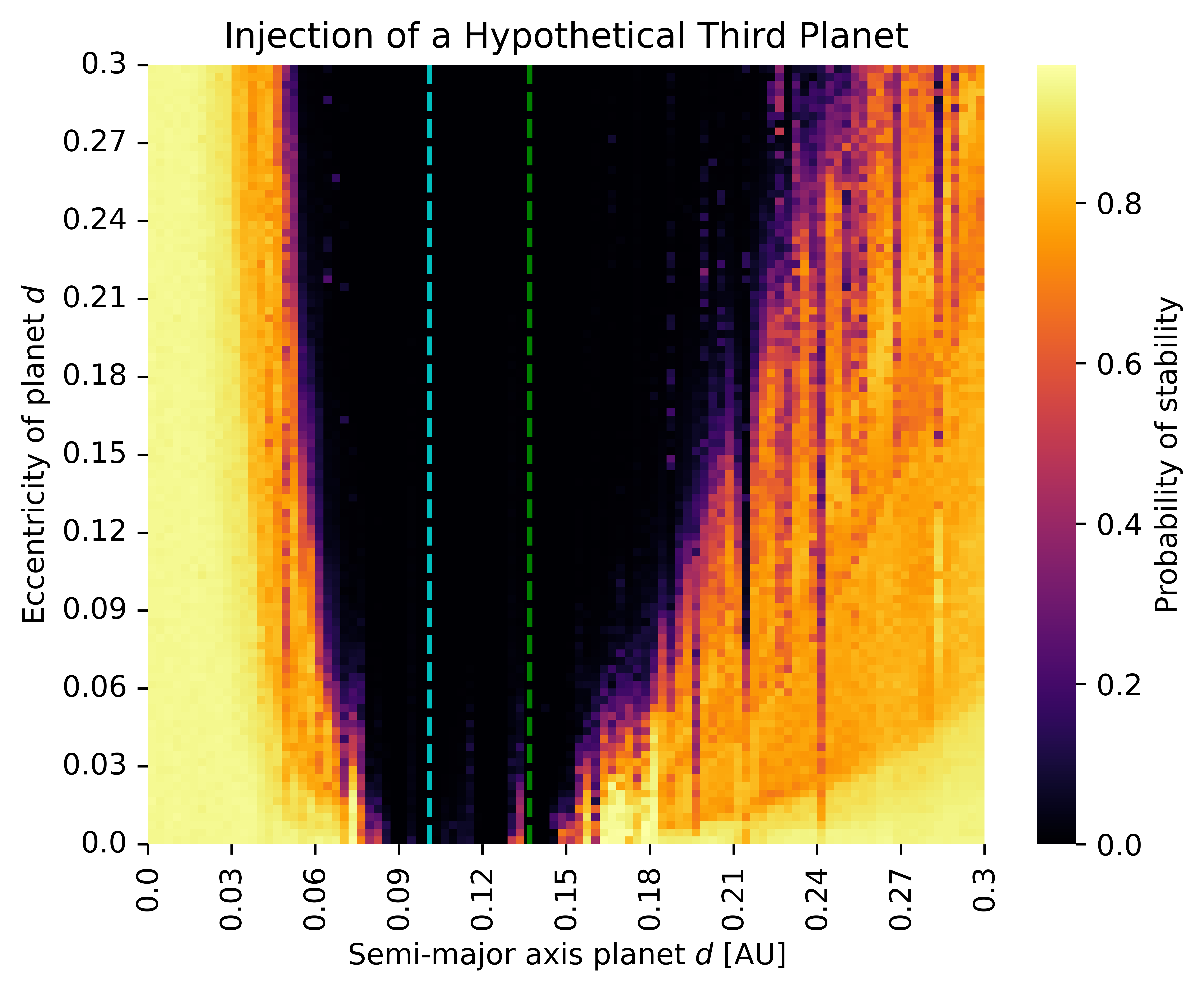}
\caption{Stability map of TOI-2095 for a hypothetical third planet in the system. \textit{Top panel:} Stability map for a third planet in a circular orbit case. Each cell of the $200 \times 200$ grid considers a different combination of $a_d$ and $M_d$ and represents the average value of $10$ randomized realizations of the system's initial conditions ($i$, $\omega$, $M$). \textit{Bottom panel:} Stability map for a third planet in a eccentric orbit case. The map consists of $100\times100$ grid cells plotted in the $a_d - e_d$ space. In both panels, the colorbar indicates the probability of stability from {\tt SPOCK} \citep{2020PNAS..11718194T}. The vertical dashed lines denote the locations of TOI-2095b (cyan) and TOI-2095c (green).}
\label{Fig:TOI2095_Injected_Planet}
\end{figure}

We considered two types of simulations for the TOI-2095 system with an injected third planet (``planet d''). In the first version, we assumed a circular orbit for planet d. To start, we also considered planets b and c to have circular orbits. We explored the $a_d - M_d$ space of the additional planet within a $200 \times 200$ grid in which $a_d$ was varied from $0.01-0.3$\, au and $M_d$ was varied from $1-10 \ M_{\oplus}$. Each grid cell represents the average of ten realizations of the system with randomized values of the inclinations, mean anomalies, and longitudes of ascending node ($i \sim \mathrm{Rayleigh}(1^{\circ})$, $M, \Omega \sim \mathrm{Unif}[0^{\circ}, 360^{\circ}]$).  All other parameters in the system were set to their nominal values. The resulting parameter space map is shown in the top panel of Fig.~\ref{Fig:TOI2095_Injected_Planet}. We note various bands of instability throughout a range of values of $a_d$. There are negligible variations with the mass, $M_d$. We also explored how these results change when the orbits of planets b and c are eccentric. We ran the same suite of simulations but randomly sampled the eccentricities of TOI-2095 b and c ($e \sim \mathrm{Rayleigh}(0.05)$). We found a wider region of instability in the $a_d - M_d$ space, along with small variations with the mass $M_d$. Specifically, for $M_d = 1 \ M_{\oplus}$, the band of instability has a width of $\sim 0.085$\, au ($0.075-0.16$\, au). On the other hand, when $M_d = 10 \ M_{\oplus}$, it increases in width to $\sim 0.12$\, au ($0.065-0.185$\, au).  

For the second type of simulations, we relaxed the assumption of a circular orbit for the hypothetical third planet and varied both its eccentricity and semi-major axis. Based on the results of the circular case, which showed negligible mass variations, we considered a fixed mass for planet d, $m_d = 5 \; M_{\oplus}$, and explored the $a_d - e_d$ parameter space in a $200 \times 200$ grid in which $a_d$ was varied from $0.01-0.3$\, au and $e_d$ was varied from $0-0.3$ (bottom panel of Fig.~\ref{Fig:TOI2095_Injected_Planet}). Each pixel contains the average of $10$ realizations of the system with randomized values for the inclinations, mean anomalies, and longitudes of ascending node ($i \sim \mathrm{Rayleigh}(1^{\circ})$, $M, \Omega \sim \mathrm{Unif}[0^{\circ}, 360^{\circ}]$). The map indicates that a wide swath of parameter space is disallowed depending on the eccentricity of planet d.

\subsection{Propects for transit timing variations follow-up}
\label{sec:dyn_ttv}

Our best fit (described in Sect.~\ref{Sec:joint}) indicates that the resulting period ratio $P_{c}/P_{b}$ is 1.595,  placing the system within $\sim$10\% of the first-order 3:2 mean-motion resonance. In such a configuration, the gravitational pull exerted between the planets may lead to mutual orbital excitation, which induces measurable TTVs \citep{agol2005,holman2005}. In Sect.~\ref{Sec:joint}, we describe our search for hints of these TTVs in the TESS data; unfortunately, we concluded that the low S/N of individual transits prevents us from finding any deviation from their linear prediction. 

Hence, in this section, we explore the amenability of this system to be followed up by a dedicated observational campaign in the search for TTVs that allows for the planetary masses to be estimated. To this end, we conducted a 
detailed analysis of the expected TTVs amplitudes for TOI-2095b and c, following the strategy presented by \cite{pozuelos2023}.

This strategy consisted of generating 1000 synthetic configurations by drawing orbital periods and mid-transit times from the values reported in Table~\ref{Tab:Planet_parameters} and following normal distributions. 
We imported the planetary mass distributions from the posteriors found in the joint analysis (see Fig.~\ref{Fig:TOI2095_MassDistri}). During our fitting process, for simplicity, we assumed circular orbits. However, 
some level of eccentricity might exist. Indeed, recent studies suggest a relationship connecting the number of planets in a given system and their eccentricities, where it has been found that the greater the number of planets,
the lower the eccentricities \citep[see, e.g., ][]{limbach2015,zinzi2017,zhu2018}. For two-planet systems, the distribution of eccentricities follows a lognormal distribution 
with $\mu$=-1.98 and $\sigma$=0.67 \citep{mathias2020}. We drew the eccentricities for our synthetic systems from such a distribution, adding the additional constraint obtained in Sect.~\ref{sec:Dyn} that limits the mutual values by Eq.~\ref{eq:regionA}.   

Moreover, we sampled the argument of periastron and the mean anomaly following a random distribution from 0 to 2$\pi$. Then, to compute the TTVs amplitude for each synthetic system, we used the \texttt{TTVFast2Furious} package \citep{hadden2019}. It is important to note that in this process, there is a non-null 
probability of obtaining unstable configurations due to drawing extreme values of masses and eccentricities simultaneously (not yet considered in Sect.~\ref{sec:Dyn}). Then, to prevent
considering unrealistic architectures in our analysis, we evaluated the stability of each scenario by computing the MEGNO parameter for an integration time of 10$^{5}$ orbits of the TOI-2095b, and
took only those with $\Delta \langle Y(t) \rangle$= 2.0 $-$ $\langle Y(t) \rangle < 0.1 $, which corresponded to $\sim$92\% of the drawn systems.  

\begin{figure}
    \centering
    \includegraphics[width=\columnwidth]{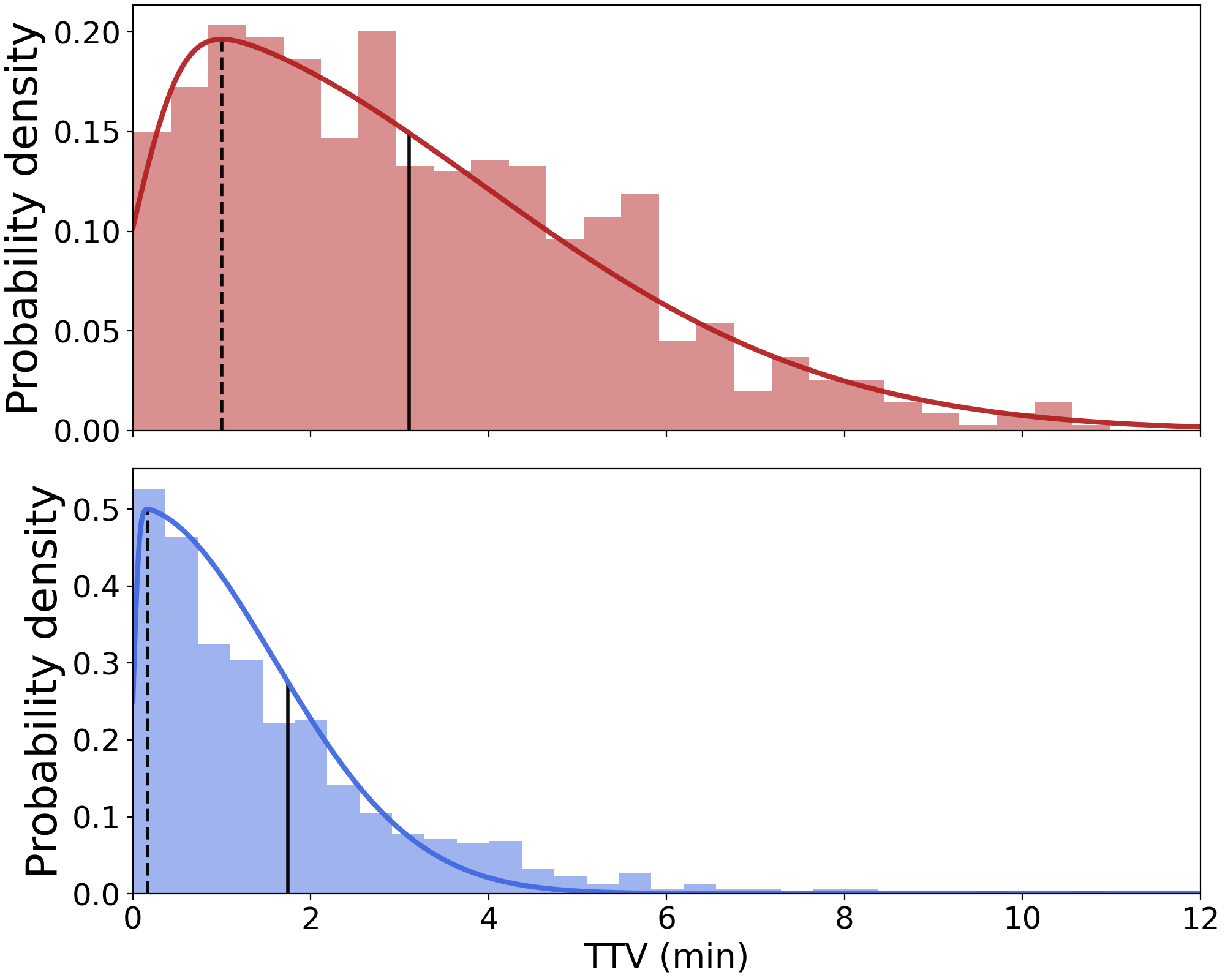}
    \caption{Expected TTVs amplitudes for planets TOI-2095b (upper panel) and TOI-2095c (lower panel). 
     Dashed and solid vertical lines correspond to the mode and the mean of the PDF, respectively.}
    \label{fig:ttv_expected}
\end{figure}

We found that the TTVs for each planet follow a nonsymmetric distribution, which we fit using the Skew-normal function from the \texttt{scipy} package \citep{scipy}. Following this procedure, we derived for each planet the  probability density function (PDF) for the TTVs (see Fig.~\ref{fig:ttv_expected}). On the one hand, we found for both planets that the modes of the PDF are in the sub-minute regime. On the other hand, the means of the PDFs are at $\sim$3.1 and $\sim$1.7~min, respectively. Hence, from these results, we concluded that measuring the planetary masses via TTVs is challenging and requires observations with mid-transit time precisions $\lesssim$1~min. 

The best individual mid-transit time precision that we achieved using \textit{TESS} for TOI-2095b and TOI-2095c are on the order of 5 to 7 minutes (Tables \ref{Tab:TTV_TOI2095b} and \ref{Tab:TTV_TOI2095c}), thus making the detection TTVs with this satellite difficult. Another space mission dedicated to study transiting planets is the CHaracterising ExOPlanet Satellite (CHEOPS, \citealp{Benz2021}). Although its photometric measurements might be precise enough ($<20$\,s precision for $G \sim 9$\,mag star, $< 2$\,min for $G \geq 11$\,mag; \citealp{Borsato2021}), to our knowledge, CHEOPS has not yet observed TOI-2095.
From the ground, the sub-millimagnitude transit depth of both planets make the detection of the transit with telescopes in the 1-2\,m range challenging; it is likely that the transits would need to be observed with telescopes larger than 4\,m to detect the transits and have accurate mid-transit time precisions. A candidate for such observations would be the 10.4\,m Gran Telescopio Canarias (GTC, \citealp{Cepa2000}) with its  HiPERCAM instrument \citep{Dhillon2021}. This is an optical camera capable of taking simultaneous images in five Sloan bands: $u$, $g$, $r$, $i$, and $z$. This instrument has reached individual mid-transit precisions of less than a minute for a faint ($V = 16.5$\,mag) \textit{TESS} transit candidate (Parviainen et al., in prep.).

\section{Discussion}
\label{Sec:Discussion}

\begin{figure*}
 \includegraphics[width=\hsize]{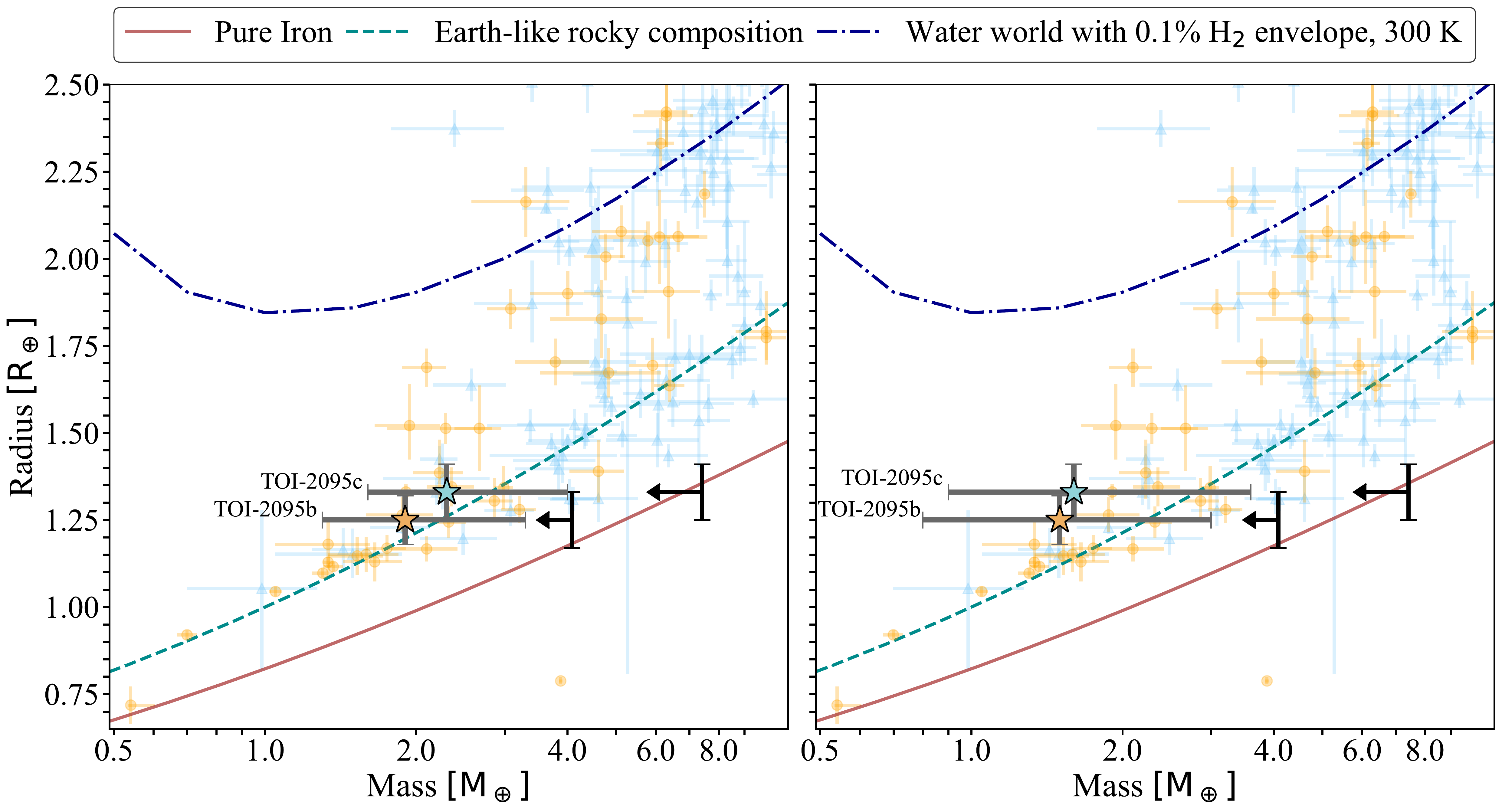}
   \caption{Mass-radius diagram of all known transiting exoplanets taken from TepCat \citep{Southworth2011}. The upper limits for the masses of TOI-2095b and TOI-2095c derived from RV measurements ($M_b < 4.1 \; M_\oplus$ and $M_c < 7.4 \; M_\oplus$ respectively) are delimited by the black arrows. The predicted masses and uncertainties for TOI-2095b and TOI-2095c computed with the mass-radius relations of \citet{Chen&Kipping2017} (left) and \citet{Kanodia2019} (right) are marked by the yellow and light blue star respectively. We show planets with $P< 30 $ days and with mass determinations with an uncertainty better than 30\%. Planets around stars with $T_{\rm eff} \leq 4000$ K and $T_{\rm eff}>4000$ K are marked by orange circles and blue triangles, respectively. The lines represent the composition models of \cite{Zeng2016, Zeng2019} for pure iron cores (100\% Fe, brown solid line), Earth-like rocky compositions (32.5\% Fe plus 67.5\% MgSiO$_3$, dash green line), and a water world (0.1\% H$_2$ envelope plus 49.95\% Earth-like rocky core plus 49.95\% H$_2$O, dash-dotted blue line).}
   \label{Fig:TOI2095_MassRadius}
\end{figure*}

TOI-2095b and TOI-2095c are a new addition to the growing list of transiting planets orbiting around M dwarfs. We derived the radii and upper mass limits for both planets ($R_b = 1.25 \pm 0.07 \; R_\oplus$, $M_b < 4.1 \; M_\oplus$; $R_c = 1.33 \pm 0.08 \; R_\oplus$, $M_c < 7.4\; M_\oplus$) and estimated their equilibrium temperatures ($T_{\mathrm{eq}\; b} = 347 \pm 9$ K, $T_{\mathrm{eq}\; c} = 297 \pm 8$ K; assuming an Earth-like Bond albedo of $A_\mathrm{Bond}=0.3$). Figure \ref{Fig:TOI2095_MassRadius} presents a mass-radius diagram of known transiting planets using data taken from TepCat (\citealp{Southworth2011}) and the composition models of \cite{Zeng2016, Zeng2019}. The models assume an isothermal atmosphere with 300 K and are truncated at a pressure level of 1 millibar (defining the radius of the planet). TOI-2095b and TOI-2095c are also shown in the diagram using their derived radii and upper mass limits. However, without a more precise mass measurement, it is hard to assess the bulk composition of TOI-2095b and TOI-2095c.

To estimate the true masses of TOI-2095b and TOI-2095c, we used the mass-radius ($M$-$R$) relations of \cite{Chen&Kipping2017} and \cite{Kanodia2019}. \cite{Chen&Kipping2017} $M$-$R$ relation was computed using a probabilistic approach based on mass and radius measurements across the parameter space starting from dwarf planets to late-type stars; the \texttt{Python} implementation of their $M$-$R$ relation, \texttt{forecaster}, is publicly available\footnote{\url{https://github.com/chenjj2/forecaster}}. \cite{Kanodia2019} implements the nonparametric $M$-$R$ relations of \cite{Ning2018}, we use \cite{Kanodia2019} nonparametric $M$-$R$ relation computed using the measurements of 24 exoplanets around M dwarfs. \cite{Kanodia2019} also provide a \texttt{Python} implementation of their $M$-$R$ relation called \texttt{MRExo}\footnote{\url{https://github.com/shbhuk/mrexo}}. To compute the masses of TOI-2095b and TOI-2095c with \texttt{forecaster} and \texttt{MRExo} we used as input the posterior distributions of the planet-to-star radius ratio multiplied by the stellar radius presented in Table \ref{Tab:Star}. Using \cite{Chen&Kipping2017} we find values of $M_b = 1.9^{+1.4}_{-0.6} \; M_\oplus$ and $M_c = 2.3^{+1.7}_{-0.7} \; M_\oplus$ for TOI-2095b and TOI-2095c, respectively. On the other hand, using \cite{Kanodia2019} we find lower median mass values of $M_b = 1.5^{+1.5}_{-0.7} \; M_\oplus$ for TOI-2095b and $M_c = 1.6^{+2.0}_{-0.7} \; M_\oplus$ for TOI-2095c. These values can only be regarded as indicative of the possible final masses. 

Despite their relatively small predicted RV amplitude, current instruments may be able to detect both planets with a more intensive follow-up campaign. For example, Wolf 1069b (\citealp{Kossakowski2023}) was discovered using CARMENES. This planet has an orbital period of 15.6 days (placing it in the HZ of their star) and a measured RV semi-amplitude of $K = 1.07 \pm 0.17 $\,m\,s$^{-1}$ (Sect.~\ref{Sec:rvs}). 
The ultra-short-period planet GJ 367b (\citealp{Lam2021}) was discovered using \textit{TESS} and has a mass measurement using HARPS. The reported RV semi-amplitude of this planet is $K = 0.79 \pm 0.11$\,m\,s$^{-1}$. In the case of 8\,m class telescopes, RV detections lower than the 1\,m\,s$^{-1}$ threshold has been achieved by ESPRESSO at the Very Large Telescope (Proxima Cen d candidate; \citealp{SuarezMascareno2020}, \citealp{Faria2022}). The stellar activity of the host star may present a challenge for the RV detection, but the combined use of contemporaneous photometric observations to monitor the stellar variability and GPs has delivered some results for the mass measurements of young planets \citep[e.g.,][]{SuarezMascareno2022}.

\cite{LuquePalle2022} recently proposed that small planets around M dwarf hosts seem to fall into two categories: an Earth-like composition or water world with equal mass fraction of ices and silicates. Using the upper limits of the planet masses, both planets in the TOI-2095 system would be compatible with both compositions. However, no water worlds were identified by \cite{LuquePalle2022} below 1.5 $R_\oplus$, which suggests that the planets may have an Earth-like composition.

\begin{figure*}
    \centering
    \includegraphics[width=\textwidth]{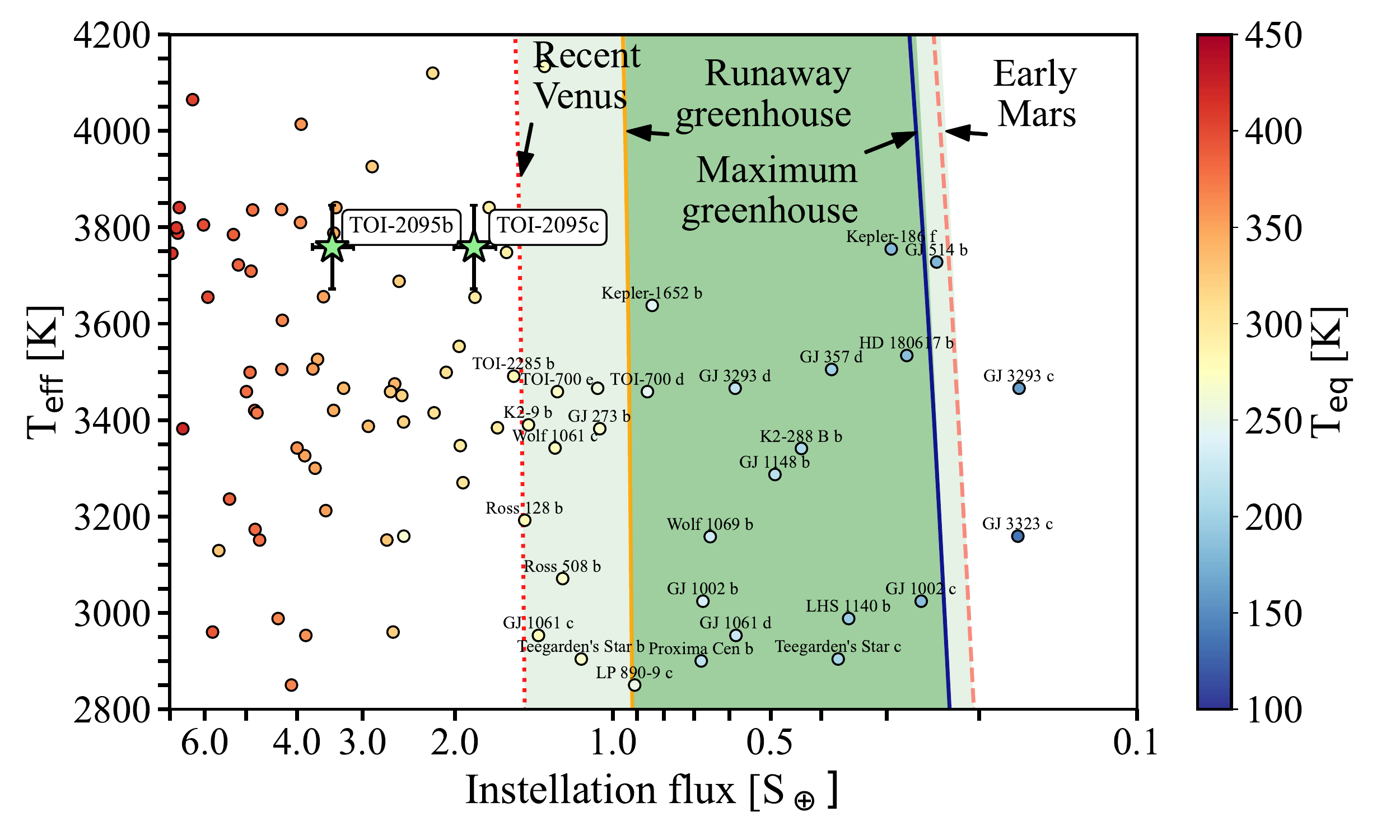}
    \caption{Incident flux in Earth units versus stellar effective temperature for known exoplanets (data taken from NASA Exoplanet Archive). The color of each circle represents the equilibrium temperature of the planet. The habitable zone limits of \cite{Kopparapu2013} are shown with lines. The position of TOI-2095b and TOI-2095c are marked by the green stars.}
    \label{Fig:TOI2095_HabZone}
\end{figure*}

Finally, Fig.~\ref{Fig:TOI2095_HabZone} shows the instellation flux in Earth units versus effective temperature of the planet host star. The vertical lines represent the limits of the habitable zone (HZ) defined by \cite{Kopparapu2013}. 
Both planets are on orbits that place them well within the runaway greenhouse limit, where planets having volatiles are expected to have atmospheres dominated by them until complete desiccation occurs~\citep{Ingersoll1969,Kasting1988}.
The question of how runaway greenhouse effects impact the volatile content and habitability of terrestrial-sized exoplanets is a subject of current research~\citep[e.g.,][]{Turbet2019,Mousis2020} and such atmospheric studies of planets within the runaway greenhouse region may provide valuable insights~\citep{Suissa2020}.
A different approach is to statistically investigate instellation-induced changes of bulk parameters such as planetary radius, whose magnitudes are predicted to exceed the resolution of state-of-the-art high precision photometry~\citep{Turbet2020,Dorn2021}.
Planets close to the runaway greenhouse limit, such as TOI-2095b and TOI-2095c, are key in such attempts to empirically test the HZ hypothesis.
This holds, in particular, if additional outer planets (whose existence we cannot test with the currently available data) allow us to probe planets on both sides of the instellation threshold within the same system~\citep{Turbet2019}.

\section{Conclusions}
\label{Sec:Conclusions}

We report the validation of two transiting planets around the M dwarf TOI-2095 discovered by \textit{TESS}. We use ground-based high-resolution imaging, \textit{TESS} photometric data, and CARMENES RVs to discard false positive scenarios, measure the planetary radii, and place stringent upper limits on the masses of the transiting candidates. 

The star is an M dwarf located at a distance $d=41.90 \pm 0.03$ pc and is relatively bright in the near-infrared ($J = 9.8$ mag, $K_s = 8.9$ mag). We derive an effective temperature of $T_{\rm eff} = 3759 \pm 87$ K and an iron abundance of [Fe/H]$=-0.24 \pm 0.04$ dex, as well as a stellar mass and radius of $M = 0.44 \pm 0.02 \; M_\odot$ and $R = 0.44 \pm 0.02 \; R_\odot$, respectively. Using ground-based photometric observations, we determined a stellar rotation period of $P_{\rm rot} = 40^{+0.2}_{-0.4}$ days, in agreement with the spectroscopic activity indices measured from CARMENES data.

In order to obtain the transit and orbital parameters of the system we fitted 22 sectors of \textit{TESS} data and CARMENES RV measurements simultaneously, while taking into account the red noise present in both time series using Gaussian processes. We assumed circular orbits for both transiting planets. We find that the inner planet, TOI-2095b, has an orbital period of $P_b=17.665$ days, a radius of $R_b = 1.25 \pm 0.07 \; R_\oplus$, an upper mass limit of $M_b < 4.1 \; M_\oplus$, and an equilibrium temperature of $T_{eq} = 347 \pm 9$ K (assuming a Bond albedo of 0.3). The other transiting planet, TOI-2095c, has an orbital period of $P_c=28.172$ days, a radius of $R_c = 1.33 \pm 0.08 \; R_\oplus$, an upper mass limit of $M_c < 7.4 \; M_\oplus$, and an equilibrium temperature of $T_{eq} = 297 \pm 8$ K (assuming a Bond albedo of 0.3). 

Both planets present interesting sizes and temperatures that makes them attractive targets for further follow-up observations. In particular, extremely precise RV follow-up observations can help to improve the mass measurements (and, hence, the bulk densities) of these planets and provide some constraints for future prospects for atmospheric characterizations.

\begin{acknowledgements}
      
    We thank the anonymous referee for the constructive report.
    
  CARMENES is an instrument at the Centro Astron\'omico Hispano en Andaluc\'ia (CAHA) at Calar Alto (Almer\'{\i}a, Spain), operated jointly by the Junta de Andaluc\'ia and the Instituto de Astrof\'isica de Andaluc\'ia (CSIC).
  CARMENES was funded by the Max-Planck-Gesellschaft (MPG), 
  the Consejo Superior de Investigaciones Cient\'{\i}ficas (CSIC),
  the Ministerio de Econom\'ia y Competitividad (MINECO) and the European Regional Development Fund (ERDF) through projects FICTS-2011-02, ICTS-2017-07-CAHA-4, and CAHA16-CE-3978, 
  and the members of the CARMENES Consortium 
  (Max-Planck-Institut f\"ur Astronomie,
  Instituto de Astrof\'{\i}sica de Andaluc\'{\i}a,
  Landessternwarte K\"onigstuhl,
  Institut de Ci\`encies de l'Espai,
  Institut f\"ur Astrophysik G\"ottingen,
  Universidad Complutense de Madrid,
  Th\"uringer Landessternwarte Tautenburg,
  Instituto de Astrof\'{\i}sica de Canarias,
  Hamburger Sternwarte,
  Centro de Astrobiolog\'{\i}a and
  Centro Astron\'omico Hispano-Alem\'an), 
  with additional contributions by the MINECO, 
  the Deutsche Forschungsgemeinschaft through the Major Research Instrumentation Programme and Research Unit FOR2544 ``Blue Planets around Red Stars'', 
  the Klaus Tschira Stiftung, 
  the states of Baden-W\"urttemberg and Niedersachsen, 
  and by the Junta de Andaluc\'{\i}a.
  
      This paper includes data collected by the \textit{TESS} mission, which are publicly available from the Mikulski Archive for Space Telescopes (MAST). Funding for the \textit{TESS} mission is provided by NASA's Science Mission directorate. Resources supporting this work were provided by the NASA High-End Computing (HEC) Program through the NASA Advanced Supercomputing (NAS) Division at Ames Research Center for the production of the SPOC data products. We acknowledge the use of public \textit{TESS} data from pipelines at the \textit{TESS} Science Office and at the \textit{TESS} Science Processing Operations Center. This research has made use of the Exoplanet Follow-up Observation Program website, which is operated by the California Institute of Technology, under contract with the National Aeronautics and Space Administration under the Exoplanet Exploration Program. 
      
      This work has made use of data from the European Space Agency (ESA) mission {\it Gaia} (\url{https://www.cosmos.esa.int/gaia}), processed by the {\it Gaia} Data Processing and Analysis Consortium (DPAC, \url{https://www.cosmos.esa.int/web/gaia/dpac/consortium}). Funding for the DPAC has been provided by national institutions, in particular the institutions participating in the {\it Gaia} Multilateral Agreement.
      
      This work makes use of observations from the LCOGT network. Part of the LCOGT telescope time was granted by NOIRLab through the Mid-Scale Innovations Program (MSIP). MSIP is funded by NSF.
      
      The Telescopi Joan Or\'o (TJO) of the Montsec Observatory (OdM) is owned by the Generalitat de Catalunya and operated by the Institute for Space Studies of Catalonia (IEEC).
      
      This paper is based on observations made with the MuSCAT3 instrument, developed by the Astrobiology Center and under financial supports by JSPS KAKENHI (JP18H05439) and JST PRESTO (JPMJPR1775), at Faulkes Telescope North on Maui, HI, operated by the Las Cumbres Observatory.
      
      Based on observations obtained with the Samuel Oschin 48-inch Telescope at the Palomar Observatory as part of the Zwicky Transient Facility project. ZTF is supported by the National Science Foundation under Grant No. AST-1440341 and a collaboration including Caltech, IPAC, the Weizmann Institute for Science, the Oskar Klein Center at Stockholm University, the University of Maryland, the University of Washington, Deutsches Elektronen-Synchrotron and Humboldt University, Los Alamos National Laboratories, the TANGO Consortium of Taiwan, the University of Wisconsin at Milwaukee, and Lawrence Berkeley National Laboratories. Operations are conducted by COO, IPAC, and UW.
      
      The results reported herein benefited from collaborations and/or information exchange within the program ``Alien Earths'' (supported by the National Aeronautics and Space Administration under agreement No. 80NSSC21K0593) for NASA’s Nexus for Exoplanet System Science (NExSS) research coordination network sponsored by NASA’s Science Mission Directorate.

      This publication makes use of data products from the Wide-field
      Infrared Survey Explorer, which is a joint project of the University of
      California, Los Angeles, and the Jet Propulsion Laboratory/California
      Institute of Technology, funded by the National Aeronautics and Space
      Administration.

    We acknowledge financial support from the Agencia Estatal de Investigaci\'on (AEI/10.13039/501100011033) of the Ministerio de Ciencia e Innovaci\'on and the ERDF ``A way of making Europe'' through projects 
    PID2021-125627OB-C31,               
    PID2019-109522GB-C5[1:4],   
    and the Centre of Excellence ``Severo Ochoa'' and ``Mar\'ia de Maeztu'' awards to the Instituto de Astrof\'isica de Canarias (CEX2019-000920-S), Instituto de Astrof\'isica de Andaluc\'ia (SEV-2017-0709) and Institut de Ci\`encies de l'Espai (CEX2020-001058-M).

    This work was also funded by the Generalitat de Catalunya/CERCA programme; 
    Swedish National Space Agency (SNSA; DNR 2020-00104) and Swedish Research Council (VR: Etableringsbidrag 2017-04945);
    Gobierno de Canarias and ERDF through project ProID2021010128;
    Japan Society for the Promotion of Science (JSPS) KAKENHI (JP17H04574, JP18H05439, JP21K20376), and Japan Science and Technology Agency (JST) CREST (JPMJCR1761);
    Ariel Postdoctoral Fellowship program of the Swedish National Space Agency (SNSA);
    Ministry of Science and Higher Education of the Russian Federation under the grant 075-15-2020-780 (N13.1902.21.0039);
    and Massachusetts Institute of Technology Undergraduate Research Opportunities Program (UROP).
      
      
\end{acknowledgements}

%
%

\bibliographystyle{aa}
\bibliography{references}

\begin{thebibliography}{166}
\expandafter\ifx\csname natexlab\endcsname\relax\def\natexlab#1{#1}\fi

\bibitem[{{Agol} {et~al.}(2005){Agol}, {Steffen}, {Sari}, \&
  {Clarkson}}]{agol2005}
{Agol}, E., {Steffen}, J., {Sari}, R., \& {Clarkson}, W. 2005, \mnras, 359, 567

\bibitem[{{Aller} {et~al.}(2020){Aller}, {Lillo-Box}, {Jones}, {Miranda}, \&
  {Barcel{\'o} Forteza}}]{Aller2020}
{Aller}, A., {Lillo-Box}, J., {Jones}, D., {Miranda}, L.~F., \& {Barcel{\'o}
  Forteza}, S. 2020, \aap, 635, A128

\bibitem[{{Ambikasaran} {et~al.}(2015){Ambikasaran}, {Foreman-Mackey},
  {Greengard}, {Hogg}, \& {O'Neil}}]{Ambikasaran2015}
{Ambikasaran}, S., {Foreman-Mackey}, D., {Greengard}, L., {Hogg}, D.~W., \&
  {O'Neil}, M. 2015, IEEE Transactions on Pattern Analysis and Machine
  Intelligence, 38, 252

\bibitem[{{Bailer-Jones} {et~al.}(2021){Bailer-Jones}, {Rybizki}, {Fouesneau},
  {Demleitner}, \& {Andrae}}]{Bailer-Jones2021}
{Bailer-Jones}, C.~A.~L., {Rybizki}, J., {Fouesneau}, M., {Demleitner}, M., \&
  {Andrae}, R. 2021, \aj, 161, 147

\bibitem[{{Batalha} {et~al.}(2018){Batalha}, {Lewis}, {Line}, {Valenti}, \&
  {Stevenson}}]{Batalha2018}
{Batalha}, N.~E., {Lewis}, N.~K., {Line}, M.~R., {Valenti}, J., \& {Stevenson},
  K. 2018, \apjl, 856, L34

\bibitem[{{Bauer} {et~al.}(2020){Bauer}, {Zechmeister}, {Kaminski},
  {Rodr{\'\i}guez L{\'o}pez}, {Caballero}, {Azzaro}, {Stahl}, {Kossakowski},
  {Quirrenbach}, {Becerril Jarque}, {Rodr{\'\i}guez}, {Amado}, {Seifert},
  {Reiners}, {Sch{\"a}fer}, {Ribas}, {B{\'e}jar}, {Cort{\'e}s-Contreras},
  {Dreizler}, {Hatzes}, {Henning}, {Jeffers}, {K{\"u}rster}, {Lafarga},
  {Montes}, {Morales}, {Schmitt}, {Schweitzer}, \& {Solano}}]{Bauer2020}
{Bauer}, F.~F., {Zechmeister}, M., {Kaminski}, A., {et~al.} 2020, \aap, 640,
  A50

\bibitem[{{Bauer} {et~al.}(2015){Bauer}, {Zechmeister}, \&
  {Reiners}}]{Bauer2015}
{Bauer}, F.~F., {Zechmeister}, M., \& {Reiners}, A. 2015, \aap, 581, A117

\bibitem[{{Benz} {et~al.}(2021){Benz}, {Broeg}, {Fortier}, {Rando}, {Beck},
  {Beck}, {Queloz}, {Ehrenreich}, {Maxted}, {Isaak}, {Billot}, {Alibert},
  {Alonso}, {Ant{\'o}nio}, {Asquier}, {Bandy}, {B{\'a}rczy}, {Barrado},
  {Barros}, {Baumjohann}, {Bekkelien}, {Bergomi}, {Biondi}, {Bonfils},
  {Borsato}, {Brandeker}, {Busch}, {Cabrera}, {Cessa}, {Charnoz}, {Chazelas},
  {Collier Cameron}, {Corral Van Damme}, {Cortes}, {Davies}, {Deleuil},
  {Deline}, {Delrez}, {Demangeon}, {Demory}, {Erikson}, {Farinato}, {Fossati},
  {Fridlund}, {Futyan}, {Gandolfi}, {Garcia Munoz}, {Gillon}, {Guterman},
  {Gutierrez}, {Hasiba}, {Heng}, {Hernandez}, {Hoyer}, {Kiss}, {Kovacs},
  {Kuntzer}, {Laskar}, {Lecavelier des Etangs}, {Lendl}, {L{\'o}pez}, {Lora},
  {Lovis}, {L{\"u}ftinger}, {Magrin}, {Malvasio}, {Marafatto}, {Michaelis}, {de
  Miguel}, {Modrego}, {Munari}, {Nascimbeni}, {Olofsson}, {Ottacher},
  {Ottensamer}, {Pagano}, {Palacios}, {Pall{\'e}}, {Peter}, {Piazza}, {Piotto},
  {Pizarro}, {Pollaco}, {Ragazzoni}, {Ratti}, {Rauer}, {Ribas}, {Rieder},
  {Rohlfs}, {Safa}, {Salatti}, {Santos}, {Scandariato}, {S{\'e}gransan},
  {Simon}, {Smith}, {Sordet}, {Sousa}, {Steller}, {Szab{\'o}}, {Szoke},
  {Thomas}, {Tschentscher}, {Udry}, {Van Grootel}, {Viotto}, {Walter},
  {Walton}, {Wildi}, \& {Wolter}}]{Benz2021}
{Benz}, W., {Broeg}, C., {Fortier}, A., {et~al.} 2021, Experimental Astronomy,
  51, 109

\bibitem[{{Bonfils} {et~al.}(2005){Bonfils}, {Delfosse}, {Udry}, {Santos},
  {Forveille}, \& {S{\'e}gransan}}]{Bonfils2005}
{Bonfils}, X., {Delfosse}, X., {Udry}, S., {et~al.} 2005, \aap, 442, 635

\bibitem[{{Borsato} {et~al.}(2021){Borsato}, {Piotto}, {Gandolfi},
  {Nascimbeni}, {Lacedelli}, {Marzari}, {Billot}, {Maxted}, {Sousa}, {Cameron},
  {Bonfanti}, {Wilson}, {Serrano}, {Garai}, {Alibert}, {Alonso}, {Asquier},
  {B{\'a}rczy}, {Bandy}, {Barrado}, {Barros}, {Baumjohann}, {Beck}, {Beck},
  {Benz}, {Bonfils}, {Brandeker}, {Broeg}, {Cabrera}, {Charnoz}, {Csizmadia},
  {Davies}, {Deleuil}, {Delrez}, {Demangeon}, {Demory}, {des Etangs},
  {Ehrenreich}, {Erikson}, {Escud{\'e}}, {Fortier}, {Fossati}, {Fridlund},
  {Gillon}, {Guedel}, {Hasiba}, {Heng}, {Hoyer}, {Isaak}, {Kiss}, {Kopp},
  {Laskar}, {Lendl}, {Lovis}, {Magrin}, {Munari}, {Olofsson}, {Ottensamer},
  {Pagano}, {Pall{\'e}}, {Peter}, {Pollacco}, {Queloz}, {Ragazzoni}, {Rando},
  {Rauer}, {Ribas}, {S{\'e}gransan}, {Santos}, {Scandariato}, {Simon}, {Smith},
  {Steller}, {Szab{\'o}}, {Thomas}, {Udry}, {Van Grootel}, \&
  {Walton}}]{Borsato2021}
{Borsato}, L., {Piotto}, G., {Gandolfi}, D., {et~al.} 2021, \mnras, 506, 3810

\bibitem[{{Brown} {et~al.}(2013){Brown}, {Baliber}, {Bianco}, {Bowman},
  {Burleson}, {Conway}, {Crellin}, {Depagne}, {De Vera}, {Dilday}, {Dragomir},
  {Dubberley}, {Eastman}, {Elphick}, {Falarski}, {Foale}, {Ford}, {Fulton},
  {Garza}, {Gomez}, {Graham}, {Greene}, {Haldeman}, {Hawkins}, {Haworth},
  {Haynes}, {Hidas}, {Hjelstrom}, {Howell}, {Hygelund}, {Lister}, {Lobdill},
  {Martinez}, {Mullins}, {Norbury}, {Parrent}, {Paulson}, {Petry}, {Pickles},
  {Posner}, {Rosing}, {Ross}, {Sand}, {Saunders}, {Shobbrook}, {Shporer},
  {Street}, {Thomas}, {Tsapras}, {Tufts}, {Valenti}, {Vander Horst}, {Walker},
  {White}, \& {Willis}}]{Brown2013}
{Brown}, T.~M., {Baliber}, N., {Bianco}, F.~B., {et~al.} 2013, \pasp, 125, 1031

\bibitem[{{Caballero} {et~al.}(2016{\natexlab{a}}){Caballero},
  {Cort{\'e}s-Contreras}, {Alonso-Floriano}, {Montes}, {Quirrenbach}, {Amado},
  {Ribas}, {Reiners}, {Abellan}, {B{\'e}jar}, {Brinkm{\"o}ller}, {Czesla},
  {Dorda}, {Gallardo}, {Gonz{\'a}lez-{\'A}lvarez}, {Hidalgo}, {Holgado},
  {Jeffers}, {Kim}, {Klutsch}, {Lamert}, {Llamas}, {L{\'o}pez-Santiago},
  {Mart{\'\i}nez-Rodr{\'\i}guez}, {Morales}, {Mundt}, {Passegger},
  {Sch{\"o}fer}, {Seifert}, \& {Zechmeister}}]{2016csss.confE.148C}
{Caballero}, J.~A., {Cort{\'e}s-Contreras}, M., {Alonso-Floriano}, F.~J.,
  {et~al.} 2016{\natexlab{a}}, in 19th Cambridge Workshop on Cool Stars,
  Stellar Systems, and the Sun (CS19), Cambridge Workshop on Cool Stars,
  Stellar Systems, and the Sun, 148

\bibitem[{{Caballero} {et~al.}(2016{\natexlab{b}}){Caballero}, {Gu{\`a}rdia},
  {L{\'o}pez del Fresno}, {Zechmeister}, {de Juan}, {Alonso-Floriano}, {Amado},
  {Colom{\'e}}, {Cort{\'e}s-Contreras}, {Garc{\'\i}a-Piquer}, {Gesa}, {de
  Guindos}, {Hagen}, {Helmling}, {Hern{\'a}ndez Casta{\~n}o}, {K{\"u}rster},
  {L{\'o}pez-Santiago}, {Montes}, {Morales Mu{\~n}oz}, {Pavlov}, {Quirrenbach},
  {Reiners}, {Ribas}, {Seifert}, \& {Solano}}]{Caballero2016}
{Caballero}, J.~A., {Gu{\`a}rdia}, J., {L{\'o}pez del Fresno}, M., {et~al.}
  2016{\natexlab{b}}, in Society of Photo-Optical Instrumentation Engineers
  (SPIE) Conference Series, Vol. 9910, Observatory Operations: Strategies,
  Processes, and Systems VI, ed. A.~B. {Peck}, R.~L. {Seaman}, \& C.~R. {Benn},
  99100E

\bibitem[{{Cabrera} {et~al.}(2017){Cabrera}, {Barros}, {Armstrong}, {Hidalgo},
  {Santos}, {Almenara}, {Alonso}, {Deleuil}, {Demangeon}, {D{\'\i}az}, {Lendl},
  {Pfaff}, {Rauer}, {Santerne}, {Serrano}, \& {Zucker}}]{2017A&A...606A..75C}
{Cabrera}, J., {Barros}, S.~C.~C., {Armstrong}, D., {et~al.} 2017, \aap, 606,
  A75

\bibitem[{{Casagrande} {et~al.}(2008){Casagrande}, {Flynn}, \&
  {Bessell}}]{Casagrande2008}
{Casagrande}, L., {Flynn}, C., \& {Bessell}, M. 2008, \mnras, 389, 585

\bibitem[{{Casagrande} {et~al.}(2011){Casagrande}, {Sch{\"o}nrich}, {Asplund},
  {Cassisi}, {Ram{\'\i}rez}, {Mel{\'e}ndez}, {Bensby}, \&
  {Feltzing}}]{Casagrande2011}
{Casagrande}, L., {Sch{\"o}nrich}, R., {Asplund}, M., {et~al.} 2011, \aap, 530,
  A138

\bibitem[{{Castro-Gonz{\'a}lez} {et~al.}(2023){Castro-Gonz{\'a}lez},
  {Demangeon}, {Lillo-Box}, {Lovis}, {Lavie}, {Adibekyan}, {Acu{\~n}a},
  {Deleuil}, {Aguichine}, {Zapatero Osorio}, {Tabernero}, {Davoult}, {Alibert},
  {Santos}, {Sousa}, {Antoniadis-Karnavas}, {Borsa}, {Winn}, {Allende Prieto},
  {Figueira}, {Jenkins}, {Sozzetti}, {Damasso}, {Silva}, {Astudillo-Defru},
  {Barros}, {Bonfils}, {Cristiani}, {Di Marcantonio}, {Gonz{\'a}lez
  Hern{\'a}ndez}, {Curto}, {Martins}, {Nunes}, {Palle}, {Pepe}, {Seager}, \&
  {Su{\'a}rez Mascare{\~n}o}}]{2023A&A...675A..52C}
{Castro-Gonz{\'a}lez}, A., {Demangeon}, O.~D.~S., {Lillo-Box}, J., {et~al.}
  2023, \aap, 675, A52

\bibitem[{{Castro-Gonz{\'a}lez} {et~al.}(2022){Castro-Gonz{\'a}lez}, {D{\'\i}ez
  Alonso}, {Men{\'e}ndez Blanco}, {Livingston}, {de Leon}, {Lillo-Box},
  {Korth}, {Fern{\'a}ndez Men{\'e}ndez}, {Recio}, {Izquierdo-Ruiz}, {Coya
  Lozano}, {Garc{\'\i}a de la Cuesta}, {G{\'o}mez Hern{\'a}ndez}, {Vidal
  Blanco}, {Hevia D{\'\i}az}, {Pardo Silva}, {P{\'e}rez Acevedo}, {Polancos
  Ruiz}, {Padilla Tijer{\'\i}n}, {V{\'a}zquez Garc{\'\i}a}, {Su{\'a}rez
  G{\'o}mez}, {Garc{\'\i}a Riesgo}, {Gonz{\'a}lez Guti{\'e}rrez}, {Bonavera},
  {Gonz{\'a}lez-Nuevo}, {Rodr{\'\i}guez Pereira}, {S{\'a}nchez Lasheras},
  {S{\'a}nchez Rodr{\'\i}guez}, {Mu{\~n}iz}, {Santos Rodr{\'\i}guez}, \& {de
  Cos Juez}}]{2022MNRAS.509.1075C}
{Castro-Gonz{\'a}lez}, A., {D{\'\i}ez Alonso}, E., {Men{\'e}ndez Blanco}, J.,
  {et~al.} 2022, \mnras, 509, 1075

\bibitem[{{Castro Gonz{\'a}lez} {et~al.}(2020){Castro Gonz{\'a}lez}, {D{\'\i}ez
  Alonso}, {Men{\'e}ndez Blanco}, {Livingston}, {de Leon}, {Su{\'a}rez
  G{\'o}mez}, {Gonz{\'a}lez Guti{\'e}rrez}, {Garc{\'\i}a Riesgo}, {Bonavera},
  {Iglesias Rodr{\'\i}guez}, {Mu{\~n}iz}, {Everett}, {Scott}, {Howell},
  {Ciardi}, {Gonzales}, {Schlieder}, \& {de Cos Juez}}]{2020MNRAS.499.5416C}
{Castro Gonz{\'a}lez}, A., {D{\'\i}ez Alonso}, E., {Men{\'e}ndez Blanco}, J.,
  {et~al.} 2020, \mnras, 499, 5416

\bibitem[{{Cepa} {et~al.}(2000){Cepa}, {Aguiar}, {Escalera},
  {Gonzalez-Serrano}, {Joven-Alvarez}, {Peraza}, {Rasilla}, {Rodriguez-Ramos},
  {Gonzalez}, {Cobos Duenas}, {Sanchez}, {Tejada}, {Bland-Hawthorn},
  {Militello}, \& {Rosa}}]{Cepa2000}
{Cepa}, J., {Aguiar}, M., {Escalera}, V.~G., {et~al.} 2000, in Society of
  Photo-Optical Instrumentation Engineers (SPIE) Conference Series, Vol. 4008,
  Optical and IR Telescope Instrumentation and Detectors, ed. M.~{Iye} \& A.~F.
  {Moorwood}, 623--631

\bibitem[{{Chen} \& {Kipping}(2017)}]{Chen&Kipping2017}
{Chen}, J. \& {Kipping}, D. 2017, \apj, 834, 17

\bibitem[{{Christiansen} {et~al.}(2022){Christiansen}, {Bhure}, {Zink},
  {Hardegree-Ullman}, {Adkins}, {Hedges}, {Morton}, {Bieryla}, {Ciardi},
  {Cochran}, {Dressing}, {Everett}, {Isaacson}, {Livingston}, {Ziegler},
  {Berlind}, {Calkins}, {Esquerdo}, {Latham}, {Endl}, {MacQueen}, {Fulton},
  {Hirsch}, {Howard}, {Weiss}, {Allen}, {Berberyann}, {Ciardi}, {Dunlavy},
  {Glassford}, {Dai}, {Hirano}, {Tamura}, {Beichman}, {Gonzales}, {Schlieder},
  {Barclay}, {Crossfield}, {Gilbert}, {Matthews}, {Giacalone}, \&
  {Petigura}}]{2022AJ....163..244C}
{Christiansen}, J.~L., {Bhure}, S., {Zink}, J.~K., {et~al.} 2022, \aj, 163, 244

\bibitem[{{Cifuentes} {et~al.}(2020){Cifuentes}, {Caballero},
  {Cort{\'e}s-Contreras}, {Montes}, {Abell{\'a}n}, {Dorda}, {Holgado},
  {Zapatero Osorio}, {Morales}, {Amado}, {Passegger}, {Quirrenbach}, {Reiners},
  {Ribas}, {Sanz-Forcada}, {Schweitzer}, {Seifert}, \&
  {Solano}}]{Cifuentes2020}
{Cifuentes}, C., {Caballero}, J.~A., {Cort{\'e}s-Contreras}, M., {et~al.} 2020,
  \aap, 642, A115

\bibitem[{{Cincotta} {et~al.}(2003){Cincotta}, {Giordano}, \&
  {Sim{\'o}}}]{2003PhyD..182..151C}
{Cincotta}, P.~M., {Giordano}, C.~M., \& {Sim{\'o}}, C. 2003, Physica D
  Nonlinear Phenomena, 182, 151

\bibitem[{{Cincotta} \& {Sim{\'o}}(2000)}]{2000A&AS..147..205C}
{Cincotta}, P.~M. \& {Sim{\'o}}, C. 2000, \aaps, 147, 205

\bibitem[{{Collins} {et~al.}(2017){Collins}, {Kielkopf}, {Stassun}, \&
  {Hessman}}]{Collins2017}
{Collins}, K.~A., {Kielkopf}, J.~F., {Stassun}, K.~G., \& {Hessman}, F.~V.
  2017, \aj, 153, 77

\bibitem[{{Colome} \& {Ribas}(2006)}]{Colome2006}
{Colome}, J. \& {Ribas}, I. 2006, IAU Special Session, 6, 11

\bibitem[{{Cort{\'e}s Contreras}(2017)}]{CortesContreras2017}
{Cort{\'e}s Contreras}, M. 2017, PhD thesis, Complutense University of Madrid,
  Spain

\bibitem[{{Cumming} {et~al.}(1999){Cumming}, {Marcy}, \&
  {Butler}}]{Cumming1999}
{Cumming}, A., {Marcy}, G.~W., \& {Butler}, R.~P. 1999, \apj, 526, 890

\bibitem[{{Curtis} {et~al.}(2020){Curtis}, {Ag{\"u}eros}, {Matt}, {Covey},
  {Douglas}, {Angus}, {Saar}, {Cody}, {Vanderburg}, {Law}, {Kraus}, {Latham},
  {Baranec}, {Riddle}, {Ziegler}, {Lund}, {Torres}, {Meibom}, {Aguirre}, \&
  {Wright}}]{Curtis2020}
{Curtis}, J.~L., {Ag{\"u}eros}, M.~A., {Matt}, S.~P., {et~al.} 2020, \apj, 904,
  140

\bibitem[{{Cutri} {et~al.}(2021){Cutri}, {Wright}, {Conrow}, {Fowler},
  {Eisenhardt}, {Grillmair}, {Kirkpatrick}, {Masci}, {McCallon}, {Wheelock},
  {Fajardo-Acosta}, {Yan}, {Benford}, {Harbut}, {Jarrett}, {Lake}, {Leisawitz},
  {Ressler}, {Stanford}, {Tsai}, {Liu}, {Helou}, {Mainzer}, {Gettngs},
  {Gonzalez}, {Hoffman}, {Marsh}, {Padgett}, {Skrutskie}, {Beck}, {Papin}, \&
  {Wittman}}]{CutriWISECat2014}
{Cutri}, R.~M., {Wright}, E.~L., {Conrow}, T., {et~al.} 2021, VizieR Online
  Data Catalog, II/328

\bibitem[{{de Leon} {et~al.}(2021){de Leon}, {Livingston}, {Endl}, {Cochran},
  {Hirano}, {Garc{\'\i}a}, {Mathur}, {Lam}, {Korth}, {Trani}, {Dai}, {D{\'\i}ez
  Alonso}, {Castro-Gonz{\'a}lez}, {Fridlund}, {Fukui}, {Gandolfi}, {Kabath},
  {Kuzuhara}, {Luque}, {Savel}, {Gill}, {Dressing}, {Giacalone}, {Narita},
  {Palle}, {Van Eylen}, \& {Tamura}}]{2021MNRAS.508..195D}
{de Leon}, J.~P., {Livingston}, J., {Endl}, M., {et~al.} 2021, \mnras, 508, 195

\bibitem[{{Delrez} {et~al.}(2021){Delrez}, {Ehrenreich}, {Alibert}, {Bonfanti},
  {Borsato}, {Fossati}, {Hooton}, {Hoyer}, {Pozuelos}, {Salmon}, {Sulis},
  {Wilson}, {Adibekyan}, {Bourrier}, {Brandeker}, {Charnoz}, {Deline},
  {Guterman}, {Haldemann}, {Hara}, {Oshagh}, {Sousa}, {Van Grootel}, {Alonso},
  {Anglada-Escud{\'e}}, {B{\'a}rczy}, {Barrado}, {Barros}, {Baumjohann},
  {Beck}, {Bekkelien}, {Benz}, {Billot}, {Bonfils}, {Broeg}, {Cabrera},
  {Collier Cameron}, {Davies}, {Deleuil}, {Delisle}, {Demangeon}, {Demory},
  {Erikson}, {Fortier}, {Fridlund}, {Futyan}, {Gandolfi}, {Garcia Mu{\~n}oz},
  {Gillon}, {Guedel}, {Heng}, {Kiss}, {Laskar}, {Lecavelier des Etangs},
  {Lendl}, {Lovis}, {Maxted}, {Nascimbeni}, {Olofsson}, {Osborn}, {Pagano},
  {Pall{\'e}}, {Piotto}, {Pollacco}, {Queloz}, {Rauer}, {Ragazzoni}, {Ribas},
  {Santos}, {Scandariato}, {S{\'e}gransan}, {Simon}, {Smith}, {Steller},
  {Szab{\'o}}, {Thomas}, {Udry}, \& {Walton}}]{delrez2021}
{Delrez}, L., {Ehrenreich}, D., {Alibert}, Y., {et~al.} 2021, Nature Astronomy,
  5, 775

\bibitem[{{Delrez} {et~al.}(2022){Delrez}, {Murray}, {Pozuelos}, {Narita},
  {Ducrot}, {Timmermans}, {Watanabe}, {Burgasser}, {Hirano}, {Rackham},
  {Stassun}, {Van Grootel}, {Aganze}, {Cointepas}, {Howell}, {Kaltenegger},
  {Niraula}, {Sebastian}, {Almenara}, {Barkaoui}, {Baycroft}, {Bonfils},
  {Bouchy}, {Burdanov}, {Caldwell}, {Charbonneau}, {Ciardi}, {Collins},
  {Daylan}, {Demory}, {de Wit}, {Dransfield}, {Fajardo-Acosta}, {Fausnaugh},
  {Fukui}, {Furlan}, {Garcia}, {Gnilka}, {G{\'o}mez Maqueo Chew},
  {G{\'o}mez-Mu{\~n}oz}, {G{\"u}nther}, {Harakawa}, {Heng}, {Hooton}, {Hori},
  {Ikoma}, {Jehin}, {Jenkins}, {Kagetani}, {Kawauchi}, {Kimura}, {Kodama},
  {Kotani}, {Krishnamurthy}, {Kudo}, {Kunovac}, {Kusakabe}, {Latham},
  {Littlefield}, {McCormac}, {Melis}, {Mori}, {Murgas}, {Palle}, {Pedersen},
  {Queloz}, {Ricker}, {Sabin}, {Schanche}, {Schroffenegger}, {Seager}, {Shiao},
  {Sohy}, {Standing}, {Tamura}, {Theissen}, {Thompson}, {Triaud}, {Vanderspek},
  {Vievard}, {Wells}, {Winn}, {Zou}, {Z{\'u}{\~n}iga-Fern{\'a}ndez}, \&
  {Gillon}}]{delrez2022}
{Delrez}, L., {Murray}, C.~A., {Pozuelos}, F.~J., {et~al.} 2022, \aap, 667, A59

\bibitem[{{Demory} {et~al.}(2020){Demory}, {Pozuelos}, {G{\'o}mez Maqueo Chew},
  {Sabin}, {Petrucci}, {Schroffenegger}, {Grimm}, {Sestovic}, {Gillon},
  {McCormac}, {Barkaoui}, {Benz}, {Bieryla}, {Bouchy}, {Burdanov}, {Collins},
  {de Wit}, {Dressing}, {Garcia}, {Giacalone}, {Guerra}, {Haldemann}, {Heng},
  {Jehin}, {Jofr{\'e}}, {Kane}, {Lillo-Box}, {Maign{\'e}}, {Mordasini},
  {Morris}, {Niraula}, {Queloz}, {Rackham}, {Savel}, {Soubkiou}, {Srdoc},
  {Stassun}, {Triaud}, {Zambelli}, {Ricker}, {Latham}, {Seager}, {Winn},
  {Jenkins}, {Calvario-Vel{\'a}squez}, {Franco Herrera}, {Colorado}, {Cadena
  Zepeda}, {Figueroa}, {Watson}, {Lugo-Ibarra}, {Carigi}, {Guisa}, {Herrera},
  {Sierra D{\'\i}az}, {Su{\'a}rez}, {Barrado}, {Batalha}, {Benkhaldoun},
  {Chontos}, {Dai}, {Essack}, {Ghachoui}, {Huang}, {Huber}, {Isaacson},
  {Lissauer}, {Morales-Calder{\'o}n}, {Robertson}, {Roy}, {Twicken},
  {Vanderburg}, \& {Weiss}}]{demory2020}
{Demory}, B.~O., {Pozuelos}, F.~J., {G{\'o}mez Maqueo Chew}, Y., {et~al.} 2020,
  \aap, 642, A49

\bibitem[{{D{\'e}vora-Pajares} \& {Pozuelos}(2022)}]{matrix}
{D{\'e}vora-Pajares}, M. \& {Pozuelos}, F.~J. 2022, {MATRIX: Multi-phAse
  Transits Recovery from Injected eXoplanets}, Zenodo

\bibitem[{{Dhillon} {et~al.}(2021){Dhillon}, {Bezawada}, {Black}, {Dixon},
  {Gamble}, {Gao}, {Henry}, {Kerry}, {Littlefair}, {Lunney}, {Marsh}, {Miller},
  {Parsons}, {Ashley}, {Breedt}, {Brown}, {Dyer}, {Green}, {Pelisoli},
  {Sahman}, {Wild}, {Ives}, {Mehrgan}, {Stegmeier}, {Dubbeldam}, {Morris},
  {Osborn}, {Wilson}, {Casares}, {Mu{\~n}oz-Darias}, {Pall{\'e}},
  {Rodr{\'\i}guez-Gil}, {Shahbaz}, {Torres}, {de Ugarte Postigo},
  {Cabrera-Lavers}, {Corradi}, {Dom{\'\i}nguez}, \&
  {Garc{\'\i}a-Alvarez}}]{Dhillon2021}
{Dhillon}, V.~S., {Bezawada}, N., {Black}, M., {et~al.} 2021, \mnras, 507, 350

\bibitem[{{Dittmann} {et~al.}(2017){Dittmann}, {Irwin}, {Charbonneau},
  {Bonfils}, {Astudillo-Defru}, {Haywood}, {Berta-Thompson}, {Newton},
  {Rodriguez}, {Winters}, {Tan}, {Almenara}, {Bouchy}, {Delfosse}, {Forveille},
  {Lovis}, {Murgas}, {Pepe}, {Santos}, {Udry}, {W{\"u}nsche}, {Esquerdo},
  {Latham}, \& {Dressing}}]{Dittmann2017}
{Dittmann}, J.~A., {Irwin}, J.~M., {Charbonneau}, D., {et~al.} 2017, \nat, 544,
  333

\bibitem[{Dorn \& Lichtenberg(2021)}]{Dorn2021}
Dorn, C. \& Lichtenberg, T. 2021, The Astrophysical Journal Letters, 922, L4

\bibitem[{{Dressing} \& {Charbonneau}(2015)}]{2015ApJ...807...45D}
{Dressing}, C.~D. \& {Charbonneau}, D. 2015, \apj, 807, 45

\bibitem[{{Fabrycky} {et~al.}(2014){Fabrycky}, {Lissauer}, {Ragozzine}, {Rowe},
  {Steffen}, {Agol}, {Barclay}, {Batalha}, {Borucki}, {Ciardi}, {Ford},
  {Gautier}, {Geary}, {Holman}, {Jenkins}, {Li}, {Morehead}, {Morris},
  {Shporer}, {Smith}, {Still}, \& {Van Cleve}}]{2014ApJ...790..146F}
{Fabrycky}, D.~C., {Lissauer}, J.~J., {Ragozzine}, D., {et~al.} 2014, \apj,
  790, 146

\bibitem[{{Faria} {et~al.}(2022){Faria}, {Su{\'a}rez Mascare{\~n}o},
  {Figueira}, {Silva}, {Damasso}, {Demangeon}, {Pepe}, {Santos}, {Rebolo},
  {Cristiani}, {Adibekyan}, {Alibert}, {Allart}, {Barros}, {Cabral},
  {D'Odorico}, {Di Marcantonio}, {Dumusque}, {Ehrenreich}, {Gonz{\'a}lez
  Hern{\'a}ndez}, {Hara}, {Lillo-Box}, {Lo Curto}, {Lovis}, {Martins},
  {M{\'e}gevand}, {Mehner}, {Micela}, {Molaro}, {Nunes}, {Pall{\'e}},
  {Poretti}, {Sousa}, {Sozzetti}, {Tabernero}, {Udry}, \& {Zapatero
  Osorio}}]{Faria2022}
{Faria}, J.~P., {Su{\'a}rez Mascare{\~n}o}, A., {Figueira}, P., {et~al.} 2022,
  \aap, 658, A115

\bibitem[{{Feinstein} {et~al.}(2019){Feinstein}, {Schlieder}, {Livingston},
  {Ciardi}, {Howard}, {Arnold}, {Barentsen}, {Bristow}, {Christiansen},
  {Crossfield}, {Dressing}, {Gonzales}, {Kosiarek}, {Lintott}, {Miller},
  {Morales}, {Petigura}, {Thackeray}, {Ault}, {Baeten}, {Jonkeren}, {Langley},
  {Moshinaly}, {Pearson}, {Tanner}, \& {Treasure}}]{Feinstein2019}
{Feinstein}, A.~D., {Schlieder}, J.~E., {Livingston}, J.~H., {et~al.} 2019,
  \aj, 157, 40

\bibitem[{{Foreman-Mackey} {et~al.}(2013){Foreman-Mackey}, {Hogg}, {Lang}, \&
  {Goodman}}]{ForemanMackey2013}
{Foreman-Mackey}, D., {Hogg}, D.~W., {Lang}, D., \& {Goodman}, J. 2013, \pasp,
  125, 306

\bibitem[{{Frith} {et~al.}(2013){Frith}, {Pinfield}, {Jones}, {Barnes},
  {Pavlenko}, {Martin}, {Brown}, {Kuznetsov}, {Marocco}, {Tata}, \&
  {Cappetta}}]{2013MNRAS.435.2161F}
{Frith}, J., {Pinfield}, D.~J., {Jones}, H.~R.~A., {et~al.} 2013, \mnras, 435,
  2161

\bibitem[{{Fulton} {et~al.}(2018){Fulton}, {Petigura}, {Blunt}, \&
  {Sinukoff}}]{Fulton2018}
{Fulton}, B.~J., {Petigura}, E.~A., {Blunt}, S., \& {Sinukoff}, E. 2018, \pasp,
  130, 044504

\bibitem[{{Fulton} {et~al.}(2017){Fulton}, {Petigura}, {Howard}, {Isaacson},
  {Marcy}, {Cargile}, {Hebb}, {Weiss}, {Johnson}, {Morton}, {Sinukoff},
  {Crossfield}, \& {Hirsch}}]{Fulton2017}
{Fulton}, B.~J., {Petigura}, E.~A., {Howard}, A.~W., {et~al.} 2017, \aj, 154,
  109

\bibitem[{{Furlan} {et~al.}(2017){Furlan}, {Ciardi}, {Everett}, {Saylors},
  {Teske}, {Horch}, {Howell}, {van Belle}, {Hirsch}, {Gautier}, {Adams},
  {Barrado}, {Cartier}, {Dressing}, {Dupree}, {Gilliland}, {Lillo-Box},
  {Lucas}, \& {Wang}}]{furlan2017}
{Furlan}, E., {Ciardi}, D.~R., {Everett}, M.~E., {et~al.} 2017, \aj, 153, 71

\bibitem[{{Gagn{\'e}} {et~al.}(2018){Gagn{\'e}}, {Mamajek}, {Malo}, {Riedel},
  {Rodriguez}, {Lafreni{\`e}re}, {Faherty}, {Roy-Loubier}, {Pueyo}, {Robin}, \&
  {Doyon}}]{Gagne2018}
{Gagn{\'e}}, J., {Mamajek}, E.~E., {Malo}, L., {et~al.} 2018, \apj, 856, 23

\bibitem[{{Gaia Collaboration} {et~al.}(2022){Gaia Collaboration}, {Vallenari},
  {Brown}, {Prusti}, {de Bruijne}, {Arenou}, {Babusiaux}, {Biermann},
  {Creevey}, {Ducourant}, {Evans}, {Eyer}, {Guerra}, {Hutton}, {Jordi},
  {Klioner}, {Lammers}, {Lindegren}, {Luri}, {Mignard}, {Panem}, {Pourbaix},
  {Randich}, {Sartoretti}, {Soubiran}, {Tanga}, {Walton}, {Bailer-Jones},
  {Bastian}, {Drimmel}, {Jansen}, {Katz}, {Lattanzi}, {van Leeuwen}, {Bakker},
  {Cacciari}, {Casta{\~n}eda}, {De Angeli}, {Fabricius}, {Fouesneau},
  {Fr{\'e}mat}, {Galluccio}, {Guerrier}, {Heiter}, {Masana}, {Messineo},
  {Mowlavi}, {Nicolas}, {Nienartowicz}, {Pailler}, {Panuzzo}, {Riclet}, {Roux},
  {Seabroke}, {Sordo{\o}rcit}, {Th{\'e}venin}, {Gracia-Abril}, {Portell},
  {Teyssier}, {Altmann}, {Andrae}, {Audard}, {Bellas-Velidis}, {Benson},
  {Berthier}, {Blomme}, {Burgess}, {Busonero}, {Busso}, {C{\'a}novas}, {Carry},
  {Cellino}, {Cheek}, {Clementini}, {Damerdji}, {Davidson}, {de Teodoro},
  {Nu{\~n}ez Campos}, {Delchambre}, {Dell'Oro}, {Esquej},
  {Fern{\'a}ndez-Hern{\'a}ndez}, {Fraile}, {Garabato}, {Garc{\'\i}a-Lario},
  {Gosset}, {Haigron}, {Halbwachs}, {Hambly}, {Harrison}, {Hern{\'a}ndez},
  {Hestroffer}, {Hodgkin}, {Holl}, {Jan{\ss}en}, {Jevardat de Fombelle},
  {Jordan}, {Krone-Martins}, {Lanzafame}, {L{\"o}ffler}, {Marchal}, {Marrese},
  {Moitinho}, {Muinonen}, {Osborne}, {Pancino}, {Pauwels}, {Recio-Blanco},
  {Reyl{\'e}}, {Riello}, {Rimoldini}, {Roegiers}, {Rybizki}, {Sarro}, {Siopis},
  {Smith}, {Sozzetti}, {Utrilla}, {van Leeuwen}, {Abbas}, {{\'A}brah{\'a}m},
  {Abreu Aramburu}, {Aerts}, {Aguado}, {Ajaj}, {Aldea-Montero}, {Altavilla},
  {{\'A}lvarez}, {Alves}, {Anders}, {Anderson}, {Anglada Varela}, {Antoja},
  {Baines}, {Baker}, {Balaguer-N{\'u}{\~n}ez}, {Balbinot}, {Balog}, {Barache},
  {Barbato}, {Barros}, {Barstow}, {Bartolom{\'e}}, {Bassilana}, {Bauchet},
  {Becciani}, {Bellazzini}, {Berihuete}, {Bernet}, {Bertone}, {Bianchi},
  {Binnenfeld}, {Blanco-Cuaresma}, {Blazere}, {Boch}, {Bombrun}, {Bossini},
  {Bouquillon}, {Bragaglia}, {Bramante}, {Breedt}, {Bressan}, {Brouillet},
  {Brugaletta}, {Bucciarelli}, {Burlacu}, {Butkevich}, {Buzzi}, {Caffau},
  {Cancelliere}, {Cantat-Gaudin}, {Carballo}, {Carlucci}, {Carnerero},
  {Carrasco}, {Casamiquela}, {Castellani}, {Castro-Ginard}, {Chaoul},
  {Charlot}, {Chemin}, {Chiaramida}, {Chiavassa}, {Chornay}, {Comoretto},
  {Contursi}, {Cooper}, {Cornez}, {Cowell}, {Crifo}, {Cropper}, {Crosta},
  {Crowley}, {Dafonte}, {Dapergolas}, {David}, {David}, {de Laverny}, {De
  Luise}, {De March}, {De Ridder}, {de Souza}, {de Torres}, {del Peloso}, {del
  Pozo}, {Delbo}, {Delgado}, {Delisle}, {Demouchy}, {Dharmawardena}, {Di
  Matteo}, {Diakite}, {Diener}, {Distefano}, {Dolding}, {Edvardsson}, {Enke},
  {Fabre}, {Fabrizio}, {Faigler}, {Fedorets}, {Fernique}, {Fienga}, {Figueras},
  {Fournier}, {Fouron}, {Fragkoudi}, {Gai}, {Garcia-Gutierrez},
  {Garcia-Reinaldos}, {Garc{\'\i}a-Torres}, {Garofalo}, {Gavel}, {Gavras},
  {Gerlach}, {Geyer}, {Giacobbe}, {Gilmore}, {Girona}, {Giuffrida}, {Gomel},
  {Gomez}, {Gonz{\'a}lez-N{\'u}{\~n}ez}, {Gonz{\'a}lez-Santamar{\'\i}a},
  {Gonz{\'a}lez-Vidal}, {Granvik}, {Guillout}, {Guiraud},
  {Guti{\'e}rrez-S{\'a}nchez}, {Guy}, {Hatzidimitriou}, {Hauser}, {Haywood},
  {Helmer}, {Helmi}, {Sarmiento}, {Hidalgo}, {Hilger}, {H{\l}adczuk}, {Hobbs},
  {Holland}, {Huckle}, {Jardine}, {Jasniewicz}, {Jean-Antoine Piccolo},
  {Jim{\'e}nez-Arranz}, {Jorissen}, {Juaristi Campillo}, {Julbe}, {Karbevska},
  {Kervella}, {Khanna}, {Kontizas}, {Kordopatis}, {Korn}, {K{\'o}sp{\'a}l},
  {Kostrzewa-Rutkowska}, {Kruszy{\'n}ska}, {Kun}, {Laizeau}, {Lambert},
  {Lanza}, {Lasne}, {Le Campion}, {Lebreton}, {Lebzelter}, {Leccia}, {Leclerc},
  {Lecoeur-Taibi}, {Liao}, {Licata}, {Lindstr{\o}m}, {Lister}, {Livanou},
  {Lobel}, {Lorca}, {Loup}, {Madrero Pardo}, {Magdaleno Romeo}, {Managau},
  {Mann}, {Manteiga}, {Marchant}, {Marconi}, {Marcos}, {Marcos Santos},
  {Mar{\'\i}n Pina}, {Marinoni}, {Marocco}, {Marshall}, {Polo},
  {Mart{\'\i}n-Fleitas}, {Marton}, {Mary}, {Masip}, {Massari},
  {Mastrobuono-Battisti}, {Mazeh}, {McMillan}, {Messina}, {Michalik}, {Millar},
  {Mints}, {Molina}, {Molinaro}, {Moln{\'a}r}, {Monari}, {Mongui{\'o}},
  {Montegriffo}, {Montero}, {Mor}, {Mora}, {Morbidelli}, {Morel}, {Morris},
  {Muraveva}, {Murphy}, {Musella}, {Nagy}, {Noval}, {Oca{\~n}a}, {Ogden},
  {Ordenovic}, {Osinde}, {Pagani}, {Pagano}, {Palaversa}, {Palicio},
  {Pallas-Quintela}, {Panahi}, {Payne-Wardenaar}, {Pe{\~n}alosa Esteller},
  {Penttil{\"a}}, {Pichon}, {Piersimoni}, {Pineau}, {Plachy}, {Plum}, {Poggio},
  {Pr{\v{s}}a}, {Pulone}, {Racero}, {Ragaini}, {Rainer}, {Raiteri}, {Rambaux},
  {Ramos}, {Ramos-Lerate}, {Re Fiorentin}, {Regibo}, {Richards}, {Rios Diaz},
  {Ripepi}, {Riva}, {Rix}, {Rixon}, {Robichon}, {Robin}, {Robin}, {Roelens},
  {Rogues}, {Rohrbasser}, {Romero-G{\'o}mez}, {Rowell}, {Royer}, {Ruz Mieres},
  {Rybicki}, {Sadowski}, {S{\'a}ez N{\'u}{\~n}ez}, {Sagrist{\`a} Sell{\'e}s},
  {Sahlmann}, {Salguero}, {Samaras}, {Sanchez Gimenez}, {Sanna},
  {Santove{\~n}a}, {Sarasso}, {Schultheis}, {Sciacca}, {Segol}, {Segovia},
  {S{\'e}gransan}, {Semeux}, {Shahaf}, {Siddiqui}, {Siebert}, {Siltala},
  {Silvelo}, {Slezak}, {Slezak}, {Smart}, {Snaith}, {Solano}, {Solitro},
  {Souami}, {Souchay}, {Spagna}, {Spina}, {Spoto}, {Steele},
  {Steidelm{\"u}ller}, {Stephenson}, {S{\"u}veges}, {Surdej}, {Szabados},
  {Szegedi-Elek}, {Taris}, {Taylo}, {Teixeira}, {Tolomei}, {Tonello}, {Torra},
  {Torra}, {Torralba Elipe}, {Trabucchi}, {Tsounis}, {Turon}, {Ulla}, {Unger},
  {Vaillant}, {van Dillen}, {van Reeven}, {Vanel}, {Vecchiato}, {Viala},
  {Vicente}, {Voutsinas}, {Weiler}, {Wevers}, {Wyrzykowski}, {Yoldas}, {Yvard},
  {Zhao}, {Zorec}, {Zucker}, \& {Zwitter}}]{GaiaDR3}
{Gaia Collaboration}, {Vallenari}, A., {Brown}, A.~G.~A., {et~al.} 2022, arXiv
  e-prints, arXiv:2208.00211

\bibitem[{{Gardner} {et~al.}(2006){Gardner}, {Mather}, {Clampin}, {Doyon},
  {Greenhouse}, {Hammel}, {Hutchings}, {Jakobsen}, {Lilly}, {Long}, {Lunine},
  {McCaughrean}, {Mountain}, {Nella}, {Rieke}, {Rieke}, {Rix}, {Smith},
  {Sonneborn}, {Stiavelli}, {Stockman}, {Windhorst}, \& {Wright}}]{Gardner2006}
{Gardner}, J.~P., {Mather}, J.~C., {Clampin}, M., {et~al.} 2006, \ssr, 123, 485

\bibitem[{{Giacalone} {et~al.}(2021){Giacalone}, {Dressing}, {Jensen},
  {Collins}, {Ricker}, {Vanderspek}, {Seager}, {Winn}, {Jenkins}, {Barclay},
  {Barkaoui}, {Cadieux}, {Charbonneau}, {Collins}, {Conti}, {Doyon}, {Evans},
  {Ghachoui}, {Gillon}, {Guerrero}, {Hart}, {Jehin}, {Kielkopf}, {McLean},
  {Murgas}, {Palle}, {Parviainen}, {Pozuelos}, {Relles}, {Shporer}, {Socia},
  {Stockdale}, {Tan}, {Torres}, {Twicken}, {Waalkes}, \& {Waite}}]{triceratops}
{Giacalone}, S., {Dressing}, C.~D., {Jensen}, E. L.~N., {et~al.} 2021, \aj,
  161, 24

\bibitem[{{Gillon} {et~al.}(2016){Gillon}, {Jehin}, {Lederer}, {Delrez}, {de
  Wit}, {Burdanov}, {Van Grootel}, {Burgasser}, {Triaud}, {Opitom}, {Demory},
  {Sahu}, {Bardalez Gagliuffi}, {Magain}, \& {Queloz}}]{Gillon2016}
{Gillon}, M., {Jehin}, E., {Lederer}, S.~M., {et~al.} 2016, \nat, 533, 221

\bibitem[{{Gillon} {et~al.}(2017){Gillon}, {Triaud}, {Demory}, {Jehin}, {Agol},
  {Deck}, {Lederer}, {de Wit}, {Burdanov}, {Ingalls}, {Bolmont}, {Leconte},
  {Raymond}, {Selsis}, {Turbet}, {Barkaoui}, {Burgasser}, {Burleigh}, {Carey},
  {Chaushev}, {Copperwheat}, {Delrez}, {Fernandes}, {Holdsworth}, {Kotze}, {Van
  Grootel}, {Almleaky}, {Benkhaldoun}, {Magain}, \& {Queloz}}]{Gillon2017}
{Gillon}, M., {Triaud}, A. H.~M.~J., {Demory}, B.-O., {et~al.} 2017, \nat, 542,
  456

\bibitem[{{Guerrero} {et~al.}(2021){Guerrero}, {Seager}, {Huang}, {Vanderburg},
  {Garcia Soto}, {Mireles}, {Hesse}, {Fong}, {Glidden}, {Shporer}, {Latham},
  {Collins}, {Quinn}, {Burt}, {Dragomir}, {Crossfield}, {Vanderspek},
  {Fausnaugh}, {Burke}, {Ricker}, {Daylan}, {Essack}, {G{\"u}nther}, {Osborn},
  {Pepper}, {Rowden}, {Sha}, {Villanueva}, {Yahalomi}, {Yu}, {Ballard},
  {Batalha}, {Berardo}, {Chontos}, {Dittmann}, {Esquerdo}, {Mikal-Evans},
  {Jayaraman}, {Krishnamurthy}, {Louie}, {Mehrle}, {Niraula}, {Rackham},
  {Rodriguez}, {Rowden}, {Sousa-Silva}, {Watanabe}, {Wong}, {Zhan},
  {Zivanovic}, {Christiansen}, {Ciardi}, {Swain}, {Lund}, {Mullally},
  {Fleming}, {Rodriguez}, {Boyd}, {Quintana}, {Barclay}, {Col{\'o}n},
  {Rinehart}, {Schlieder}, {Clampin}, {Jenkins}, {Twicken}, {Caldwell},
  {Coughlin}, {Henze}, {Lissauer}, {Morris}, {Rose}, {Smith}, {Tenenbaum},
  {Ting}, {Wohler}, {Bakos}, {Bean}, {Berta-Thompson}, {Bieryla}, {Bouma},
  {Buchhave}, {Butler}, {Charbonneau}, {Doty}, {Ge}, {Holman}, {Howard},
  {Kaltenegger}, {Kane}, {Kjeldsen}, {Kreidberg}, {Lin}, {Minsky}, {Narita},
  {Paegert}, {P{\'a}l}, {Palle}, {Sasselov}, {Spencer}, {Sozzetti}, {Stassun},
  {Torres}, {Udry}, \& {Winn}}]{2021ApJS..254...39G}
{Guerrero}, N.~M., {Seager}, S., {Huang}, C.~X., {et~al.} 2021, \apjs, 254, 39

\bibitem[{{G{\"u}nther} \& {Daylan}(2021)}]{allesfitter}
{G{\"u}nther}, M.~N. \& {Daylan}, T. 2021, \apjs, 254, 13

\bibitem[{{G{\"u}nther} {et~al.}(2020){G{\"u}nther}, {Zhan}, {Seager},
  {Rimmer}, {Ranjan}, {Stassun}, {Oelkers}, {Daylan}, {Newton}, {Kristiansen},
  {Olah}, {Gillen}, {Rappaport}, {Ricker}, {Vanderspek}, {Latham}, {Winn},
  {Jenkins}, {Glidden}, {Fausnaugh}, {Levine}, {Dittmann}, {Quinn},
  {Krishnamurthy}, \& {Ting}}]{Gunther2020}
{G{\"u}nther}, M.~N., {Zhan}, Z., {Seager}, S., {et~al.} 2020, \aj, 159, 60

\bibitem[{{Hadden}(2019)}]{hadden2019}
{Hadden}, S. 2019, {shadden/TTV2Fast2Furious: First release of
  TTV2Fast2Furious}

\bibitem[{{He} {et~al.}(2020){He}, {Ford}, {Ragozzine}, \&
  {Carrera}}]{mathias2020}
{He}, M.~Y., {Ford}, E.~B., {Ragozzine}, D., \& {Carrera}, D. 2020, \aj, 160,
  276

\bibitem[{{Hilton} {et~al.}(2010){Hilton}, {West}, {Hawley}, \&
  {Kowalski}}]{2010AJ....140.1402H}
{Hilton}, E.~J., {West}, A.~A., {Hawley}, S.~L., \& {Kowalski}, A.~F. 2010,
  \aj, 140, 1402

\bibitem[{{Hippke} {et~al.}(2019){Hippke}, {David}, {Mulders}, \&
  {Heller}}]{wotan}
{Hippke}, M., {David}, T.~J., {Mulders}, G.~D., \& {Heller}, R. 2019, \aj, 158,
  143

\bibitem[{{Hippke} \& {Heller}(2019)}]{tls}
{Hippke}, M. \& {Heller}, R. 2019, \aap, 623, A39

\bibitem[{{Hobson} {et~al.}(2018){Hobson}, {Jofr{\'e}}, {Garc{\'\i}a},
  {Petrucci}, \& {G{\'o}mez}}]{Hobson2018}
{Hobson}, M.~J., {Jofr{\'e}}, E., {Garc{\'\i}a}, L., {Petrucci}, R., \&
  {G{\'o}mez}, M. 2018, \rmxaa, 54, 65

\bibitem[{{Holman} \& {Murray}(2005)}]{holman2005}
{Holman}, M.~J. \& {Murray}, N.~W. 2005, Science, 307, 1288

\bibitem[{Ingersoll(1969)}]{Ingersoll1969}
Ingersoll, A.~P. 1969, Journal of Atmospheric Sciences, 26, 1191

\bibitem[{{Jeffers} {et~al.}(2018){Jeffers}, {Sch{\"o}fer}, {Lamert},
  {Reiners}, {Montes}, {Caballero}, {Cort{\'e}s-Contreras}, {Marvin},
  {Passegger}, {Zechmeister}, {Quirrenbach}, {Alonso-Floriano}, {Amado},
  {Bauer}, {Casal}, {Diez Alonso}, {Herrero}, {Morales}, {Mundt}, {Ribas}, \&
  {Sarmiento}}]{Jeffers2018}
{Jeffers}, S.~V., {Sch{\"o}fer}, P., {Lamert}, A., {et~al.} 2018, \aap, 614,
  A76

\bibitem[{{Jenkins}(2002)}]{Jenkins2002}
{Jenkins}, J.~M. 2002, \apj, 575, 493

\bibitem[{{Jenkins} {et~al.}(2020){Jenkins}, {Tenenbaum}, {Seader}, {Burke},
  {McCauliff}, {Smith}, {Twicken}, \& {Chandrasekaran}}]{JenkinsJM2020}
{Jenkins}, J.~M., {Tenenbaum}, P., {Seader}, S., {et~al.} 2020, {Kepler Data
  Processing Handbook: Transiting Planet Search}, Kepler Science Document
  KSCI-19081-003

\bibitem[{{Jenkins} {et~al.}(2016){Jenkins}, {Twicken}, {McCauliff},
  {Campbell}, {Sanderfer}, {Lung}, {Mansouri-Samani}, {Girouard}, {Tenenbaum},
  {Klaus}, {Smith}, {Caldwell}, {Chacon}, {Henze}, {Heiges}, {Latham},
  {Morgan}, {Swade}, {Rinehart}, \& {Vanderspek}}]{Jenkins2016}
{Jenkins}, J.~M., {Twicken}, J.~D., {McCauliff}, S., {et~al.} 2016, in Society
  of Photo-Optical Instrumentation Engineers (SPIE) Conference Series, Vol.
  9913, Software and Cyberinfrastructure for Astronomy IV, ed. G.~{Chiozzi} \&
  J.~C. {Guzman}, 99133E

\bibitem[{{Jenkins} {et~al.}(2019){Jenkins}, {Pozuelos}, {Tuomi},
  {Berdi{\~n}as}, {D{\'\i}az}, {Vines}, {Su{\'a}rez}, \& {Pe{\~n}a
  Rojas}}]{jenkins2019}
{Jenkins}, J.~S., {Pozuelos}, F.~J., {Tuomi}, M., {et~al.} 2019, \mnras, 490,
  5585

\bibitem[{{Jensen}(2013)}]{Jensen2013}
{Jensen}, E. 2013, {Tapir: A web interface for transit/eclipse observability},
  Astrophysics Source Code Library, record ascl:1306.007

\bibitem[{{Kaltenegger} {et~al.}(2019){Kaltenegger}, {Pepper}, {Stassun}, \&
  {Oelkers}}]{2019ApJ...874L...8K}
{Kaltenegger}, L., {Pepper}, J., {Stassun}, K., \& {Oelkers}, R. 2019, \apjl,
  874, L8

\bibitem[{{Kanodia} {et~al.}(2019){Kanodia}, {Wolfgang}, {Stefansson}, {Ning},
  \& {Mahadevan}}]{Kanodia2019}
{Kanodia}, S., {Wolfgang}, A., {Stefansson}, G.~K., {Ning}, B., \& {Mahadevan},
  S. 2019, \apj, 882, 38

\bibitem[{Kasting(1988)}]{Kasting1988}
Kasting, J.~F. 1988, Icarus, 74, 472

\bibitem[{{Kipping}(2013)}]{Kipping2013}
{Kipping}, D.~M. 2013, \mnras, 435, 2152

\bibitem[{{Kopparapu} {et~al.}(2013){Kopparapu}, {Ramirez}, {Kasting}, {Eymet},
  {Robinson}, {Mahadevan}, {Terrien}, {Domagal-Goldman}, {Meadows}, \&
  {Deshpande}}]{Kopparapu2013}
{Kopparapu}, R.~K., {Ramirez}, R., {Kasting}, J.~F., {et~al.} 2013, \apj, 765,
  131

\bibitem[{{Kossakowski} {et~al.}(2023){Kossakowski}, {K{\"u}rster}, {Trifonov},
  {Henning}, {Kemmer}, {Caballero}, {Burn}, {Sabotta}, {Crouse}, {Fauchez},
  {Nagel}, {Kaminski}, {Herrero}, {Rodr{\'\i}guez}, {Gonz{\'a}lez-{\'A}lvarez},
  {Quirrenbach}, {Amado}, {Ribas}, {Reiners}, {Aceituno}, {B{\'e}jar},
  {Baroch}, {Bastelberger}, {Chaturvedi}, {Cifuentes}, {Dreizler}, {Jeffers},
  {Kopparapu}, {Lafarga}, {L{\'o}pez-Gonz{\'a}lez}, {Mart{\'\i}n-Ruiz},
  {Montes}, {Morales}, {Pall{\'e}}, {Pavlov}, {Pedraz}, {Perdelwitz},
  {P{\'e}rez-Torres}, {Perger}, {Reffert}, {Rodr{\'\i}guez L{\'o}pez},
  {Schlecker}, {Sch{\"o}fer}, {Schweitzer}, {Shan}, {Shields}, {Stock}, {Wolf},
  {Zapatero Osorio}, \& {Zechmeister}}]{Kossakowski2023}
{Kossakowski}, D., {K{\"u}rster}, M., {Trifonov}, T., {et~al.} 2023, \aap, 670,
  A84

\bibitem[{{Lafarga} {et~al.}(2020){Lafarga}, {Ribas}, {Lovis}, {Perger},
  {Zechmeister}, {Bauer}, {K{\"u}rster}, {Cort{\'e}s-Contreras}, {Morales},
  {Herrero}, {Rosich}, {Baroch}, {Reiners}, {Caballero}, {Quirrenbach},
  {Amado}, {Alacid}, {B{\'e}jar}, {Dreizler}, {Hatzes}, {Henning}, {Jeffers},
  {Kaminski}, {Montes}, {Pedraz}, {Rodr{\'\i}guez-L{\'o}pez}, \&
  {Schmitt}}]{Lafarga2020}
{Lafarga}, M., {Ribas}, I., {Lovis}, C., {et~al.} 2020, \aap, 636, A36

\bibitem[{{Lam} {et~al.}(2021){Lam}, {Csizmadia}, {Astudillo-Defru}, {Bonfils},
  {Gandolfi}, {Padovan}, {Esposito}, {Hellier}, {Hirano}, {Livingston},
  {Murgas}, {Smith}, {Collins}, {Mathur}, {Garcia}, {Howell}, {Santos}, {Dai},
  {Ricker}, {Vanderspek}, {Latham}, {Seager}, {Winn}, {Jenkins}, {Albrecht},
  {Almenara}, {Artigau}, {Barrag{\'a}n}, {Bouchy}, {Cabrera}, {Charbonneau},
  {Chaturvedi}, {Chaushev}, {Christiansen}, {Cochran}, {De Meideiros},
  {Delfosse}, {D{\'\i}az}, {Doyon}, {Eigm{\"u}ller}, {Figueira}, {Forveille},
  {Fridlund}, {Gaisn{\'e}}, {Goffo}, {Georgieva}, {Grziwa}, {Guenther},
  {Hatzes}, {Johnson}, {Kab{\'a}th}, {Knudstrup}, {Korth}, {Lewin}, {Lissauer},
  {Lovis}, {Luque}, {Melo}, {Morgan}, {Morris}, {Mayor}, {Narita}, {Osborne},
  {Palle}, {Pepe}, {Persson}, {Quinn}, {Rauer}, {Redfield}, {Schlieder},
  {S{\'e}gransan}, {Serrano}, {Smith}, {{\v{S}}ubjak}, {Twicken}, {Udry}, {Van
  Eylen}, \& {Vezie}}]{Lam2021}
{Lam}, K. W.~F., {Csizmadia}, S., {Astudillo-Defru}, N., {et~al.} 2021,
  Science, 374, 1271

\bibitem[{{Latham} {et~al.}(2011){Latham}, {Rowe}, {Quinn}, {Batalha},
  {Borucki}, {Brown}, {Bryson}, {Buchhave}, {Caldwell}, {Carter},
  {Christiansen}, {Ciardi}, {Cochran}, {Dunham}, {Fabrycky}, {Ford}, {Gautier},
  {Gilliland}, {Holman}, {Howell}, {Ibrahim}, {Isaacson}, {Jenkins}, {Koch},
  {Lissauer}, {Marcy}, {Quintana}, {Ragozzine}, {Sasselov}, {Shporer},
  {Steffen}, {Welsh}, \& {Wohler}}]{2011ApJ...732L..24L}
{Latham}, D.~W., {Rowe}, J.~F., {Quinn}, S.~N., {et~al.} 2011, \apjl, 732, L24

\bibitem[{{L{\'e}pine} \& {Gaidos}(2011)}]{2011AJ....142..138L}
{L{\'e}pine}, S. \& {Gaidos}, E. 2011, \aj, 142, 138

\bibitem[{{L{\'e}pine} \& {Shara}(2005{\natexlab{a}})}]{Lepine2005}
{L{\'e}pine}, S. \& {Shara}, M.~M. 2005{\natexlab{a}}, \aj, 129, 1483

\bibitem[{{L{\'e}pine} \& {Shara}(2005{\natexlab{b}})}]{2005AJ....129.1483L}
{L{\'e}pine}, S. \& {Shara}, M.~M. 2005{\natexlab{b}}, \aj, 129, 1483

\bibitem[{{Li} {et~al.}(2019){Li}, {Tenenbaum}, {Twicken}, {Burke}, {Jenkins},
  {Quintana}, {Rowe}, \& {Seader}}]{Li2019}
{Li}, J., {Tenenbaum}, P., {Twicken}, J.~D., {et~al.} 2019, \pasp, 131, 024506

\bibitem[{{Lillo-Box} {et~al.}(2020){Lillo-Box}, {Figueira}, {Leleu},
  {Acu{\~n}a}, {Faria}, {Hara}, {Santos}, {Correia}, {Robutel}, {Deleuil},
  {Barrado}, {Sousa}, {Bonfils}, {Mousis}, {Almenara}, {Astudillo-Defru},
  {Marcq}, {Udry}, {Lovis}, \& {Pepe}}]{LilloBox2020}
{Lillo-Box}, J., {Figueira}, P., {Leleu}, A., {et~al.} 2020, \aap, 642, A121

\bibitem[{{Lillo-Box} {et~al.}(2022){Lillo-Box}, {Santos}, {Santerne}, {Silva},
  {Barrado}, {Faria}, {Castro-Gonz{\'a}lez}, {Balsalobre-Ruza},
  {Morales-Calder{\'o}n}, {Saavedra}, {Marfil}, {Sousa}, {Adibekyan},
  {Berihuete}, {Barros}, {Delgado-Mena}, {Hu{\'e}lamo}, {Deleuil}, {Demangeon},
  {Figueira}, {Grouffal}, {Aceituno}, {Azzaro}, {Bergond},
  {Fern{\'a}ndez-Mart{\'\i}n}, {Galad{\'\i}}, {Gallego}, {Gardini},
  {G{\'o}ngora}, {Guijarro}, {Hermelo}, {Mart{\'\i}n}, {M{\'\i}nguez},
  {Montoya}, {Pedraz}, \& {Vico Linares}}]{LilloBox2022}
{Lillo-Box}, J., {Santos}, N.~C., {Santerne}, A., {et~al.} 2022, \aap, 667,
  A102

\bibitem[{{Limbach} \& {Turner}(2015)}]{limbach2015}
{Limbach}, M.~A. \& {Turner}, E.~L. 2015, Proceedings of the National Academy
  of Science, 112, 20

\bibitem[{{Lissauer} {et~al.}(2012){Lissauer}, {Marcy}, {Rowe}, {Bryson},
  {Adams}, {Buchhave}, {Ciardi}, {Cochran}, {Fabrycky}, {Ford}, {Fressin},
  {Geary}, {Gilliland}, {Holman}, {Howell}, {Jenkins}, {Kinemuchi}, {Koch},
  {Morehead}, {Ragozzine}, {Seader}, {Tanenbaum}, {Torres}, \&
  {Twicken}}]{2012ApJ...750..112L}
{Lissauer}, J.~J., {Marcy}, G.~W., {Rowe}, J.~F., {et~al.} 2012, \apj, 750, 112

\bibitem[{{Lissauer} {et~al.}(2011){Lissauer}, {Ragozzine}, {Fabrycky},
  {Steffen}, {Ford}, {Jenkins}, {Shporer}, {Holman}, {Rowe}, {Quintana},
  {Batalha}, {Borucki}, {Bryson}, {Caldwell}, {Carter}, {Ciardi}, {Dunham},
  {Fortney}, {Gautier}, {Howell}, {Koch}, {Latham}, {Marcy}, {Morehead}, \&
  {Sasselov}}]{2011ApJS..197....8L}
{Lissauer}, J.~J., {Ragozzine}, D., {Fabrycky}, D.~C., {et~al.} 2011, \apjs,
  197, 8

\bibitem[{{Livingston} {et~al.}(2018){Livingston}, {Crossfield}, {Petigura},
  {Gonzales}, {Ciardi}, {Beichman}, {Christiansen}, {Dressing}, {Henning},
  {Howard}, {Isaacson}, {Fulton}, {Kosiarek}, {Schlieder}, {Sinukoff}, \&
  {Tamura}}]{2018AJ....156..277L}
{Livingston}, J.~H., {Crossfield}, I. J.~M., {Petigura}, E.~A., {et~al.} 2018,
  \aj, 156, 277

\bibitem[{{Lopez} \& {Rice}(2018)}]{Lopez&Rice2018}
{Lopez}, E.~D. \& {Rice}, K. 2018, \mnras, 479, 5303

\bibitem[{{Luque} {et~al.}(2018){Luque}, {Nowak}, {Pall{\'e}}, {Kossakowski},
  {Trifonov}, {Zechmeister}, {B{\'e}jar}, {Cardona Guill{\'e}n}, {Tal-Or},
  {Hidalgo}, {Ribas}, {Reiners}, {Caballero}, {Amado}, {Quirrenbach},
  {Aceituno}, {Cort{\'e}s-Contreras}, {D{\'\i}ez-Alonso}, {Dreizler},
  {Guenther}, {Henning}, {Jeffers}, {Kaminski}, {K{\"u}rster}, {Lafarga},
  {Montes}, {Morales}, {Passegger}, {Schmitt}, \& {Schweitzer}}]{Luque2018}
{Luque}, R., {Nowak}, G., {Pall{\'e}}, E., {et~al.} 2018, \aap, 620, A171

\bibitem[{{Luque} \& {Pall{\'e}}(2022)}]{LuquePalle2022}
{Luque}, R. \& {Pall{\'e}}, E. 2022, Science, 377, 1211

\bibitem[{{Marfil} {et~al.}(2021){Marfil}, {Tabernero}, {Montes}, {Caballero},
  {L{\'a}zaro}, {Gonz{\'a}lez Hern{\'a}ndez}, {Nagel}, {Passegger},
  {Schweitzer}, {Ribas}, {Reiners}, {Quirrenbach}, {Amado}, {Cifuentes},
  {Cort{\'e}s-Contreras}, {Dreizler}, {Duque-Arribas},
  {Galad{\'\i}-Enr{\'\i}quez}, {Henning}, {Jeffers}, {Kaminski}, {K{\"u}rster},
  {Lafarga}, {L{\'o}pez-Gallifa}, {Morales}, {Shan}, \&
  {Zechmeister}}]{Marfil2021}
{Marfil}, E., {Tabernero}, H.~M., {Montes}, D., {et~al.} 2021, \aap, 656, A162

\bibitem[{{Martinez} {et~al.}(2019){Martinez}, {Cunha}, {Ghezzi}, \&
  {Smith}}]{Martinez2019}
{Martinez}, C.~F., {Cunha}, K., {Ghezzi}, L., \& {Smith}, V.~V. 2019, \apj,
  875, 29

\bibitem[{{Mart{\'\i}nez-Rodr{\'\i}guez}
  {et~al.}(2019){Mart{\'\i}nez-Rodr{\'\i}guez}, {Caballero}, {Cifuentes},
  {Piro}, \& {Barnes}}]{MartinezRodriguez2019}
{Mart{\'\i}nez-Rodr{\'\i}guez}, H., {Caballero}, J.~A., {Cifuentes}, C.,
  {Piro}, A.~L., \& {Barnes}, R. 2019, \apj, 887, 261

\bibitem[{{Masci} {et~al.}(2019){Masci}, {Laher}, {Rusholme}, {Shupe}, {Groom},
  {Surace}, {Jackson}, {Monkewitz}, {Beck}, {Flynn}, {Terek}, {Landry},
  {Hacopians}, {Desai}, {Howell}, {Brooke}, {Imel}, {Wachter}, {Ye}, {Lin},
  {Cenko}, {Cunningham}, {Rebbapragada}, {Bue}, {Miller}, {Mahabal}, {Bellm},
  {Patterson}, {Juri{\'c}}, {Golkhou}, {Ofek}, {Walters}, {Graham}, {Kasliwal},
  {Dekany}, {Kupfer}, {Burdge}, {Cannella}, {Barlow}, {Van Sistine}, {Giomi},
  {Fremling}, {Blagorodnova}, {Levitan}, {Riddle}, {Smith}, {Helou}, {Prince},
  \& {Kulkarni}}]{Masci2019}
{Masci}, F.~J., {Laher}, R.~R., {Rusholme}, B., {et~al.} 2019, \pasp, 131,
  018003

\bibitem[{{Mayo} {et~al.}(2018){Mayo}, {Vanderburg}, {Latham}, {Bieryla},
  {Morton}, {Buchhave}, {Dressing}, {Beichman}, {Berlind}, {Calkins}, {Ciardi},
  {Crossfield}, {Esquerdo}, {Everett}, {Gonzales}, {Hirsch}, {Horch}, {Howard},
  {Howell}, {Livingston}, {Patel}, {Petigura}, {Schlieder}, {Scott}, {Schumer},
  {Sinukoff}, {Teske}, \& {Winters}}]{2018AJ....155..136M}
{Mayo}, A.~W., {Vanderburg}, A., {Latham}, D.~W., {et~al.} 2018, \aj, 155, 136

\bibitem[{{McCully} {et~al.}(2018){McCully}, {Volgenau}, {Harbeck}, {Lister},
  {Saunders}, {Turner}, {Siiverd}, \& {Bowman}}]{McCully2018}
{McCully}, C., {Volgenau}, N.~H., {Harbeck}, D.-R., {et~al.} 2018, in Society
  of Photo-Optical Instrumentation Engineers (SPIE) Conference Series, Vol.
  10707, Software and Cyberinfrastructure for Astronomy V, ed. J.~C. {Guzman}
  \& J.~{Ibsen}, 107070K

\bibitem[{{Montes} {et~al.}(2001){Montes}, {L{\'o}pez-Santiago}, {G{\'a}lvez},
  {Fern{\'a}ndez-Figueroa}, {De Castro}, \& {Cornide}}]{Montes2001}
{Montes}, D., {L{\'o}pez-Santiago}, J., {G{\'a}lvez}, M.~C., {et~al.} 2001,
  \mnras, 328, 45

\bibitem[{{Montet} {et~al.}(2015){Montet}, {Morton}, {Foreman-Mackey},
  {Johnson}, {Hogg}, {Bowler}, {Latham}, {Bieryla}, \&
  {Mann}}]{2015ApJ...809...25M}
{Montet}, B.~T., {Morton}, T.~D., {Foreman-Mackey}, D., {et~al.} 2015, \apj,
  809, 25

\bibitem[{{Mordasini}(2020)}]{Mordasini2020}
{Mordasini}, C. 2020, \aap, 638, A52

\bibitem[{{Morris} {et~al.}(2020){Morris}, {Twicken}, {Smith}, {Clarke},
  {Jenkins}, {Bryson}, {Girouard}, \& {Klaus}}]{Morris2020}
{Morris}, R.~L., {Twicken}, J.~D., {Smith}, J.~C., {et~al.} 2020, {Kepler Data
  Processing Handbook: Photometric Analysis}, Kepler Science Document
  KSCI-19081-003

\bibitem[{{Morton}(2012)}]{2012ApJ...761....6M}
{Morton}, T.~D. 2012, \apj, 761, 6

\bibitem[{{Morton}(2015)}]{2015ascl.soft03011M}
{Morton}, T.~D. 2015, {VESPA: False positive probabilities calculator},
  Astrophysics Source Code Library

\bibitem[{Mousis {et~al.}(2020)Mousis, Deleuil, Aguichine, Marcq, Naar,
  Aguirre, Brugger, \& Gon{\c c}alves}]{Mousis2020}
Mousis, O., Deleuil, M., Aguichine, A., {et~al.} 2020, The Astrophysical
  Journal Letters, 896, L22

\bibitem[{{Mugrauer} \& {Michel}(2020)}]{mugrauer2020}
{Mugrauer}, M. \& {Michel}, K.-U. 2020, Astronomische Nachrichten, 341, 996

\bibitem[{{Mugrauer} \& {Michel}(2021)}]{mugrauer2021}
{Mugrauer}, M. \& {Michel}, K.-U. 2021, Astronomische Nachrichten, 342, 840

\bibitem[{{Murgas} {et~al.}(2021){Murgas}, {Astudillo-Defru}, {Bonfils},
  {Crossfield}, {Almenara}, {Livingston}, {Stassun}, {Korth}, {Orell-Miquel},
  {Morello}, {Eastman}, {Lissauer}, {Kane}, {Morales}, {Werner}, {Gorjian},
  {Benneke}, {Dragomir}, {Matthews}, {Howell}, {Ciardi}, {Gonzales}, {Matson},
  {Beichman}, {Schlieder}, {Collins}, {Collins}, {Jensen}, {Evans}, {Pozuelos},
  {Gillon}, {Jehin}, {Barkaoui}, {Artigau}, {Bouchy}, {Charbonneau},
  {Delfosse}, {D{\'\i}az}, {Doyon}, {Figueira}, {Forveille}, {Lovis}, {Melo},
  {Gaisn{\'e}}, {Pepe}, {Santos}, {S{\'e}gransan}, {Udry}, {Goeke}, {Levine},
  {Quintana}, {Guerrero}, {Mireles}, {Caldwell}, {Tenenbaum}, {Brasseur},
  {Ricker}, {Vanderspek}, {Latham}, {Seager}, {Winn}, \&
  {Jenkins}}]{Murgas2021}
{Murgas}, F., {Astudillo-Defru}, N., {Bonfils}, X., {et~al.} 2021, \aap, 653,
  A60

\bibitem[{{Narita} {et~al.}(2020){Narita}, {Fukui}, {Yamamuro}, {Harbeck},
  {Bowman}, {Elphick}, {Nation}, {Armstrong}, {Han}, {Abe}, {Ikoma}, {Isogai},
  {Kawauchi}, {Kurita}, {Kusakabe}, {de Leon}, {Livingston}, {Mori},
  {Nishiumi}, {Tamura}, {Watanabe}, {Volgenau}, {Heinrich-Josties}, {Foale},
  {Daily}, {McCully}, {Kirby}, {Smith}, {Haworth}, {Conway},
  {Storrie-Lombardi}, {Rosing}, {Chatelain}, {Bachelet}, {Johnson}, \&
  {Rabus}}]{Narita2020}
{Narita}, N., {Fukui}, A., {Yamamuro}, T., {et~al.} 2020, in Society of
  Photo-Optical Instrumentation Engineers (SPIE) Conference Series, Vol. 11447,
  Society of Photo-Optical Instrumentation Engineers (SPIE) Conference Series,
  114475K

\bibitem[{{Ning} {et~al.}(2018){Ning}, {Wolfgang}, \& {Ghosh}}]{Ning2018}
{Ning}, B., {Wolfgang}, A., \& {Ghosh}, S. 2018, \apj, 869, 5

\bibitem[{{Owen} \& {Wu}(2013)}]{Owen&Wu2013}
{Owen}, J.~E. \& {Wu}, Y. 2013, \apj, 775, 105

\bibitem[{{Parviainen}(2015)}]{Parviainen2015}
{Parviainen}, H. 2015, \mnras, 450, 3233

\bibitem[{{Parviainen} \& {Aigrain}(2015)}]{Parviainen2015b}
{Parviainen}, H. \& {Aigrain}, S. 2015, \mnras, 453, 3821

\bibitem[{{Passegger} {et~al.}(2018){Passegger}, {Reiners}, {Jeffers},
  {Wende-von Berg}, {Sch{\"o}fer}, {Caballero}, {Schweitzer}, {Amado},
  {B{\'e}jar}, {Cort{\'e}s-Contreras}, {Hatzes}, {K{\"u}rster}, {Montes},
  {Pedraz}, {Quirrenbach}, {Ribas}, \& {Seifert}}]{Passegger2018}
{Passegger}, V.~M., {Reiners}, A., {Jeffers}, S.~V., {et~al.} 2018, \aap, 615,
  A6

\bibitem[{{Perryman}(2011)}]{Perryman2011}
{Perryman}, M. 2011, {The Exoplanet Handbook}

\bibitem[{{Pozuelos} {et~al.}(2020){Pozuelos}, {Su{\'a}rez}, {de El{\'\i}a},
  {Berdi{\~n}as}, {Bonfanti}, {Dugaro}, {Gillon}, {Jehin}, {G{\"u}nther}, {Van
  Grootel}, {Garcia}, {Thuillier}, {Delrez}, \& {Rod{\'o}n}}]{pozuelos2020}
{Pozuelos}, F.~J., {Su{\'a}rez}, J.~C., {de El{\'\i}a}, G.~C., {et~al.} 2020,
  \aap, 641, A23

\bibitem[{{Pozuelos} {et~al.}(2023){Pozuelos}, {Timmermans}, {Rackham},
  {Garcia}, {Burgasser}, {Kane}, {G{\"u}nther}, {Stassun}, {Van Grootel},
  {D{\'e}vora-Pajares}, {Luque}, {Edwards}, {Niraula}, {Schanche}, {Wells},
  {Ducrot}, {Howell}, {Sebastian}, {Barkaoui}, {Waalkes}, {Cadieux}, {Doyon},
  {Boyle}, {Dietrich}, {Burdanov}, {Delrez}, {Demory}, {de Wit}, {Dransfield},
  {Gillon}, {G{\'o}mez Maqueo Chew}, {Hooton}, {Jehin}, {Murray}, {Pedersen},
  {Queloz}, {Thompson}, {Triaud}, {Z{\'u}{\~n}iga-Fern{\'a}ndez}, {Collins},
  {Fausnaugh}, {Hedges}, {Hesse}, {Jenkins}, {Kunimoto}, {Latham}, {Shporer},
  {Ting}, {Torres}, {Amado}, {Rod{\'o}n}, {Rodr{\'\i}guez-L{\'o}pez},
  {Su{\'a}rez}, {Alonso}, {Benkhaldoun}, {Berta-Thompson}, {Chinchilla},
  {Ghachoui}, {G{\'o}mez-Mu{\~n}oz}, {Rebolo}, {Sabin}, {Schroffenegger},
  {Furlan}, {Gnilka}, {Lester}, {Scott}, {Aganze}, {Gerasimov}, {Hsu},
  {Theissen}, {Apai}, {Chen}, {Gabor}, {Henning}, \& {Mancini}}]{pozuelos2023}
{Pozuelos}, F.~J., {Timmermans}, M., {Rackham}, B.~V., {et~al.} 2023, arXiv
  e-prints, arXiv:2303.08174

\bibitem[{{Quintana} {et~al.}(2014){Quintana}, {Barclay}, {Raymond}, {Rowe},
  {Bolmont}, {Caldwell}, {Howell}, {Kane}, {Huber}, {Crepp}, {Lissauer},
  {Ciardi}, {Coughlin}, {Everett}, {Henze}, {Horch}, {Isaacson}, {Ford},
  {Adams}, {Still}, {Hunter}, {Quarles}, \& {Selsis}}]{Quintana2014}
{Quintana}, E.~V., {Barclay}, T., {Raymond}, S.~N., {et~al.} 2014, Science,
  344, 277

\bibitem[{{Quirrenbach} {et~al.}(2014){Quirrenbach}, {Amado}, {Caballero},
  {Mundt}, {Reiners}, {Ribas}, {Seifert}, {Abril}, {Aceituno},
  {Alonso-Floriano}, {Ammler-von Eiff}, {Antona Jim{\'e}nez},
  {Anwand-Heerwart}, {Azzaro}, {Bauer}, {Barrado}, {Becerril}, {B{\'e}jar},
  {Ben{\'{\i}}tez}, {Berdi{\~n}as}, {C{\'a}rdenas}, {Casal}, {Claret},
  {Colom{\'e}}, {Cort{\'e}s-Contreras}, {Czesla}, {Doellinger}, {Dreizler},
  {Feiz}, {Fern{\'a}ndez}, {Galad{\'{\i}}}, {G{\'a}lvez-Ortiz},
  {Garc{\'{\i}}a-Piquer}, {Garc{\'{\i}}a-Vargas}, {Garrido}, {Gesa}, {G{\'o}mez
  Galera}, {Gonz{\'a}lez {\'A}lvarez}, {Gonz{\'a}lez Hern{\'a}ndez},
  {Gr{\"o}zinger}, {Gu{\`a}rdia}, {Guenther}, {de Guindos},
  {Guti{\'e}rrez-Soto}, {Hagen}, {Hatzes}, {Hauschildt}, {Helmling}, {Henning},
  {Hermann}, {Hern{\'a}ndez Casta{\~n}o}, {Herrero}, {Hidalgo}, {Holgado},
  {Huber}, {Huber}, {Jeffers}, {Joergens}, {de Juan}, {Kehr}, {Klein},
  {K{\"u}rster}, {Lamert}, {Lalitha}, {Laun}, {Lemke}, {Lenzen}, {L{\'o}pez del
  Fresno}, {L{\'o}pez Mart{\'{\i}}}, {L{\'o}pez-Santiago}, {Mall}, {Mandel},
  {Mart{\'{\i}}n}, {Mart{\'{\i}}n-Ruiz}, {Mart{\'{\i}}nez-Rodr{\'{\i}}guez},
  {Marvin}, {Mathar}, {Mirabet}, {Montes}, {Morales Mu{\~n}oz}, {Moya},
  {Naranjo}, {Ofir}, {Oreiro}, {Pall{\'e}}, {Panduro}, {Passegger},
  {P{\'e}rez-Calpena}, {P{\'e}rez Medialdea}, {Perger}, {Pluto}, {Ram{\'o}n},
  {Rebolo}, {Redondo}, {Reffert}, {Reinhardt}, {Rhode}, {Rix}, {Rodler},
  {Rodr{\'{\i}}guez}, {Rodr{\'{\i}}guez-L{\'o}pez},
  {Rodr{\'{\i}}guez-P{\'e}rez}, {Rohloff}, {Rosich}, {S{\'a}nchez-Blanco},
  {S{\'a}nchez Carrasco}, {Sanz-Forcada}, {Sarmiento}, {Sch{\"a}fer},
  {Schiller}, {Schmidt}, {Schmitt}, {Solano}, {Stahl}, {Storz}, {St{\"u}rmer},
  {Su{\'a}rez}, {Ulbrich}, {Veredas}, {Wagner}, {Winkler}, {Zapatero Osorio},
  {Zechmeister}, {Abell{\'a}n de Paco}, {Anglada-Escud{\'e}}, {del Burgo},
  {Klutsch}, {Lizon}, {L{\'o}pez-Morales}, {Morales}, {Perryman}, {Tulloch}, \&
  {Xu}}]{CARMENES}
{Quirrenbach}, A., {Amado}, P.~J., {Caballero}, J.~A., {et~al.} 2014, in
  \procspie, Vol. 9147, Ground-based and Airborne Instrumentation for Astronomy
  V, 91471F

\bibitem[{{Quirrenbach} {et~al.}(2018){Quirrenbach}, {Amado}, {Ribas},
  {Reiners}, {Caballero}, {Seifert}, {Aceituno}, {Azzaro}, {Baroch}, {Barrado},
  \& et~al.}]{CARMENES18}
{Quirrenbach}, A., {Amado}, P.~J., {Ribas}, I., {et~al.} 2018, in Society of
  Photo-Optical Instrumentation Engineers (SPIE) Conference Series, Vol. 10702,
  Ground-based and Airborne Instrumentation for Astronomy VII, 107020W

\bibitem[{{Rein} \& {Liu}(2012)}]{2012A&A...537A.128R}
{Rein}, H. \& {Liu}, S.~F. 2012, \aap, 537, A128

\bibitem[{{Rein} \& {Tamayo}(2015)}]{2015MNRAS.452..376R}
{Rein}, H. \& {Tamayo}, D. 2015, \mnras, 452, 376

\bibitem[{{Reiners} {et~al.}(2018){Reiners}, {Zechmeister}, {Caballero},
  {Ribas}, {Morales}, {Jeffers}, {Sch{\"o}fer}, {Tal-Or}, {Quirrenbach},
  {Amado}, {Kaminski}, {Seifert}, {Abril}, {Aceituno}, {Alonso-Floriano},
  {Ammler-von Eiff}, {Antona}, {Anglada-Escud{\'e}}, {Anwand-Heerwart},
  {Arroyo-Torres}, {Azzaro}, {Baroch}, {Barrado}, {Bauer}, {Becerril},
  {B{\'e}jar}, {Ben{\'\i}tez}, {Berdinas}, {Bergond}, {Bl{\"u}mcke},
  {Brinkm{\"o}ller}, {del Burgo}, {Cano}, {C{\'a}rdenas V{\'a}zquez}, {Casal},
  {Cifuentes}, {Claret}, {Colom{\'e}}, {Cort{\'e}s-Contreras}, {Czesla},
  {D{\'\i}ez-Alonso}, {Dreizler}, {Feiz}, {Fern{\'a}ndez}, {Ferro},
  {Fuhrmeister}, {Galad{\'\i}-Enr{\'\i}quez}, {Garcia-Piquer}, {Garc{\'\i}a
  Vargas}, {Gesa}, {G{\'o}mez Galera}, {Gonz{\'a}lez Hern{\'a}ndez},
  {Gonz{\'a}lez-Peinado}, {Gr{\"o}zinger}, {Grohnert}, {Gu{\`a}rdia},
  {Guenther}, {Guijarro}, {de Guindos}, {Guti{\'e}rrez-Soto}, {Hagen},
  {Hatzes}, {Hauschildt}, {Hedrosa}, {Helmling}, {Henning}, {Hermelo},
  {Hern{\'a}ndez Arab{\'\i}}, {Hern{\'a}ndez Casta{\~n}o}, {Hern{\'a}ndez
  Hernando}, {Herrero}, {Huber}, {Huke}, {Johnson}, {de Juan}, {Kim}, {Klein},
  {Kl{\"u}ter}, {Klutsch}, {K{\"u}rster}, {Lafarga}, {Lamert}, {Lamp{\'o}n},
  {Lara}, {Laun}, {Lemke}, {Lenzen}, {Launhardt}, {L{\'o}pez del Fresno},
  {L{\'o}pez-Gonz{\'a}lez}, {L{\'o}pez-Puertas}, {L{\'o}pez Salas},
  {L{\'o}pez-Santiago}, {Luque}, {Mag{\'a}n Madinabeitia}, {Mall}, {Mancini},
  {Mandel}, {Marfil}, {Mar{\'\i}n Molina}, {Maroto Fern{\'a}ndez},
  {Mart{\'\i}n}, {Mart{\'\i}n-Ruiz}, {Marvin}, {Mathar}, {Mirabet}, {Montes},
  {Moreno-Raya}, {Moya}, {Mundt}, {Nagel}, {Naranjo}, {Nortmann}, {Nowak},
  {Ofir}, {Oreiro}, {Pall{\'e}}, {Panduro}, {Pascual}, {Passegger}, {Pavlov},
  {Pedraz}, {P{\'e}rez-Calpena}, {P{\'e}rez Medialdea}, {Perger}, {Perryman},
  {Pluto}, {Rabaza}, {Ram{\'o}n}, {Rebolo}, {Redondo}, {Reffert}, {Reinhart},
  {Rhode}, {Rix}, {Rodler}, {Rodr{\'\i}guez}, {Rodr{\'\i}guez-L{\'o}pez},
  {Rodr{\'\i}guez Trinidad}, {Rohloff}, {Rosich}, {Sadegi},
  {S{\'a}nchez-Blanco}, {S{\'a}nchez Carrasco}, {S{\'a}nchez-L{\'o}pez},
  {Sanz-Forcada}, {Sarkis}, {Sarmiento}, {Sch{\"a}fer}, {Schmitt}, {Schiller},
  {Schweitzer}, {Solano}, {Stahl}, {Strachan}, {St{\"u}rmer}, {Su{\'a}rez},
  {Tabernero}, {Tala}, {Trifonov}, {Tulloch}, {Ulbrich}, {Veredas}, {Vico
  Linares}, {Vilardell}, {Wagner}, {Winkler}, {Wolthoff}, {Xu}, {Yan}, \&
  {Zapatero Osorio}}]{Reiners2018}
{Reiners}, A., {Zechmeister}, M., {Caballero}, J.~A., {et~al.} 2018, \aap, 612,
  A49

\bibitem[{{Ricker} {et~al.}(2014){Ricker}, {Winn}, {Vanderspek}, {Latham},
  {Bakos}, {Bean}, {Berta-Thompson}, {Brown}, {Buchhave}, {Butler}, {Butler},
  {Chaplin}, {Charbonneau}, {Christensen-Dalsgaard}, {Clampin}, {Deming},
  {Doty}, {De Lee}, {Dressing}, {Dunham}, {Endl}, {Fressin}, {Ge}, {Henning},
  {Holman}, {Howard}, {Ida}, {Jenkins}, {Jernigan}, {Johnson}, {Kaltenegger},
  {Kawai}, {Kjeldsen}, {Laughlin}, {Levine}, {Lin}, {Lissauer}, {MacQueen},
  {Marcy}, {McCullough}, {Morton}, {Narita}, {Paegert}, {Palle}, {Pepe},
  {Pepper}, {Quirrenbach}, {Rinehart}, {Sasselov}, {Sato}, {Seager},
  {Sozzetti}, {Stassun}, {Sullivan}, {Szentgyorgyi}, {Torres}, {Udry}, \&
  {Villasenor}}]{Ricker2014}
{Ricker}, G.~R., {Winn}, J.~N., {Vanderspek}, R., {et~al.} 2014, in Society of
  Photo-Optical Instrumentation Engineers (SPIE) Conference Series, Vol. 9143,
  Space Telescopes and Instrumentation 2014: Optical, Infrared, and Millimeter
  Wave, ed. J.~{Oschmann}, Jacobus~M., M.~{Clampin}, G.~G. {Fazio}, \& H.~A.
  {MacEwen}, 914320

\bibitem[{{Rodriguez} {et~al.}(2020){Rodriguez}, {Vanderburg}, {Zieba},
  {Kreidberg}, {Morley}, {Eastman}, {Kane}, {Spencer}, {Quinn}, {Cloutier},
  {Huang}, {Collins}, {Mann}, {Gilbert}, {Schlieder}, {Quintana}, {Barclay},
  {Suissa}, {Kopparapu}, {Dressing}, {Ricker}, {Vanderspek}, {Latham},
  {Seager}, {Winn}, {Jenkins}, {Berta-Thompson}, {Boyd}, {Charbonneau},
  {Caldwell}, {Chiang}, {Christiansen}, {Ciardi}, {Col{\'o}n}, {Doty}, {Gan},
  {Guerrero}, {G{\"u}nther}, {Lee}, {Levine}, {Lopez}, {Muirhead}, {Newton},
  {Rose}, {Twicken}, \& {Villase{\~n}or}}]{Rodriguez2020}
{Rodriguez}, J.~E., {Vanderburg}, A., {Zieba}, S., {et~al.} 2020, \aj, 160, 117

\bibitem[{{Rojas-Ayala} {et~al.}(2010){Rojas-Ayala}, {Covey}, {Muirhead}, \&
  {Lloyd}}]{RojasAyala2010}
{Rojas-Ayala}, B., {Covey}, K.~R., {Muirhead}, P.~S., \& {Lloyd}, J.~P. 2010,
  \apjl, 720, L113

\bibitem[{{Rowe} {et~al.}(2014){Rowe}, {Bryson}, {Marcy}, {Lissauer},
  {Jontof-Hutter}, {Mullally}, {Gilliland}, {Issacson}, {Ford}, {Howell},
  {Borucki}, {Haas}, {Huber}, {Steffen}, {Thompson}, {Quintana}, {Barclay},
  {Still}, {Fortney}, {Gautier}, {Hunter}, {Caldwell}, {Ciardi}, {Devore},
  {Cochran}, {Jenkins}, {Agol}, {Carter}, \& {Geary}}]{2014ApJ...784...45R}
{Rowe}, J.~F., {Bryson}, S.~T., {Marcy}, G.~W., {et~al.} 2014, \apj, 784, 45

\bibitem[{{Safonov} {et~al.}(2017){Safonov}, {Lysenko}, \&
  {Dodin}}]{Safonov2017}
{Safonov}, B.~S., {Lysenko}, P.~A., \& {Dodin}, A.~V. 2017, Astronomy Letters,
  43, 344

\bibitem[{{Schanche} {et~al.}(2022){Schanche}, {Pozuelos}, {G{\"u}nther},
  {Wells}, {Burgasser}, {Chinchilla}, {Delrez}, {Ducrot}, {Garcia}, {G{\'o}mez
  Maqueo Chew}, {Jofr{\'e}}, {Rackham}, {Sebastian}, {Stassun}, {Stern},
  {Timmermans}, {Barkaoui}, {Belinski}, {Benkhaldoun}, {Benz}, {Bieryla},
  {Bouchy}, {Burdanov}, {Charbonneau}, {Christiansen}, {Collins}, {Demory},
  {D{\'e}vora-Pajares}, {de Wit}, {Dragomir}, {Dransfield}, {Furlan},
  {Ghachoui}, {Gillon}, {Gnilka}, {G{\'o}mez-Mu{\~n}oz}, {Guerrero}, {Harris},
  {Heng}, {Henze}, {Hesse}, {Howell}, {Jehin}, {Jenkins}, {Jensen}, {Kunimoto},
  {Latham}, {Lester}, {McLeod}, {Mireles}, {Murray}, {Niraula}, {Pedersen},
  {Queloz}, {Quintana}, {Ricker}, {Rudat}, {Sabin}, {Safonov},
  {Schroffenegger}, {Scott}, {Seager}, {Strakhov}, {Triaud}, {Vanderspek},
  {Vezie}, \& {Winn}}]{schanche2022}
{Schanche}, N., {Pozuelos}, F.~J., {G{\"u}nther}, M.~N., {et~al.} 2022, \aap,
  657, A45

\bibitem[{{Schweitzer} {et~al.}(2019){Schweitzer}, {Passegger}, {Cifuentes},
  {B{\'e}jar}, {Cort{\'e}s-Contreras}, {Caballero}, {del Burgo}, {Czesla},
  {K{\"u}rster}, {Montes}, {Zapatero Osorio}, {Ribas}, {Reiners},
  {Quirrenbach}, {Amado}, {Aceituno}, {Anglada-Escud{\'e}}, {Bauer},
  {Dreizler}, {Jeffers}, {Guenther}, {Henning}, {Kaminski}, {Lafarga},
  {Marfil}, {Morales}, {Schmitt}, {Seifert}, {Solano}, {Tabernero}, \&
  {Zechmeister}}]{Schweitzer2019}
{Schweitzer}, A., {Passegger}, V.~M., {Cifuentes}, C., {et~al.} 2019, \aap,
  625, A68

\bibitem[{{Shields} {et~al.}(2016){Shields}, {Ballard}, \&
  {Johnson}}]{2016PhR...663....1S}
{Shields}, A.~L., {Ballard}, S., \& {Johnson}, J.~A. 2016, \physrep, 663, 1

\bibitem[{{Skrutskie} {et~al.}(2006){Skrutskie}, {Cutri}, {Stiening},
  {Weinberg}, {Schneider}, {Carpenter}, {Beichman}, {Capps}, {Chester},
  {Elias}, {Huchra}, {Liebert}, {Lonsdale}, {Monet}, {Price}, {Seitzer},
  {Jarrett}, {Kirkpatrick}, {Gizis}, {Howard}, {Evans}, {Fowler}, {Fullmer},
  {Hurt}, {Light}, {Kopan}, {Marsh}, {McCallon}, {Tam}, {Van Dyk}, \&
  {Wheelock}}]{2MASS}
{Skrutskie}, M.~F., {Cutri}, R.~M., {Stiening}, R., {et~al.} 2006, \aj, 131,
  1163

\bibitem[{{Smith} {et~al.}(2012){Smith}, {Stumpe}, {Van Cleve}, {Jenkins},
  {Barclay}, {Fanelli}, {Girouard}, {Kolodziejczak}, {McCauliff}, {Morris}, \&
  {Twicken}}]{Smith2012}
{Smith}, J.~C., {Stumpe}, M.~C., {Van Cleve}, J.~E., {et~al.} 2012, \pasp, 124,
  1000

\bibitem[{{Southworth}(2011)}]{Southworth2011}
{Southworth}, J. 2011, \mnras, 417, 2166

\bibitem[{{Stassun} {et~al.}(2018){Stassun}, {Oelkers}, {Pepper}, {Paegert},
  {De Lee}, {Torres}, {Latham}, {Charpinet}, {Dressing}, {Huber}, {Kane},
  {L{\'e}pine}, {Mann}, {Muirhead}, {Rojas-Ayala}, {Silvotti}, {Fleming},
  {Levine}, \& {Plavchan}}]{Stassun2018}
{Stassun}, K.~G., {Oelkers}, R.~J., {Pepper}, J., {et~al.} 2018, \aj, 156, 102

\bibitem[{{Stef{\'a}nsson} {et~al.}(2020){Stef{\'a}nsson}, {Kopparapu}, {Lin},
  {Mahadevan}, {Ca{\~n}as}, {Kanodia}, {Ninan}, {Cochran}, {Endl}, {Hebb},
  {Wisniewski}, {Gupta}, {Everett}, {Bender}, {Diddams}, {Ford}, {Fredrick},
  {Halverson}, {Hearty}, {Levi}, {Maney}, {Metcalf}, {Monson}, {Ramsey},
  {Robertson}, {Roy}, {Schwab}, {Terrien}, \& {Wright}}]{2020AJ....160..259S}
{Stef{\'a}nsson}, G., {Kopparapu}, R., {Lin}, A., {et~al.} 2020, \aj, 160, 259

\bibitem[{{Stelzer} {et~al.}(2022){Stelzer}, {Bogner}, {Magaudda}, \&
  {Raetz}}]{2022A&A...665A..30S}
{Stelzer}, B., {Bogner}, M., {Magaudda}, E., \& {Raetz}, S. 2022, \aap, 665,
  A30

\bibitem[{{Stumpe} {et~al.}(2014){Stumpe}, {Smith}, {Catanzarite}, {Van Cleve},
  {Jenkins}, {Twicken}, \& {Girouard}}]{Stumpe2014}
{Stumpe}, M.~C., {Smith}, J.~C., {Catanzarite}, J.~H., {et~al.} 2014, \pasp,
  126, 100

\bibitem[{{Stumpe} {et~al.}(2012){Stumpe}, {Smith}, {Van Cleve}, {Twicken},
  {Barclay}, {Fanelli}, {Girouard}, {Jenkins}, {Kolodziejczak}, {McCauliff}, \&
  {Morris}}]{Stumpe2012}
{Stumpe}, M.~C., {Smith}, J.~C., {Van Cleve}, J.~E., {et~al.} 2012, \pasp, 124,
  985

\bibitem[{{Su{\'a}rez Mascare{\~n}o} {et~al.}(2021){Su{\'a}rez Mascare{\~n}o},
  {Damasso}, {Lodieu}, {Sozzetti}, {B{\'e}jar}, {Benatti}, {Zapatero Osorio},
  {Micela}, {Rebolo}, {Desidera}, {Murgas}, {Claudi}, {Gonz{\'a}lez
  Hern{\'a}ndez}, {Malavolta}, {del Burgo}, {D'Orazi}, {Amado}, {Locci},
  {Tabernero}, {Marzari}, {Aguado}, {Turrini}, {Cardona Guill{\'e}n},
  {Toledo-Padr{\'o}n}, {Maggio}, {Aceituno}, {Bauer}, {Caballero},
  {Chinchilla}, {Esparza-Borges}, {Gonz{\'a}lez-{\'A}lvarez}, {Granzer},
  {Luque}, {Mart{\'\i}n}, {Nowak}, {Oshagh}, {Pall{\'e}}, {Parviainen},
  {Quirrenbach}, {Reiners}, {Ribas}, {Strassmeier}, {Weber}, \&
  {Mallonn}}]{SuarezMascareno2022}
{Su{\'a}rez Mascare{\~n}o}, A., {Damasso}, M., {Lodieu}, N., {et~al.} 2021,
  Nature Astronomy, 6, 232

\bibitem[{{Su{\'a}rez Mascare{\~n}o} {et~al.}(2020){Su{\'a}rez Mascare{\~n}o},
  {Faria}, {Figueira}, {Lovis}, {Damasso}, {Gonz{\'a}lez Hern{\'a}ndez},
  {Rebolo}, {Cristiani}, {Pepe}, {Santos}, {Zapatero Osorio}, {Adibekyan},
  {Hojjatpanah}, {Sozzetti}, {Murgas}, {Abreu}, {Affolter}, {Alibert},
  {Aliverti}, {Allart}, {Allende Prieto}, {Alves}, {Amate}, {Avila}, {Baldini},
  {Bandi}, {Barros}, {Bianco}, {Benz}, {Bouchy}, {Broeng}, {Cabral},
  {Calderone}, {Cirami}, {Coelho}, {Conconi}, {Coretti}, {Cumani}, {Cupani},
  {D'Odorico}, {Deiries}, {Delabre}, {Di Marcantonio}, {Dumusque},
  {Ehrenreich}, {Fragoso}, {Genolet}, {Genoni}, {G{\'e}nova Santos}, {Hughes},
  {Iwert}, {Kerber}, {Knusdstrup}, {Landoni}, {Lavie}, {Lillo-Box}, {Lizon},
  {Lo Curto}, {Maire}, {Manescau}, {Martins}, {M{\'e}gevand}, {Mehner},
  {Micela}, {Modigliani}, {Molaro}, {Monteiro}, {Monteiro}, {Moschetti},
  {Mueller}, {Nunes}, {Oggioni}, {Oliveira}, {Pall{\'e}}, {Pariani},
  {Pasquini}, {Poretti}, {Rasilla}, {Redaelli}, {Riva}, {Santana Tschudi},
  {Santin}, {Santos}, {Segovia}, {Sosnowska}, {Sousa}, {Span{\`o}}, {Tenegi},
  {Udry}, {Zanutta}, \& {Zerbi}}]{SuarezMascareno2020}
{Su{\'a}rez Mascare{\~n}o}, A., {Faria}, J.~P., {Figueira}, P., {et~al.} 2020,
  \aap, 639, A77

\bibitem[{Suissa {et~al.}(2020)Suissa, Mandell, Wolf, Villanueva, Fauchez, \&
  kumar Kopparapu}]{Suissa2020}
Suissa, G., Mandell, A.~M., Wolf, E.~T., {et~al.} 2020, The Astrophysical
  Journal, 891, 58

\bibitem[{{Tabernero} {et~al.}(2022){Tabernero}, {Marfil}, {Montes}, \&
  {Gonz{\'a}lez Hern{\'a}ndez}}]{Tabernero2022}
{Tabernero}, H.~M., {Marfil}, E., {Montes}, D., \& {Gonz{\'a}lez
  Hern{\'a}ndez}, J.~I. 2022, \aap, 657, A66

\bibitem[{{Tamayo} {et~al.}(2020){Tamayo}, {Cranmer}, {Hadden}, {Rein},
  {Battaglia}, {Obertas}, {Armitage}, {Ho}, {Spergel}, {Gilbertson}, {Hussain},
  {Silburt}, {Jontof-Hutter}, \& {Menou}}]{2020PNAS..11718194T}
{Tamayo}, D., {Cranmer}, M., {Hadden}, S., {et~al.} 2020, Proceedings of the
  National Academy of Science, 117, 18194

\bibitem[{{Terrien} {et~al.}(2012){Terrien}, {Mahadevan}, {Bender},
  {Deshpande}, {Ramsey}, \& {Bochanski}}]{Terrien2012}
{Terrien}, R.~C., {Mahadevan}, S., {Bender}, C.~F., {et~al.} 2012, \apjl, 747,
  L38

\bibitem[{{Torres} {et~al.}(2017){Torres}, {Kane}, {Rowe}, {Batalha}, {Henze},
  {Ciardi}, {Barclay}, {Borucki}, {Buchhave}, {Crepp}, {Everett}, {Horch},
  {Howard}, {Howell}, {Isaacson}, {Jenkins}, {Latham}, {Petigura}, \&
  {Quintana}}]{Torres2017}
{Torres}, G., {Kane}, S.~R., {Rowe}, J.~F., {et~al.} 2017, \aj, 154, 264

\bibitem[{{Trifonov} {et~al.}(2018){Trifonov}, {K{\"u}rster}, {Zechmeister},
  {Tal-Or}, {Caballero}, {Quirrenbach}, {Amado}, {Ribas}, {Reiners}, {Reffert},
  {Dreizler}, {Hatzes}, {Kaminski}, {Launhardt}, {Henning}, {Montes},
  {B{\'e}jar}, {Mundt}, {Pavlov}, {Schmitt}, {Seifert}, {Morales}, {Nowak},
  {Jeffers}, {Rodr{\'\i}guez-L{\'o}pez}, {del Burgo}, {Anglada-Escud{\'e}},
  {L{\'o}pez-Santiago}, {Mathar}, {Ammler-von Eiff}, {Guenther}, {Barrado},
  {Gonz{\'a}lez Hern{\'a}ndez}, {Mancini}, {St{\"u}rmer}, {Abril}, {Aceituno},
  {Alonso-Floriano}, {Antona}, {Anwand-Heerwart}, {Arroyo-Torres}, {Azzaro},
  {Baroch}, {Bauer}, {Becerril}, {Ben{\'\i}tez}, {Berdi{\~n}as}, {Bergond},
  {Bl{\"u}mcke}, {Brinkm{\"o}ller}, {Cano}, {C{\'a}rdenas V{\'a}zquez},
  {Casal}, {Cifuentes}, {Claret}, {Colom{\'e}}, {Cort{\'e}s-Contreras},
  {Czesla}, {D{\'\i}ez-Alonso}, {Feiz}, {Fern{\'a}ndez}, {Ferro},
  {Fuhrmeister}, {Galad{\'\i}-Enr{\'\i}quez}, {Garcia-Piquer}, {Garc{\'\i}a
  Vargas}, {Gesa}, {G{\'o}mez Galera}, {Gonz{\'a}lez-Peinado}, {Gr{\"o}zinger},
  {Grohnert}, {Gu{\`a}rdia}, {Guijarro}, {de Guindos}, {Guti{\'e}rrez-Soto},
  {Hagen}, {Hauschildt}, {Hedrosa}, {Helmling}, {Hermelo}, {Hern{\'a}ndez
  Arab{\'\i}}, {Hern{\'a}ndez Casta{\~n}o}, {Hern{\'a}ndez Hernando},
  {Herrero}, {Huber}, {Huke}, {Johnson}, {de Juan}, {Kim}, {Klein},
  {Kl{\"u}ter}, {Klutsch}, {Lafarga}, {Lamp{\'o}n}, {Lara}, {Laun}, {Lemke},
  {Lenzen}, {L{\'o}pez del Fresno}, {L{\'o}pez-Gonz{\'a}lez},
  {L{\'o}pez-Puertas}, {L{\'o}pez Salas}, {Luque}, {Mag{\'a}n Madinabeitia},
  {Mall}, {Mandel}, {Marfil}, {Mar{\'\i}n Molina}, {Maroto Fern{\'a}ndez},
  {Mart{\'\i}n}, {Mart{\'\i}n-Ruiz}, {Marvin}, {Mirabet}, {Moya},
  {Moreno-Raya}, {Nagel}, {Naranjo}, {Nortmann}, {Ofir}, {Oreiro}, {Pall{\'e}},
  {Panduro}, {Pascual}, {Passegger}, {Pedraz}, {P{\'e}rez-Calpena}, {P{\'e}rez
  Medialdea}, {Perger}, {Perryman}, {Pluto}, {Rabaza}, {Ram{\'o}n}, {Rebolo},
  {Redondo}, {Reinhardt}, {Rhode}, {Rix}, {Rodler}, {Rodr{\'\i}guez},
  {Rodr{\'\i}guez Trinidad}, {Rohloff}, {Rosich}, {Sadegi},
  {S{\'a}nchez-Blanco}, {S{\'a}nchez Carrasco}, {S{\'a}nchez-L{\'o}pez},
  {Sanz-Forcada}, {Sarkis}, {Sarmiento}, {Sch{\"a}fer}, {Schiller},
  {Sch{\"o}fer}, {Schweitzer}, {Solano}, {Stahl}, {Strachan}, {Su{\'a}rez},
  {Tabernero}, {Tala}, {Tulloch}, {Veredas}, {Vico Linares}, {Vilardell},
  {Wagner}, {Winkler}, {Wolthoff}, {Xu}, {Yan}, \& {Zapatero
  Osorio}}]{Trifonov2018}
{Trifonov}, T., {K{\"u}rster}, M., {Zechmeister}, M., {et~al.} 2018, \aap, 609,
  A117

\bibitem[{Turbet {et~al.}(2020)Turbet, Bolmont, Ehrenreich, Gratier, Leconte,
  Selsis, Hara, \& Lovis}]{Turbet2020}
Turbet, M., Bolmont, E., Ehrenreich, D., {et~al.} 2020, Astronomy \&amp;
  Astrophysics, Volume 638, id.A41, {$<$}NUMPAGES{$>$}10{$<$}/NUMPAGES{$>$}
  pp., 638, A41

\bibitem[{{Turbet} {et~al.}(2019){Turbet}, {Ehrenreich}, {Lovis}, {Bolmont}, \&
  {Fauchez}}]{Turbet2019}
{Turbet}, M., {Ehrenreich}, D., {Lovis}, C., {Bolmont}, E., \& {Fauchez}, T.
  2019, \aap, 628, A12

\bibitem[{{Twicken} {et~al.}(2018){Twicken}, {Catanzarite}, {Clarke},
  {Girouard}, {Jenkins}, {Klaus}, {Li}, {McCauliff}, {Seader}, {Tenenbaum},
  {Wohler}, {Bryson}, {Burke}, {Caldwell}, {Haas}, {Henze}, \&
  {Sanderfer}}]{Twicken2018}
{Twicken}, J.~D., {Catanzarite}, J.~H., {Clarke}, B.~D., {et~al.} 2018, \pasp,
  130, 064502

\bibitem[{{Virtanen} {et~al.}(2020){Virtanen}, {Gommers}, {Oliphant},
  {Haberland}, {Reddy}, {Cournapeau}, {Burovski}, {Peterson}, {Weckesser},
  {Bright}, {van der Walt}, {Brett}, {Wilson}, {Millman}, {Mayorov}, {Nelson},
  {Jones}, {Kern}, {Larson}, {Carey}, {Polat}, {Feng}, {Moore}, {VanderPlas},
  {Laxalde}, {Perktold}, {Cimrman}, {Henriksen}, {Quintero}, {Harris},
  {Archibald}, {Ribeiro}, {Pedregosa}, {van Mulbregt}, \& {SciPy 1. 0
  Contributors}}]{scipy}
{Virtanen}, P., {Gommers}, R., {Oliphant}, T.~E., {et~al.} 2020, Nature
  Methods, 17, 261

\bibitem[{{Wells} {et~al.}(2021){Wells}, {Rackham}, {Schanche}, {Petrucci},
  {G{\'o}mez Maqueo Chew}, {Demory}, {Burgasser}, {Burn}, {Pozuelos},
  {G{\"u}nther}, {Sabin}, {Schroffenegger}, {G{\'o}mez-Mu{\~n}oz}, {Stassun},
  {Van Grootel}, {Howell}, {Sebastian}, {Triaud}, {Apai}, {Plauchu-Frayn},
  {Guerrero}, {Guill{\'e}n}, {Landa}, {Melgoza}, {Montalvo}, {Serrano},
  {Riesgo}, {Barkaoui}, {Bixel}, {Burdanov}, {Chen}, {Chinchilla}, {Collins},
  {Daylan}, {de Wit}, {Delrez}, {D{\'e}vora-Pajares}, {Dietrich}, {Dransfield},
  {Ducrot}, {Fausnaugh}, {Furlan}, {Gabor}, {Gan}, {Garcia}, {Ghachoui},
  {Giacalone}, {Gibbs}, {Gillon}, {Gnilka}, {Gore}, {Guerrero}, {Henning},
  {Hesse}, {Jehin}, {Jenkins}, {Latham}, {Lester}, {McCormac}, {Murray},
  {Niraula}, {Pedersen}, {Queloz}, {Ricker}, {Rodriguez}, {Schroeder},
  {Schwarz}, {Scott}, {Seager}, {Theissen}, {Thompson}, {Timmermans},
  {Twicken}, \& {Winn}}]{wells2021}
{Wells}, R.~D., {Rackham}, B.~V., {Schanche}, N., {et~al.} 2021, \aap, 653, A97

\bibitem[{{Wisdom} \& {Holman}(1991)}]{1991AJ....102.1528W}
{Wisdom}, J. \& {Holman}, M. 1991, \aj, 102, 1528

\bibitem[{{Wizinowich} {et~al.}(2000){Wizinowich}, {Acton}, {Shelton},
  {Stomski}, {Gathright}, {Ho}, {Lupton}, {Tsubota}, {Lai}, {Max}, {Brase},
  {An}, {Avicola}, {Olivier}, {Gavel}, {Macintosh}, {Ghez}, \&
  {Larkin}}]{wizinowich2000}
{Wizinowich}, P., {Acton}, D.~S., {Shelton}, C., {et~al.} 2000, \pasp, 112, 315

\bibitem[{{Wu}(2019)}]{Wu2019}
{Wu}, Y. 2019, \apj, 874, 91

\bibitem[{{Zacharias} {et~al.}(2013){Zacharias}, {Finch}, {Girard}, {Henden},
  {Bartlett}, {Monet}, \& {Zacharias}}]{UCAC4}
{Zacharias}, N., {Finch}, C.~T., {Girard}, T.~M., {et~al.} 2013, \aj, 145, 44

\bibitem[{{Zechmeister} {et~al.}(2014){Zechmeister}, {Anglada-Escud{\'e}}, \&
  {Reiners}}]{Zechmeister2014}
{Zechmeister}, M., {Anglada-Escud{\'e}}, G., \& {Reiners}, A. 2014, \aap, 561,
  A59

\bibitem[{{Zechmeister} \& {K{\"u}rster}(2009)}]{Zechmeister2009}
{Zechmeister}, M. \& {K{\"u}rster}, M. 2009, \aap, 496, 577

\bibitem[{{Zechmeister} {et~al.}(2018){Zechmeister}, {Reiners}, {Amado},
  {Azzaro}, {Bauer}, {B{\'e}jar}, {Caballero}, {Guenther}, {Hagen}, {Jeffers},
  {Kaminski}, {K{\"u}rster}, {Launhardt}, {Montes}, {Morales}, {Quirrenbach},
  {Reffert}, {Ribas}, {Seifert}, {Tal-Or}, \& {Wolthoff}}]{Zechmeister2018}
{Zechmeister}, M., {Reiners}, A., {Amado}, P.~J., {et~al.} 2018, \aap, 609, A12

\bibitem[{{Zeng} {et~al.}(2019){Zeng}, {Jacobsen}, {Sasselov}, {Petaev},
  {Vanderburg}, {Lopez-Morales}, {Perez-Mercader}, {Mattsson}, {Li}, {Heising},
  {Bonomo}, {Damasso}, {Berger}, {Cao}, {Levi}, \& {Wordsworth}}]{Zeng2019}
{Zeng}, L., {Jacobsen}, S.~B., {Sasselov}, D.~D., {et~al.} 2019, Proceedings of
  the National Academy of Science, 116, 9723

\bibitem[{{Zeng} {et~al.}(2016){Zeng}, {Sasselov}, \& {Jacobsen}}]{Zeng2016}
{Zeng}, L., {Sasselov}, D.~D., \& {Jacobsen}, S.~B. 2016, \apj, 819, 127

\bibitem[{{Zhu} {et~al.}(2018){Zhu}, {Petrovich}, {Wu}, {Dong}, \&
  {Xie}}]{zhu2018}
{Zhu}, W., {Petrovich}, C., {Wu}, Y., {Dong}, S., \& {Xie}, J. 2018, \apj, 860,
  101

\bibitem[{{Ziegler} {et~al.}(2020){Ziegler}, {Tokovinin}, {Brice{\~n}o},
  {Mang}, {Law}, \& {Mann}}]{ziegler2020}
{Ziegler}, C., {Tokovinin}, A., {Brice{\~n}o}, C., {et~al.} 2020, \aj, 159, 19

\bibitem[{{Zinzi} \& {Turrini}(2017)}]{zinzi2017}
{Zinzi}, A. \& {Turrini}, D. 2017, \aap, 605, L4

\bibitem[{{Zsom} {et~al.}(2013){Zsom}, {Seager}, {de Wit}, \&
  {Stamenkovi{\'c}}}]{Zsom2013}
{Zsom}, A., {Seager}, S., {de Wit}, J., \& {Stamenkovi{\'c}}, V. 2013, \apj,
  778, 109

\end{thebibliography}

\begin{appendix}

\section{\textit{TESS} target pixel file images}
\begin{figure*}[htbp]
   \centering
   \includegraphics[width=\hsize]{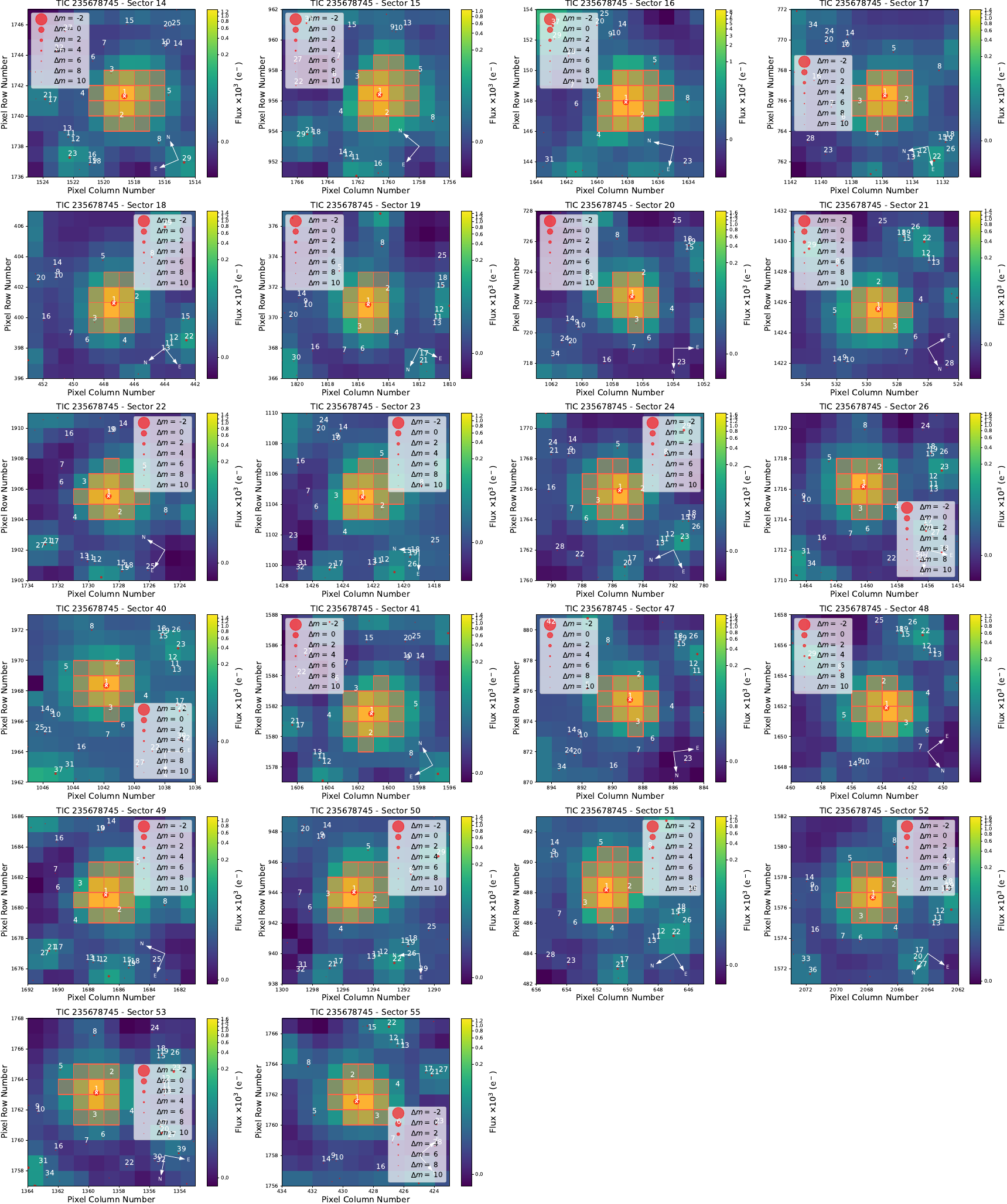}
   \caption{\textit{TESS} target pixel file images of TOI-2095 for the 22 sectors analyzed in this work (made with \texttt{tpfplotter}, \citealp{Aller2020}).}
    \label{Fig:TESS_TPFs}
\end{figure*}

In Figure \ref{Fig:TESS_TPFs}, we present the \textit{TESS} target pixel file images of all the sectors analyzed in this work. 

\section{Ground-based light curves}
\begin{figure*}[htbp]
   \centering
   \includegraphics[width=\hsize]{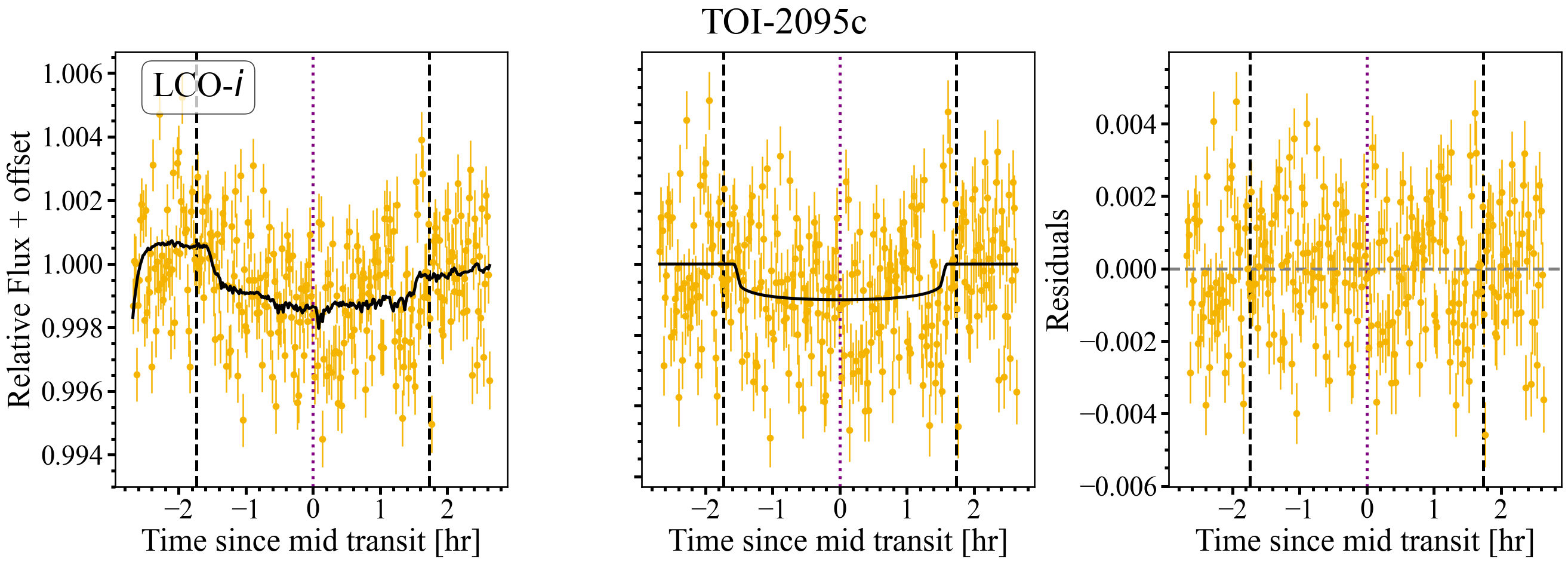}
   \caption{LCOGT ground-based transit observations of TOI-2095c. \textit{Left panel:} Transit observations and best fitted model (black line). \textit{Middle panel:} Transit model and model after removing systematics. \textit{Right panel:} Residuals of the fit. The central time of the transit is marked by the purple dotted line and the predicted ingress and egress by the black dashed line.}
    \label{Fig:Transit_TOI2095c_GroundBased}
\end{figure*}

\begin{figure*}[htbp]
   \centering
   \includegraphics[width=\hsize]{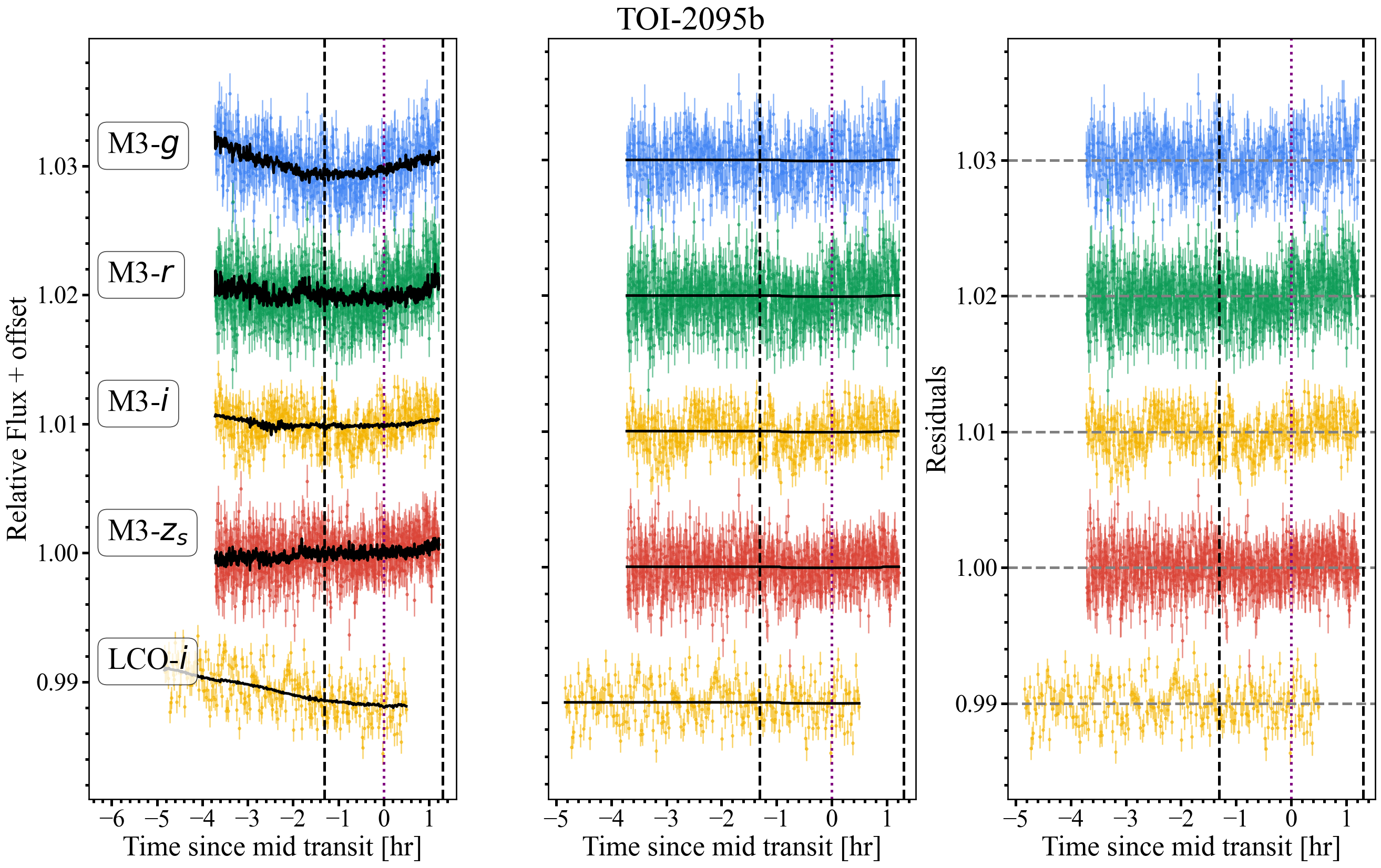}
   \caption{MuSCAT3 and LCOGT ground-based transit observations of TOI-2095b. \textit{Middle panel:} Transit data and model after removing systematics. \textit{Right panel:} Residuals of the fit. The central time of the transit is marked by the purple dotted line and the predicted ingress and egress by the black-dashed line.}
    \label{Fig:Transit_TOI2095b_GroundBased}
\end{figure*}

In this section we present the ground-based light curves for TOI-2095b and TOI-2095c taken with LCOGT and MuSCAT3 (see Figs. \ref{Fig:Transit_TOI2095b_GroundBased} and \ref{Fig:Transit_TOI2095c_GroundBased}).

\section{Radial velocity data}
\begin{figure}
 \centering
 \includegraphics[width=\hsize]{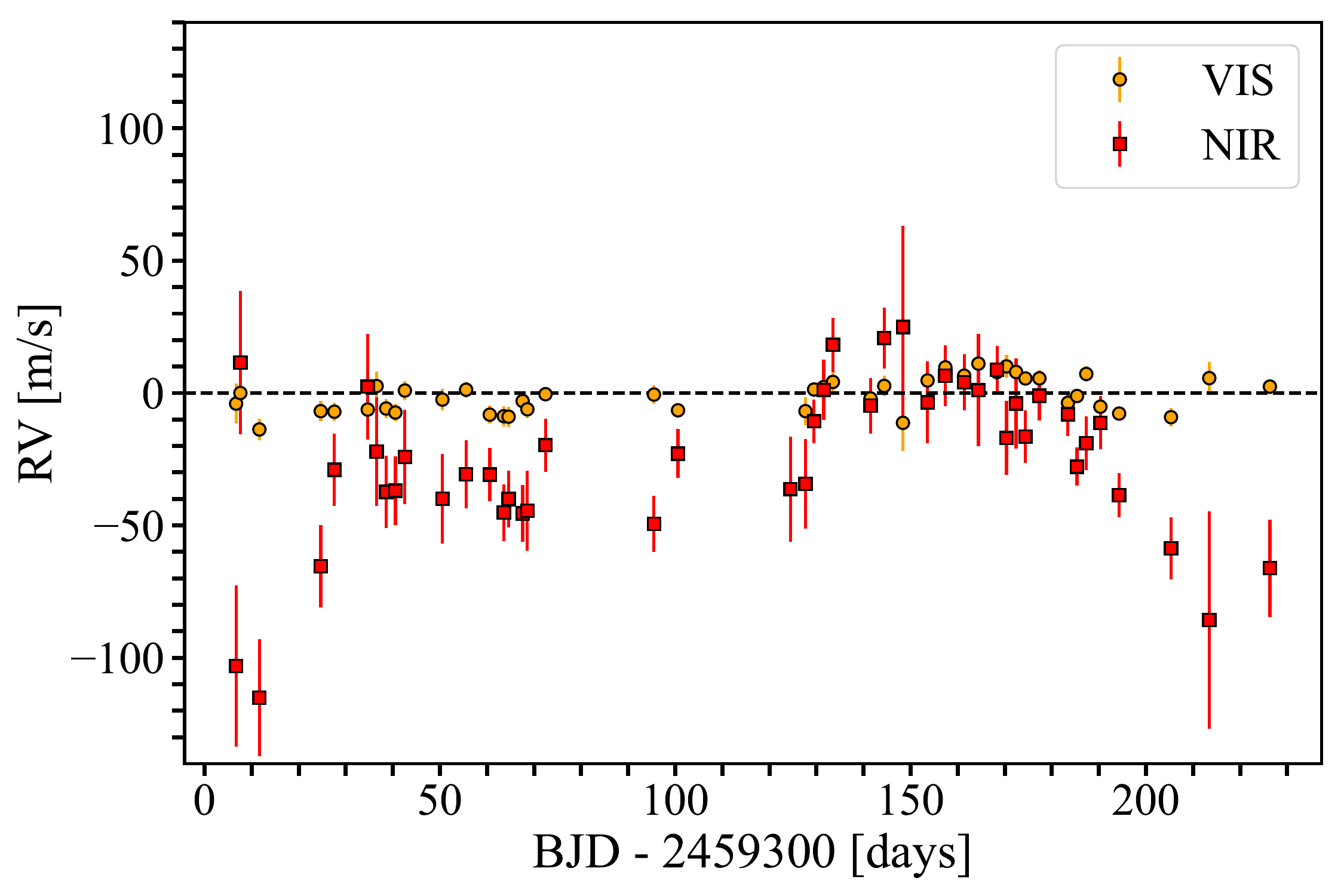}
   \caption{Radial velocity measurements obtained using \texttt{SERVAL} for the visible (VIS) and near-infrared (NIR) CARMENES arms.}
   \label{Fig:TOI2095_RVs_VIS_NIR}
\end{figure}

In Tables \ref{Tab:CARMENES_RVs} and \ref{Tab:CARMENES_Activ}, we present RV measurements taken with CARMENES visible channel and line activity indicators of TOI-2095 computed using \texttt{SERVAL}. Table \ref{Tab:CARMENES_Activ_CCF} presents the CCF activity indicators computed with \texttt{raccoon}. Figure \ref{Fig:TOI2095_RVs_VIS_NIR} shows the CARMENES RV measurements for the visible and near-infrared channels.

\begin{table}[]
\caption{CARMENES visual channel radial velocity measurements of TOI-2095 computed by \texttt{SERVAL}.}
\label{Tab:CARMENES_RVs}
\centering
\begin{tabular}{c c}
\hline 
\hline
BJD -- 2\,459\,000 & RV [m\,s$^{-1}$] \\
\hline
\noalign{\smallskip}

306.700877 & $-4.005 \pm 7.545$ \\ 
307.606073 & $0.012 \pm 6.346$ \\ 
311.640836 & $-13.752 \pm 3.995$ \\ 
324.673929 & $-6.845 \pm 3.803$ \\ 
327.574854 & $-7.055 \pm 3.549$ \\ 
334.661366 & $-6.253 \pm 6.847$ \\ 
336.547887 & $2.628 \pm 5.431$ \\ 
338.573383 & $-5.838 \pm 3.529$ \\ 
340.533658 & $-7.365 \pm 3.234$ \\ 
342.553046 & $0.948 \pm 3.608$ \\ 
350.531737 & $-2.477 \pm 3.963$ \\ 
355.561823 & $1.257 \pm 2.967$ \\ 
360.583611 & $-8.166 \pm 3.430$ \\ 
363.581988 & $-8.756 \pm 3.776$ \\ 
364.584887 & $-8.938 \pm 3.822$ \\ 
367.570299 & $-3.023 \pm 3.655$ \\ 
368.575744 & $-6.141 \pm 3.280$ \\ 
372.427499 & $-0.438 \pm 2.724$ \\ 
395.472621 & $-0.580 \pm 3.578$ \\ 
400.577946 & $-6.528 \pm 2.760$ \\ 
427.650559 & $-6.839 \pm 5.270$ \\ 
429.484273 & $1.333 \pm 2.769$ \\ 
431.528152 & $2.265 \pm 3.125$ \\ 
433.487500 & $4.190 \pm 3.996$ \\ 
441.477336 & $-2.048 \pm 2.964$ \\ 
444.393281 & $2.678 \pm 3.874$ \\ 
448.367263 & $-11.256 \pm 10.641$ \\ 
453.597582 & $4.742 \pm 3.437$ \\ 
457.375964 & $9.710 \pm 2.724$ \\ 
461.373561 & $6.554 \pm 3.030$ \\ 
464.398055 & $11.107 \pm 4.866$ \\ 
468.347662 & $7.933 \pm 3.030$ \\ 
470.359520 & $10.064 \pm 4.262$ \\ 
472.395563 & $7.952 \pm 4.414$ \\ 
474.389298 & $5.508 \pm 2.365$ \\ 
477.356478 & $5.543 \pm 3.050$ \\ 
483.423217 & $-3.681 \pm 3.161$ \\ 
485.346034 & $-1.122 \pm 2.776$ \\ 
487.313429 & $7.212 \pm 2.938$ \\ 
490.311946 & $-5.165 \pm 2.799$ \\ 
494.296097 & $-7.754 \pm 2.556$ \\ 
505.314133 & $-9.123 \pm 3.528$ \\ 
513.442386 & $5.636 \pm 6.078$ \\ 
526.324052 & $2.439 \pm 2.785$ \\ 
    
\noalign{\smallskip}
\hline
\end{tabular}
\end{table}

\begin{sidewaystable*}
\caption{CARMENES spectroscopic line activity indicators for TOI-2095 computed by \texttt{SERVAL}.}             
\label{Tab:CARMENES_Activ}      
\centering          
\begin{tabular}{l c c c c c c c c}     
\hline 
\hline
\noalign{\smallskip}
BJD -- 2\,459\,000 & CRX [m\,s$^{-1}$N$_\mathrm{p}^{-1}$] & dLW [m$^2$s$^{-2}$] & H$_\alpha$ [m\,s$^{-1}$] & Na~{\sc i} D$_1$ [m\,s$^{-1}$] & Na~{\sc i} D$_2$ [m\,s$^{-1}$] & Ca~{\sc ii} IRTa [m\,s$^{-1}$] & Ca~{\sc ii} IRTb [m\,s$^{-1}$] & Ca~{\sc ii} IRTc [m\,s$^{-1}$]  \\
\hline
\noalign{\smallskip}

306.700877 & $-19.4 \pm 67.6$ & $26.6 \pm 7.1$ & $0.815 \pm 0.006$ & $0.326 \pm 0.018$ & $0.273 \pm 0.017$ & $0.541 \pm 0.005$ & $0.409 \pm 0.006$ & $0.397 \pm 0.006$ \\ 
307.606073 & $-74.8 \pm 46.6$ & $7.3 \pm 4.6$ & $0.807 \pm 0.004$ & $0.209 \pm 0.012$ & $0.191 \pm 0.012$ & $0.565 \pm 0.004$ & $0.407 \pm 0.004$ & $0.399 \pm 0.004$ \\ 
311.640836 & $-20.7 \pm 52.5$ & $11.6 \pm 6.4$ & $0.827 \pm 0.004$ & $0.206 \pm 0.012$ & $0.224 \pm 0.011$ & $0.545 \pm 0.004$ & $0.409 \pm 0.004$ & $0.396 \pm 0.004$ \\ 
324.673929 & $-19.7 \pm 39.3$ & $5.9 \pm 4.7$ & $0.816 \pm 0.003$ & $0.172 \pm 0.007$ & $0.178 \pm 0.007$ & $0.547 \pm 0.003$ & $0.406 \pm 0.003$ & $0.395 \pm 0.003$ \\ 
327.574854 & $-79.1 \pm 37.2$ & $14.2 \pm 4.9$ & $0.819 \pm 0.004$ & $0.172 \pm 0.009$ & $0.187 \pm 0.009$ & $0.556 \pm 0.004$ & $0.404 \pm 0.004$ & $0.399 \pm 0.004$ \\ 
334.661366 & $-72.0 \pm 77.3$ & $24.6 \pm 9.5$ & $0.821 \pm 0.007$ & $0.192 \pm 0.019$ & $0.210 \pm 0.019$ & $0.560 \pm 0.006$ & $0.403 \pm 0.007$ & $0.390 \pm 0.007$ \\ 
336.547887 & $-57.9 \pm 57.4$ & $2.6 \pm 5.6$ & $0.806 \pm 0.005$ & $0.204 \pm 0.013$ & $0.191 \pm 0.013$ & $0.557 \pm 0.005$ & $0.408 \pm 0.005$ & $0.398 \pm 0.005$ \\ 
338.573383 & $-59.8 \pm 37.7$ & $12.2 \pm 3.2$ & $0.813 \pm 0.003$ & $0.181 \pm 0.005$ & $0.171 \pm 0.005$ & $0.560 \pm 0.003$ & $0.409 \pm 0.003$ & $0.403 \pm 0.003$ \\ 
340.533658 & $-37.4 \pm 32.8$ & $3.6 \pm 2.5$ & $0.814 \pm 0.002$ & $0.175 \pm 0.005$ & $0.175 \pm 0.005$ & $0.561 \pm 0.002$ & $0.413 \pm 0.002$ & $0.404 \pm 0.002$ \\ 
342.553046 & $-45.1 \pm 41.3$ & $3.3 \pm 4.0$ & $0.815 \pm 0.004$ & $0.185 \pm 0.008$ & $0.186 \pm 0.008$ & $0.562 \pm 0.003$ & $0.410 \pm 0.003$ & $0.410 \pm 0.003$ \\ 
350.531737 & $-93.4 \pm 39.6$ & $11.6 \pm 5.5$ & $0.824 \pm 0.004$ & $0.170 \pm 0.011$ & $0.183 \pm 0.010$ & $0.546 \pm 0.004$ & $0.416 \pm 0.004$ & $0.405 \pm 0.004$ \\ 
355.561823 & $-49.2 \pm 31.3$ & $1.9 \pm 4.4$ & $0.809 \pm 0.003$ & $0.183 \pm 0.006$ & $0.173 \pm 0.006$ & $0.559 \pm 0.003$ & $0.406 \pm 0.003$ & $0.398 \pm 0.003$ \\ 
360.583611 & $-94.4 \pm 33.2$ & $11.0 \pm 3.4$ & $0.808 \pm 0.003$ & $0.161 \pm 0.005$ & $0.183 \pm 0.005$ & $0.553 \pm 0.003$ & $0.397 \pm 0.003$ & $0.399 \pm 0.003$ \\ 
363.581988 & $-91.4 \pm 38.7$ & $-2.7 \pm 4.1$ & $0.825 \pm 0.003$ & $0.159 \pm 0.007$ & $0.185 \pm 0.007$ & $0.545 \pm 0.003$ & $0.396 \pm 0.003$ & $0.397 \pm 0.003$ \\ 
364.584887 & $-145.2 \pm 35.8$ & $10.2 \pm 3.9$ & $0.815 \pm 0.003$ & $0.150 \pm 0.006$ & $0.187 \pm 0.006$ & $0.547 \pm 0.003$ & $0.408 \pm 0.003$ & $0.394 \pm 0.003$ \\ 
367.570299 & $-96.3 \pm 32.3$ & $0.2 \pm 4.7$ & $0.825 \pm 0.003$ & $0.154 \pm 0.006$ & $0.162 \pm 0.006$ & $0.552 \pm 0.003$ & $0.404 \pm 0.003$ & $0.392 \pm 0.003$ \\ 
368.575744 & $-47.7 \pm 36.5$ & $-1.4 \pm 2.8$ & $0.820 \pm 0.003$ & $0.155 \pm 0.006$ & $0.162 \pm 0.006$ & $0.552 \pm 0.003$ & $0.400 \pm 0.003$ & $0.404 \pm 0.003$ \\ 
372.427499 & $-51.3 \pm 30.6$ & $6.7 \pm 2.9$ & $0.820 \pm 0.002$ & $0.164 \pm 0.004$ & $0.168 \pm 0.004$ & $0.554 \pm 0.002$ & $0.397 \pm 0.002$ & $0.394 \pm 0.002$ \\ 
395.472621 & $-34.7 \pm 36.0$ & $7.5 \pm 3.3$ & $0.808 \pm 0.003$ & $0.181 \pm 0.006$ & $0.178 \pm 0.006$ & $0.557 \pm 0.003$ & $0.420 \pm 0.003$ & $0.423 \pm 0.003$ \\ 
400.577946 & $-10.4 \pm 29.0$ & $11.7 \pm 2.1$ & $0.800 \pm 0.002$ & $0.172 \pm 0.004$ & $0.178 \pm 0.004$ & $0.560 \pm 0.002$ & $0.404 \pm 0.002$ & $0.406 \pm 0.002$ \\ 
427.650559 & $108.6 \pm 51.2$ & $-14.6 \pm 6.5$ & $0.829 \pm 0.005$ & $0.165 \pm 0.014$ & $0.158 \pm 0.014$ & $0.559 \pm 0.005$ & $0.424 \pm 0.005$ & $0.415 \pm 0.005$ \\ 
429.484273 & $13.6 \pm 28.0$ & $5.0 \pm 2.5$ & $0.821 \pm 0.003$ & $0.165 \pm 0.005$ & $0.187 \pm 0.005$ & $0.567 \pm 0.003$ & $0.419 \pm 0.003$ & $0.412 \pm 0.002$ \\ 
431.528152 & $0.1 \pm 31.0$ & $-5.9 \pm 3.7$ & $0.820 \pm 0.003$ & $0.177 \pm 0.006$ & $0.193 \pm 0.006$ & $0.567 \pm 0.003$ & $0.426 \pm 0.003$ & $0.416 \pm 0.003$ \\ 
433.487500 & $25.9 \pm 34.4$ & $3.3 \pm 3.4$ & $0.808 \pm 0.003$ & $0.171 \pm 0.006$ & $0.178 \pm 0.006$ & $0.565 \pm 0.003$ & $0.422 \pm 0.003$ & $0.412 \pm 0.003$ \\ 
441.477336 & $-14.2 \pm 29.8$ & $-4.6 \pm 3.1$ & $0.809 \pm 0.003$ & $0.173 \pm 0.007$ & $0.182 \pm 0.006$ & $0.560 \pm 0.003$ & $0.412 \pm 0.003$ & $0.405 \pm 0.003$ \\ 
444.393281 & $60.4 \pm 36.1$ & $-9.9 \pm 3.7$ & $0.802 \pm 0.003$ & $0.153 \pm 0.007$ & $0.175 \pm 0.007$ & $0.556 \pm 0.003$ & $0.403 \pm 0.003$ & $0.399 \pm 0.003$ \\ 
448.367263 & $141.9 \pm 109.6$ & $-69.2 \pm 17.6$ & $0.792 \pm 0.014$ & $0.155 \pm 0.053$ & $0.195 \pm 0.049$ & $0.542 \pm 0.012$ & $0.408 \pm 0.015$ & $0.403 \pm 0.014$ \\ 
453.597582 & $49.9 \pm 36.5$ & $1.6 \pm 3.8$ & $0.813 \pm 0.003$ & $0.172 \pm 0.006$ & $0.192 \pm 0.006$ & $0.551 \pm 0.003$ & $0.405 \pm 0.003$ & $0.398 \pm 0.003$ \\ 
457.375964 & $29.9 \pm 26.9$ & $-1.8 \pm 2.5$ & $0.825 \pm 0.002$ & $0.164 \pm 0.004$ & $0.173 \pm 0.004$ & $0.554 \pm 0.002$ & $0.402 \pm 0.002$ & $0.397 \pm 0.002$ \\ 
461.373561 & $77.2 \pm 28.4$ & $-6.8 \pm 2.9$ & $0.813 \pm 0.002$ & $0.173 \pm 0.004$ & $0.179 \pm 0.004$ & $0.553 \pm 0.002$ & $0.407 \pm 0.002$ & $0.395 \pm 0.002$ \\ 
464.398055 & $31.7 \pm 47.0$ & $-5.4 \pm 5.3$ & $0.817 \pm 0.005$ & $0.172 \pm 0.011$ & $0.181 \pm 0.011$ & $0.555 \pm 0.005$ & $0.407 \pm 0.005$ & $0.405 \pm 0.005$ \\ 
468.347662 & $78.2 \pm 29.1$ & $-8.1 \pm 2.4$ & $0.823 \pm 0.002$ & $0.181 \pm 0.004$ & $0.182 \pm 0.004$ & $0.560 \pm 0.002$ & $0.412 \pm 0.002$ & $0.408 \pm 0.002$ \\ 
470.359520 & $81.6 \pm 36.5$ & $-19.2 \pm 3.7$ & $0.816 \pm 0.003$ & $0.151 \pm 0.006$ & $0.155 \pm 0.006$ & $0.559 \pm 0.003$ & $0.421 \pm 0.003$ & $0.409 \pm 0.003$ \\ 
472.395563 & $115.0 \pm 42.6$ & $6.3 \pm 4.7$ & $0.838 \pm 0.004$ & $0.183 \pm 0.011$ & $0.198 \pm 0.010$ & $0.566 \pm 0.004$ & $0.422 \pm 0.004$ & $0.419 \pm 0.004$ \\ 
474.389298 & $42.3 \pm 23.4$ & $7.3 \pm 3.0$ & $0.821 \pm 0.003$ & $0.187 \pm 0.005$ & $0.182 \pm 0.005$ & $0.561 \pm 0.003$ & $0.421 \pm 0.003$ & $0.416 \pm 0.002$ \\ 
477.356478 & $45.0 \pm 27.4$ & $9.9 \pm 3.6$ & $0.814 \pm 0.003$ & $0.189 \pm 0.006$ & $0.190 \pm 0.006$ & $0.568 \pm 0.003$ & $0.427 \pm 0.003$ & $0.413 \pm 0.003$ \\ 
483.423217 & $58.7 \pm 31.6$ & $-8.1 \pm 3.3$ & $0.801 \pm 0.003$ & $0.154 \pm 0.006$ & $0.179 \pm 0.006$ & $0.558 \pm 0.003$ & $0.413 \pm 0.003$ & $0.410 \pm 0.003$ \\ 
485.346034 & $38.3 \pm 24.0$ & $-7.1 \pm 3.2$ & $0.794 \pm 0.002$ & $0.156 \pm 0.004$ & $0.172 \pm 0.004$ & $0.561 \pm 0.002$ & $0.408 \pm 0.002$ & $0.401 \pm 0.002$ \\ 
487.313429 & $-60.6 \pm 27.5$ & $-14.9 \pm 4.5$ & $0.814 \pm 0.003$ & $0.150 \pm 0.006$ & $0.156 \pm 0.006$ & $0.557 \pm 0.003$ & $0.406 \pm 0.003$ & $0.403 \pm 0.003$ \\ 
490.311946 & $31.6 \pm 23.5$ & $-13.0 \pm 2.9$ & $0.811 \pm 0.003$ & $0.149 \pm 0.005$ & $0.169 \pm 0.004$ & $0.559 \pm 0.003$ & $0.407 \pm 0.002$ & $0.403 \pm 0.002$ \\ 
494.296097 & $10.4 \pm 26.8$ & $-4.4 \pm 2.7$ & $0.812 \pm 0.002$ & $0.166 \pm 0.004$ & $0.173 \pm 0.004$ & $0.557 \pm 0.003$ & $0.404 \pm 0.003$ & $0.404 \pm 0.003$ \\ 
505.314133 & $-68.7 \pm 26.7$ & $-5.1 \pm 2.4$ & $0.811 \pm 0.003$ & $0.174 \pm 0.005$ & $0.173 \pm 0.005$ & $0.554 \pm 0.003$ & $0.415 \pm 0.003$ & $0.405 \pm 0.003$ \\ 
513.442386 & $21.1 \pm 64.1$ & $37.5 \pm 7.7$ & $0.814 \pm 0.007$ & $0.220 \pm 0.022$ & $0.229 \pm 0.022$ & $0.567 \pm 0.007$ & $0.421 \pm 0.008$ & $0.410 \pm 0.008$ \\ 
526.324052 & $10.0 \pm 27.3$ & $11.1 \pm 3.1$ & $0.794 \pm 0.003$ & $0.181 \pm 0.007$ & $0.184 \pm 0.007$ & $0.562 \pm 0.003$ & $0.409 \pm 0.003$ & $0.395 \pm 0.003$ \\

\noalign{\smallskip}
\hline                  
\end{tabular}
\end{sidewaystable*}

\begin{sidewaystable*}[]
\caption{CARMENES activity indicators from cross-correlation for TOI-2095 computed by \texttt{raccoon}.}
\label{Tab:CARMENES_Activ_CCF}
\centering
\begin{tabular}{c c c c}
\hline 
\hline
BJD -- 2\,459\,000 & FWHM [km\,s$^{-1}$] & CON [\%] & BIS [km\,s$^{-1}$] \\
\hline
\noalign{\smallskip}

306.700877 & $4.391 \pm 0.025$ & $13.60 \pm 0.05$ & $-0.023 \pm 0.031$ \\ 
307.606073 & $4.470 \pm 0.017$ & $13.57 \pm 0.04$ & $-0.020 \pm 0.024$ \\ 
311.640836 & $4.436 \pm 0.023$ & $13.65 \pm 0.05$ & $0.005 \pm 0.023$ \\ 
324.673929 & $4.442 \pm 0.025$ & $13.62 \pm 0.05$ & $-0.005 \pm 0.017$ \\ 
327.574854 & $4.460 \pm 0.024$ & $13.59 \pm 0.05$ & $-0.025 \pm 0.020$ \\ 
334.661366 & $4.450 \pm 0.021$ & $13.49 \pm 0.04$ & $-0.001 \pm 0.036$ \\ 
336.547887 & $4.410 \pm 0.024$ & $13.51 \pm 0.05$ & $0.030 \pm 0.026$ \\ 
338.573383 & $4.471 \pm 0.021$ & $13.50 \pm 0.04$ & $0.000 \pm 0.015$ \\ 
340.533658 & $4.455 \pm 0.018$ & $13.54 \pm 0.04$ & $-0.001 \pm 0.013$ \\ 
342.553046 & $4.442 \pm 0.021$ & $13.55 \pm 0.04$ & $-0.008 \pm 0.019$ \\ 
350.531737 & $4.483 \pm 0.023$ & $13.62 \pm 0.05$ & $0.005 \pm 0.023$ \\ 
355.561823 & $4.498 \pm 0.018$ & $13.54 \pm 0.04$ & $-0.016 \pm 0.017$ \\ 
360.583611 & $4.478 \pm 0.017$ & $13.50 \pm 0.04$ & $-0.012 \pm 0.014$ \\ 
363.581988 & $4.503 \pm 0.017$ & $13.49 \pm 0.03$ & $-0.011 \pm 0.018$ \\ 
364.584887 & $4.493 \pm 0.018$ & $13.53 \pm 0.04$ & $-0.007 \pm 0.016$ \\ 
367.570299 & $4.446 \pm 0.023$ & $13.66 \pm 0.05$ & $-0.016 \pm 0.016$ \\ 
368.575744 & $4.516 \pm 0.022$ & $13.57 \pm 0.05$ & $-0.013 \pm 0.016$ \\ 
372.427499 & $4.462 \pm 0.018$ & $13.56 \pm 0.04$ & $-0.012 \pm 0.013$ \\ 
395.472621 & $4.478 \pm 0.020$ & $13.46 \pm 0.04$ & $0.002 \pm 0.016$ \\ 
400.577946 & $4.450 \pm 0.017$ & $13.49 \pm 0.04$ & $0.007 \pm 0.013$ \\ 
427.650559 & $4.488 \pm 0.027$ & $13.69 \pm 0.06$ & $0.010 \pm 0.027$ \\ 
429.484273 & $4.479 \pm 0.020$ & $13.54 \pm 0.04$ & $-0.010 \pm 0.014$ \\ 
431.528152 & $4.470 \pm 0.018$ & $13.51 \pm 0.04$ & $-0.001 \pm 0.017$ \\ 
433.487500 & $4.445 \pm 0.018$ & $13.60 \pm 0.04$ & $-0.008 \pm 0.016$ \\ 
441.477336 & $4.463 \pm 0.022$ & $13.55 \pm 0.05$ & $-0.006 \pm 0.018$ \\ 
444.393281 & $4.472 \pm 0.022$ & $13.57 \pm 0.04$ & $-0.020 \pm 0.017$ \\ 
448.367263 & $4.350 \pm 0.025$ & $13.31 \pm 0.05$ & $-0.020 \pm 0.081$ \\ 
453.597582 & $4.491 \pm 0.019$ & $13.59 \pm 0.04$ & $-0.018 \pm 0.016$ \\ 
457.375964 & $4.498 \pm 0.022$ & $13.57 \pm 0.04$ & $-0.019 \pm 0.013$ \\ 
461.373561 & $4.474 \pm 0.019$ & $13.53 \pm 0.04$ & $-0.015 \pm 0.013$ \\ 
464.398055 & $4.467 \pm 0.018$ & $13.65 \pm 0.04$ & $-0.038 \pm 0.025$ \\ 
468.347662 & $4.491 \pm 0.023$ & $13.50 \pm 0.05$ & $-0.019 \pm 0.013$ \\ 
470.359520 & $4.475 \pm 0.023$ & $13.57 \pm 0.05$ & $-0.022 \pm 0.016$ \\ 
472.395563 & $4.496 \pm 0.022$ & $13.61 \pm 0.05$ & $-0.005 \pm 0.022$ \\ 
474.389298 & $4.496 \pm 0.025$ & $13.59 \pm 0.05$ & $-0.013 \pm 0.014$ \\ 
477.356478 & $4.476 \pm 0.022$ & $13.56 \pm 0.05$ & $-0.022 \pm 0.016$ \\ 
483.423217 & $4.491 \pm 0.019$ & $13.59 \pm 0.04$ & $0.000 \pm 0.017$ \\ 
485.346034 & $4.471 \pm 0.017$ & $13.56 \pm 0.04$ & $-0.006 \pm 0.013$ \\ 
487.313429 & $4.451 \pm 0.017$ & $13.57 \pm 0.04$ & $-0.017 \pm 0.016$ \\ 
490.311946 & $4.462 \pm 0.021$ & $13.61 \pm 0.04$ & $-0.010 \pm 0.014$ \\ 
494.296097 & $4.467 \pm 0.020$ & $13.56 \pm 0.04$ & $-0.005 \pm 0.014$ \\ 
505.314133 & $4.455 \pm 0.023$ & $13.65 \pm 0.05$ & $-0.015 \pm 0.015$ \\ 
513.442386 & $4.371 \pm 0.026$ & $13.15 \pm 0.05$ & $-0.001 \pm 0.040$ \\ 
526.324052 & $4.494 \pm 0.023$ & $13.53 \pm 0.05$ & $-0.021 \pm 0.018$ \\

\noalign{\smallskip}
\hline
\end{tabular}
\end{sidewaystable*}

\section{\textit{TESS} light curves and radial velocity joint fit}

\begin{figure}
    \centering
    \includegraphics[width=\hsize]{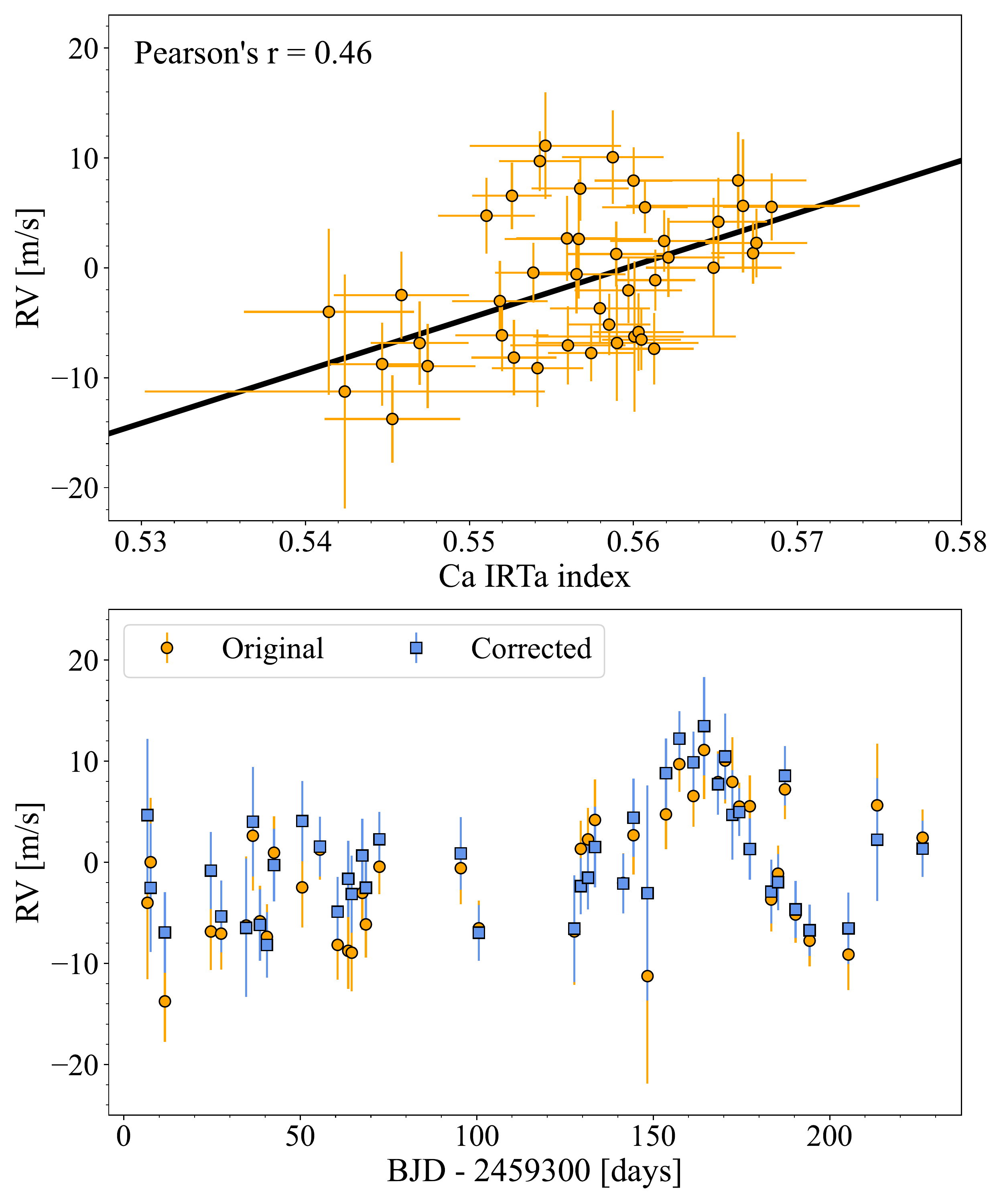}
    \caption{Radial velocity and Ca~{\sc ii} stellar activity index correction. \textit{Top panel:} RV measurements versus Ca IRTa activity index, the black line represents a linear fit to the data. \textit{Bottom panel:} Uncorrected (orange circles) and activity-corrected (blue squares) RV measurements.}
    \label{Fig:RV_vs_CaIndex}
\end{figure}

\begin{figure*}[htbp]
   \centering
   \includegraphics[width=\hsize]{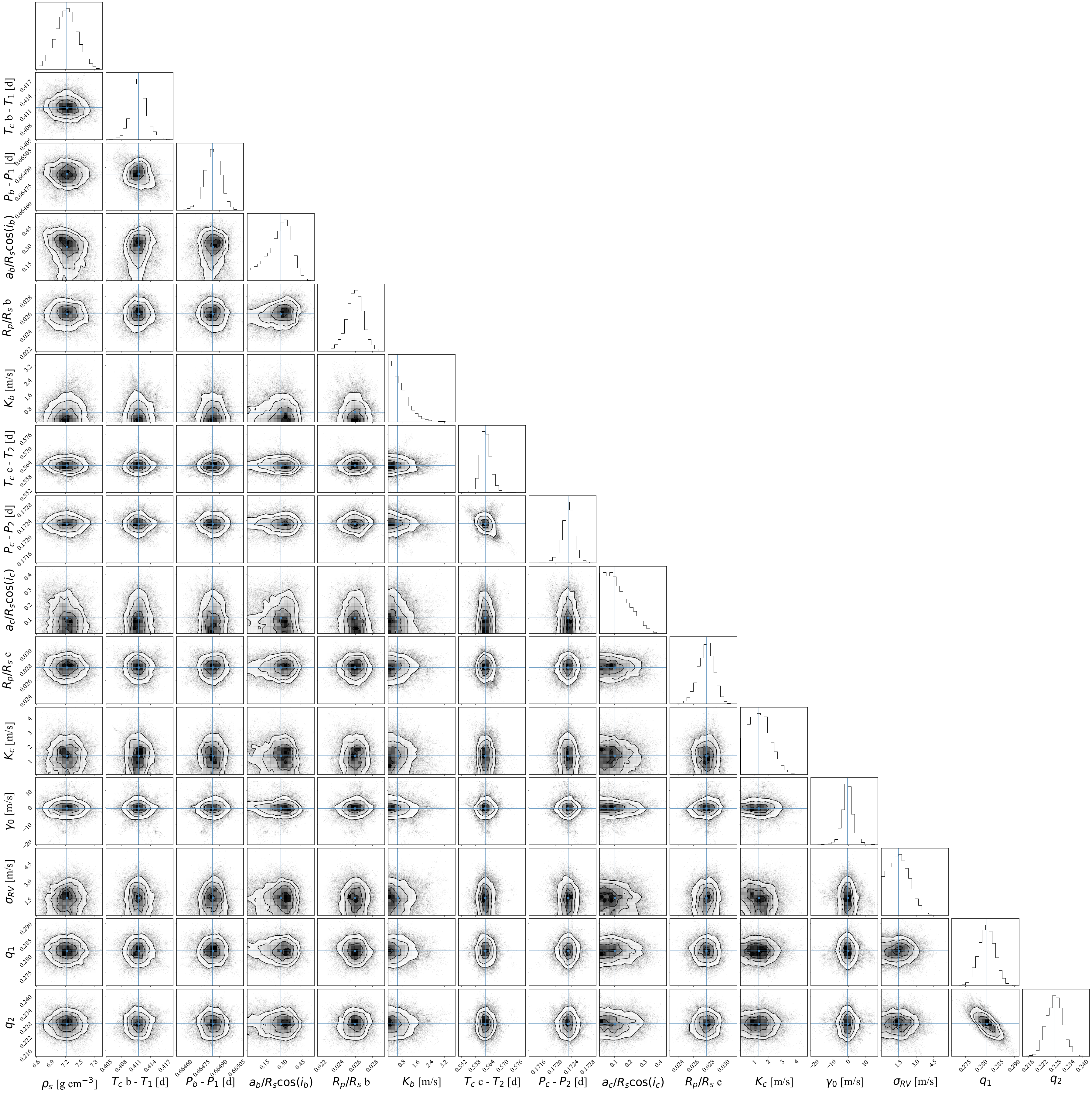}
   \caption{Correlation plot for the fitted transit and orbital parameter. The free parameters used to model the systematic effects were intentionally left out for easy viewing. The blue lines marks the median values of the distribution. For plotting purposes the distributions for the central time of the transit and orbital period were offset by $T_1 = 2459240$ days and $P_1=17$ days for TOI-2095b and $T_2 = 2459239$ days and $P_2=28$ days for TOI-2095c, respectively.}
    \label{Fig:Fit_ParamDistr_CornerPlot}
\end{figure*}

\begin{figure*}[htbp]
   \centering
   \includegraphics[width=\hsize]{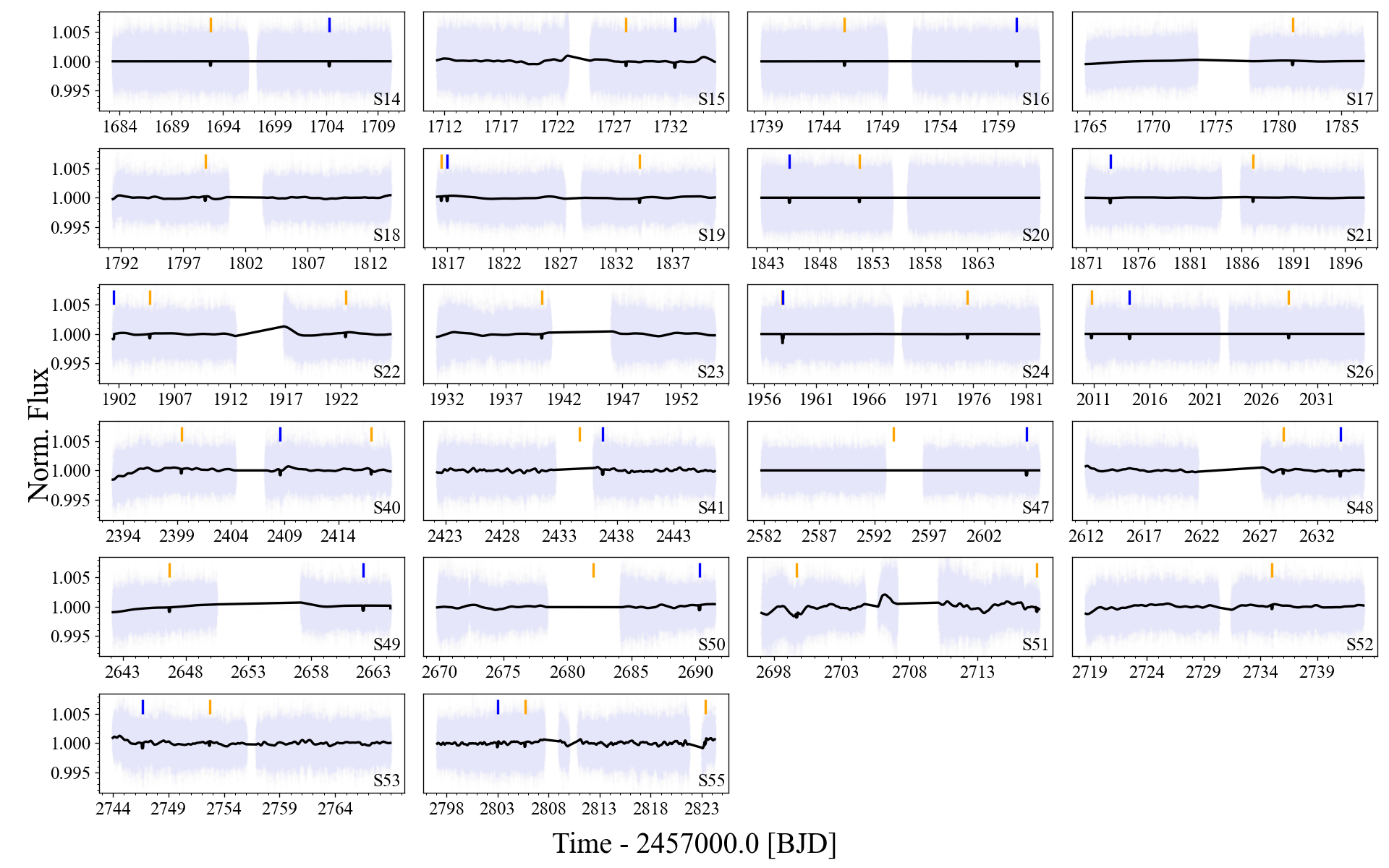}
   \caption{\textit{TESS} photometric observations of TOI-2095 and best fitting transit model including systematic effects (black line). The central transit time of TOI-2095b and TOI-2095c are marked by the orange and blue vertical lines, respectively.}
    \label{Fig:Fit_AllTESS_TransitFit}
\end{figure*}

Figure \ref{Fig:RV_vs_CaIndex} shows the correlation between the CaII activity index and the RV measurements taken with CARMENES. Figure \ref{Fig:Fit_ParamDistr_CornerPlot} shows the parameter distributions for the joint fit, where the GP parameters for the transit and RV fit were left out intentionally. Figure \ref{Fig:Fit_AllTESS_TransitFit} shows the photometry from the \textit{TESS} sectors analyzed in this work and the best transit model including systematic effects. Tables \ref{Tab:TTV_TOI2095b} and \ref{Tab:TTV_TOI2095c} present the central transit times for TOI-2095b and TOI-2095c measured with \texttt{PyTTV}, respectively.

\begin{figure*}
    \centering
    \includegraphics[width=\hsize]{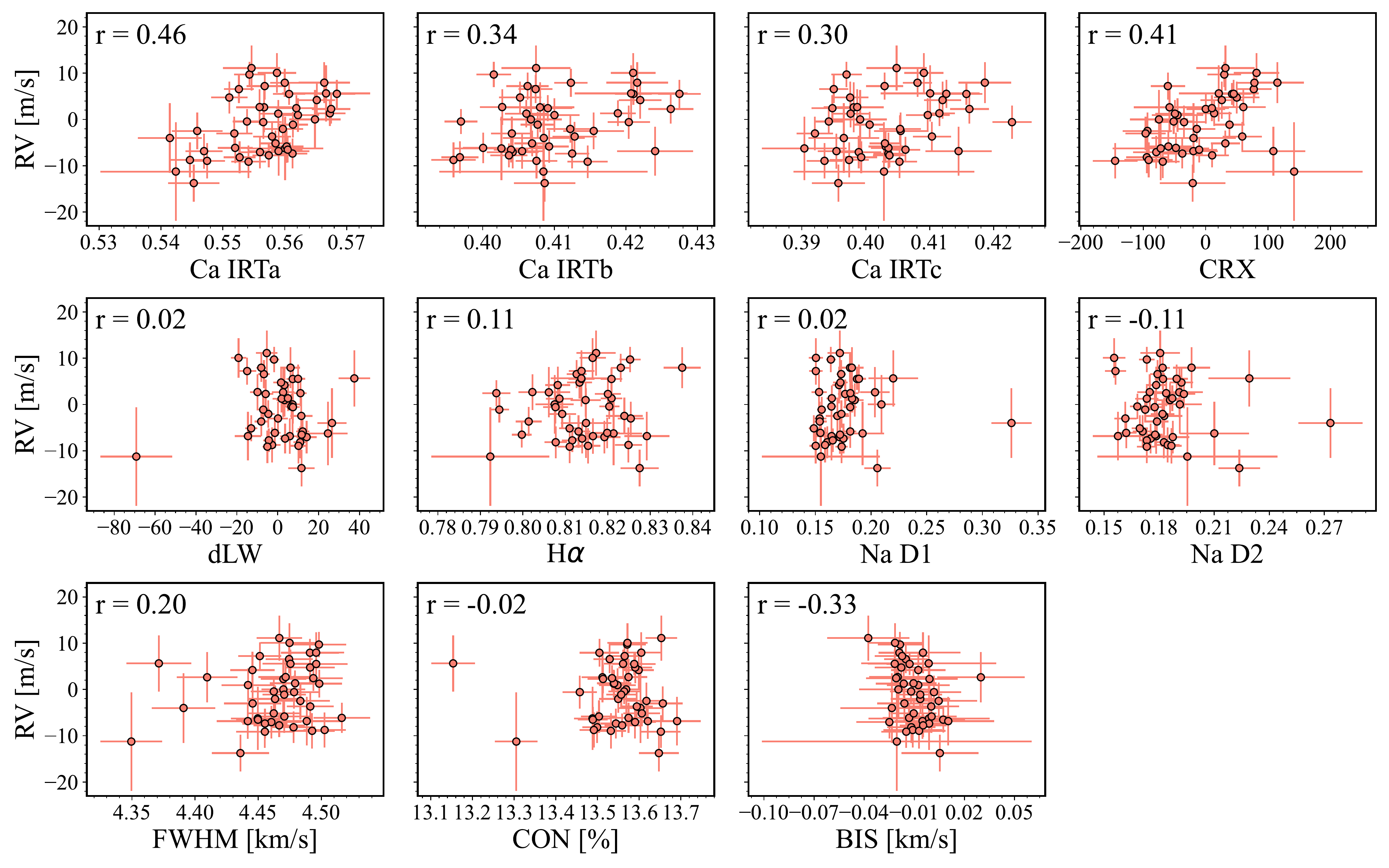}
    \caption{Radial velocity measurements versus activity indices. In each panel, the Pearson product-moment correlation coefficient $r$ is shown.}
    \label{Fig:RV_vs_ActivIndx}
\end{figure*}

\begin{table}[]
\caption{Measured central time of transit and uncertainties for \textit{TESS} data of TOI-2095b.}
\label{Tab:TTV_TOI2095b}
\centering
\begin{tabular}{c}
\hline 
\hline
\noalign{\smallskip}
$T_c -- 2\,457\,000$ [BJD]  \\
\noalign{\smallskip}
\hline
\noalign{\smallskip}

$ 1692.8287^{+0.0113}_{-0.0096} $ \\
$ 1728.1363^{+0.0053}_{-0.0104} $ \\
$ 1745.7950^{+0.0077}_{-0.0107} $ \\
$ 1781.1385^{+0.0111}_{-0.0117} $ \\
$ 1798.7990^{+0.0143}_{-0.0132} $ \\
$ 1816.4564^{+0.0119}_{-0.0201} $ \\
$ 1834.1363^{+0.0041}_{-0.0072} $ \\
$ 1851.7607^{+0.0116}_{-0.0069} $ \\
$ 1887.1231^{+0.0137}_{-0.0086} $ \\
$ 1904.7716^{+0.0055}_{-0.0063} $ \\
$ 1922.4601^{+0.0080}_{-0.0072} $ \\
$ 1940.0984^{+0.0085}_{-0.0135} $ \\
$ 1957.7628^{+0.0098}_{-0.0142} $ \\
$ 1975.4116^{+0.0110}_{-0.0069} $ \\
$ 2010.7736^{+0.0100}_{-0.0081} $ \\
$ 2028.4185^{+0.0081}_{-0.0104} $ \\
$ 2399.3933^{+0.0073}_{-0.0085} $ \\
$ 2417.0779^{+0.0096}_{-0.0124} $ \\
$ 2629.0278^{+0.0064}_{-0.0104} $ \\
$ 2646.6803^{+0.0113}_{-0.0182} $ \\
$ 2717.3620^{+0.0123}_{-0.0157} $ \\
$ 2735.0098^{+0.0102}_{-0.0050} $ \\
$ 2752.6802^{+0.0110}_{-0.0073} $ \\
$ 2805.6873^{+0.0056}_{-0.0035} $ \\

\noalign{\smallskip}
\hline
\end{tabular}
\end{table}

\begin{table}[]
\caption{Measured central time of transit and uncertainties for \textit{TESS} data of TOI-2095c.}
\label{Tab:TTV_TOI2095c}
\centering
\begin{tabular}{c}
\hline 
\hline
\noalign{\smallskip}
$T_c -- 2\,457\,000$ [BJD]  \\
\noalign{\smallskip}
\hline

$ 1704.2654^{+0.0069}_{-0.0069} $ \\
$ 1732.4526^{+0.0072}_{-0.0055} $ \\
$ 1760.6394^{+0.0146}_{-0.0096} $ \\
$ 1816.9818^{+0.0163}_{-0.0099} $ \\
$ 1845.1154^{+0.0068}_{-0.0062} $ \\
$ 1873.3285^{+0.0069}_{-0.0069} $ \\
$ 1901.4633^{+0.0148}_{-0.0126} $ \\
$ 1957.8390^{+0.0053}_{-0.0060} $ \\
$ 2014.1860^{+0.0124}_{-0.0057} $ \\
$ 2408.6072^{+0.0107}_{-0.0096} $ \\
$ 2436.7797^{+0.0118}_{-0.0100} $ \\
$ 2605.8111^{+0.0221}_{-0.0090} $ \\
$ 2633.9663^{+0.0188}_{-0.0098} $ \\
$ 2662.1473^{+0.0073}_{-0.0075} $ \\
$ 2690.3154^{+0.0095}_{-0.0144} $ \\
$ 2746.6724^{+0.0104}_{-0.0136} $ \\
$ 2802.9917^{+0.0112}_{-0.0151} $ \\

\hline
\end{tabular}
\end{table}

\end{appendix}


\end{document}